\documentclass[Style/sn-mathphys]{Style/sn-jnl}
\theoremstyle{thmstyleone}%
\theoremstyle{thmstyletwo}%

\theoremstyle{thmstylethree}%

\graphicspath{{Figures/}}
\usepackage{amsmath}

\DeclareMathOperator*{\argmin}{arg\,min}

\usepackage{etoolbox,xspace}
\newcommand{\definetextttcmd}[1]{\csdef{#1}{\texttt{#1}\xspace}}
\forcsvlist{\definetextttcmd}{PORTALS, CGYRO, TGLF, TGYRO, NEO, BoTorch, GPyTorch, ASTRA, TRANSP, GX, TANGO, GENE, TRINITY, QuaLiKiz, PyTorch, GKW, NCLASS}

\newcommand{\referee}[1]{{\color{black}#1}}

\begin{document}

\title[Surrogate-based optimization in core transport solvers]{Enhancing predictive capabilities in fusion burning plasmas through surrogate-based optimization in core transport solvers}

\author*[1]{\fnm{P.} \sur{Rodriguez-Fernandez}}\email{pablorf@mit.edu}

\author[1]{\fnm{N. T.} \sur{Howard}}

\author[1]{\fnm{A.} \sur{Saltzman}}

\author[1]{\fnm{S.} \sur{Kantamneni}}

\author[2]{\fnm{J.} \sur{Candy}}

\author[3]{\fnm{C.} \sur{Holland}}

\author[4]{\fnm{M.} \sur{Balandat}}

\author[4]{\fnm{S.} \sur{Ament}}

\author[1]{\fnm{A. E.} \sur{White}}


\affil[1]{\orgdiv{MIT Plasma Science and Fusion Center}, \city{Cambridge}, \state{MA}, \country{United States}}
\affil[2]{\orgdiv{General Atomics}, \city{San Diego}, \state{CA}, \country{United States}}
\affil[3]{\orgdiv{University of California San Diego}, \city{La Jolla}, \state{CA}, \country{United States}}
\affil[4]{\orgdiv{Meta, USA}, \city{Menlo Park}, \state{CA}, \country{United States}}

\abstract{
This work presents the \PORTALS framework \cite{rodriguez-fernandez_nonlinear_2022}, which leverages surrogate modeling and optimization techniques to enable the prediction of core plasma profiles and performance with nonlinear gyrokinetic simulations at significantly reduced cost, with no loss of accuracy.
The efficiency of \PORTALS is benchmarked against standard methods, and its full potential is demonstrated on a unique, simultaneous 5-channel (electron temperature, ion temperature, electron density, impurity density and angular rotation) prediction of steady-state profiles in a DIII-D ITER Similar Shape plasma with GPU-accelerated, nonlinear \CGYRO \cite{Candy2016}.
This paper also provides general guidelines for accurate performance predictions in burning plasmas and the impact of transport modeling in fusion pilot plants studies.
}

\keywords{
PORTALS, Bayesian, Gyrokinetics, Optimization, Multi-Scale, Surrogate
\\
\\
\\
\textbf{Submitted for publication to}: \textit{Nucl. Fusion}
}

\maketitle

\section{Introduction}
\label{sec:Intro}

As we approach the operation of magnetic confinement burning-plasma experiments \cite{Doyle2007,Rodriguez-Fernandez2022} and as reactor design studies are becoming widespread in the fusion energy community \cite{frank_radiative_2022,meschini_review_2023, Menard2022,doe_milestone}, the need for reliable, fast and accurate physics models of the confined plasma becomes essential to realize economically-attractive fusion energy. Particularly, accurate physics models for the transport of energy, particles and momentum in the confined plasma are exceptionally important. Fusion power production and reactor efficiency strongly depend on the core pressure gradients that are attained within the operational space of the fusion device. These gradients in kinetic profiles are determined by a balance of the energy, particle and torque input and turbulent and collisional transport processes.

The nonlinear gyrokinetic framework for turbulent transport \cite{Brizard2007,krommes_nonlinear_2010} is the gold standard for a rigorous description of micro-turbulence in the plasma core. Gyrokinetic theory uses the ordering parameter $\delta \doteq \rho_i/a$ (where $\rho_i$ is the thermal ion gyroradius and $a$ is the plasma minor radius) to average over the ion gyromotion thereby reducing the kinetic description from 6D to 5D. However, nonlinear gyrokinetic simulations that attempt to evolve the full distribution function, without further approximation, are computationally impractical for predictive modeling applications.  By a further time and space-scale ordering in powers of $\delta$ between turbulence and macroscopic evolution, one can separate the local ($\delta f$) gyrokinetic equations from the global transport equations \cite{sugama_nonlinear_1998}. Importantly, it is only in this limit that gyrokinetic and neoclassical fluxes are themselves separable \cite{sugama_transport_1996}. The community has developed sophisticated tools based on this spatio-temporal ordering. The gyrokinetic fluxes can be calculated with nonlinear time-dependent codes  such as \CGYRO \cite{Candy2016}, \GENE \cite{Jenko2000}, \GKW \cite{peeters_nonlinear_2009} or \GX \cite{mandell_gx_2022}, or with quasilinear models like \TGLF \cite{Staebler2007} or \QuaLiKiz \cite{Bourdelle2015a}. Similarly, the neoclassical fluxes can be computed with kinetic codes like \NEO \cite{Belli2008}, or moment-based models like \NCLASS \cite{Houlberg1997}.

The local neoclassical and gyrokinetic calculations are then embedded into global \textit{transport solvers} that compute self-consistent particle (fueling), momentum (torque) and energy (heating) sources to be balanced against the neoclassical and gyrokinetic losses, iterating until steady-state conditions are found. Even with the many simplifications provided by this embedded approach, the stiff behavior of turbulence results in the requirement of running many transport iterations to achieve steady-state (\textit{flux-matching}) conditions.
While it is hard to determine a priori how many flux-tube $\delta f$ simulations are required to simulate a given plasma condition, it is often the case that hundreds or even thousands of evaluations per radial location are required.
This number varies from one case to another, but generally it increases with the number of channels simulated (e.g. simultaneous prediction of temperatures and densities is more expensive than only temperatures prediction), the evolution of targets (e.g. self-consistent alpha heating introduces a nonlinearity) and the stiffness, discontinuities and non-monotonicity of the transport model.
In burning plasmas where heating is dominated by the profiles themselves via alpha heating, where density peaking comes mostly from transport processes and not sources, and with large gyro-Bohm units, the convergence of transport solvers becomes exceptionally difficult.
Beginning with \texttt{TGYRO-GYRO} \cite{Candy2009a} and \texttt{TRINITY-GS2} \cite{Barnes2010} \referee{whose pioneering studies set the foundation for work in this area}, efforts in the community are ongoing to develop advanced transport solvers (including \TANGO-\GENE \cite{siena_global_2022} and \TRINITY-\GX \cite{mandell_gx_2022}) with computationally efficient transport models that capture the physics of gyrokinetic turbulence brought to macroscopic steady-state conditions. 
\referee{The work presented here is complementary to these efforts, as it focuses on the acceleration of the convergence of transport solvers, rather than the development of the transport models themselves that are efficient and applicable to a wide range of plasmas (e.g. for global gyrokinetics and 3D geometry).}
\referee{The tokamak transport modeling community has also recently begun to employ machine learning-accelerated and data-driven models for transport predictions \cite{Meneghini2016,VanDePlassche2020,Ho2021,zanisi_efficient_2024,hornsby_gaussian_2024}. These efforts are primarily aimed at developing high-dimensional surrogate models to simulate transport fluxes and integrating them into standard transport solvers.
}

This paper presents the paradigm of re-defining the flux-matching problem encountered in transport solvers as a surrogate-based optimization problem, and presents, to our knowledge, the most comprehensive prediction of core kinetic profiles in a tokamak: a 5-channel, nonlinear flux-matched gyrokinetic prediction with self-consistent energy exchange and radiation.
Section~\ref{sec:Background} introduces the flux-matching formulation for steady-state prediction of core profiles and Section~\ref{sec:FM} presents the paradigm of re-writing the system as a physics-guided surrogate-based optimization problem (\PORTALS framework).
Section~\ref{sec:Guide} discusses guidelines for accurate profile predictions and Section~\ref{sec:5c} presents a 5-channel nonlinear gyrokinetic prediction enabled by the surrogate formulation at reduced cost, with \PORTALS-\CGYRO.
In Sections~\ref{sec:Discussion} and~\ref{sec:Conclusions}, general discussion and conclusions are presented.

\section{Background}
\label{sec:Background}

Core transport solvers in tokamak geometry aim at solving the fluid-like flux-surface averaged (FSA) 1D transport equations \cite{Rodriguez-Fernandez2019a}, where kinetic or gyrokinetic effects are only included \textit{externally} to the code, usually in the form of black-box modules for theory-based or first-principles models of turbulence and neoclassical transport.

When employing high-fidelity physics models for the nonlinear transport flux calculations, one encounters a non-analytic dependence on background profiles and, particularly, on their gradients.
This is due to the drift-wave nature of electromagnetic tokamak turbulence, which is driven unstable by pressure gradients with complex sensitivity to collisions and background plasma conditions.
This results in the need to develop robust numerical techniques that reduce oscillations, ensure convergence and minimize simulation errors \cite{Jardin2008,pereverzev_stable_2008}, otherwise standard numerical schemes to solve parabolic partial differential equations would require extremely small time steps of integration and would be exceptionally time consuming.
Even though the time step required to ensure convergence of transport solvers has been reduced by orders of magnitude thanks to the implementation of numerical diffusivities \cite{pereverzev_stable_2008} (e.g. in \ASTRA \cite{Pereverzev2002}) and internal Newton iterations \cite{Jardin2008} (e.g. in \TRANSP \cite{Breslau2018}), calls to the transport model are still in the order of a hundred per radial location, making the use of first-principles turbulence simulations unfeasible, at least for routine analysis and study of turbulence physics.
Furthermore, the requirement to include grid-scale numerical diffusion in these solvers requires a careful analysis of convergence, robustness and stability for each application.

When the time dynamics is not important or when the goal of the simulation exercise is to produce a steady-state solution, time independent transport solvers are capable of predicting core kinetic profiles with lower computational expense.
This is achieved by re-defining the system of equations as an inverse problem \cite{Candy2009a}, where the kinetic profiles are the free parameters and the difference between transport fluxes and target fluxes is minimized below some convergence criterion.
The system of equations that is often solved in time independent solvers is obtained by integrating over the interior of each flux surface (see Fig~\ref{Fig:rendering} for a visual representation of flux surfaces) and removing any time dependence of the macroscopic transport:
\begin{align}
    \label{eq:FSA}
    \begin{split}
    \text{electron density}\rightarrow    \langle \mathbf{\Gamma}_e\cdot\nabla r\rangle &=\frac{1}{V'} \int_{0}^{r}\langle S_e \rangle V' dr \\
    \text{electron energy}\rightarrow    \langle \mathbf{q}_e\cdot\nabla r\rangle &= \frac{1}{V'}\int_{0}^{r} \Big( \langle P_{\mathrm{aux},e}\rangle +  \langle P_{\alpha,e}\rangle-  \langle P_{ei}\rangle -  \langle P_{\mathrm{rad}} \rangle \Big) V' dr \\
    \text{ion energy}\rightarrow    \langle \mathbf{q}_i\cdot\nabla r\rangle &=  \frac{1}{V'}\int_{0}^{r} \Big(\langle P_{\mathrm{aux},i}\rangle +  \langle P_{\alpha,i}\rangle +  \langle P_{ei}\rangle \Big)V' dr \\
    \text{momentum}\rightarrow    \langle \mathbf{\Pi}\cdot\nabla r\rangle &=\frac{1}{V'} \int_{0}^{r}\langle S_\omega \rangle V' dr\\
    \text{impurity density}\rightarrow    \langle \mathbf{\Gamma}_Z\cdot\nabla r\rangle &=0
    \end{split}
\end{align}
where $r$ is the half-width of the mid-plane intercept, $\langle\cdot\rangle$ indicates flux-surface averaging, and $V'=\partial V /\partial r$ where $V$ is the flux surface volume.
$S_e$ represents particle sources, $S_\omega$ represents momentum torque input and the $P_X$ represent the different components (sources and sinks) of energy in the plasma.

Eq.~\ref{eq:FSA} clearly shows how coupled the system of FSA equations is. Every transport flux component ($\mathbf{\Gamma}_e$, $\mathbf{q}_e$, $\mathbf{q}_i$, $\mathbf{\Pi}$, $\mathbf{\Gamma}_Z$), depends on the state of background turbulence, which is in turn determined by each of the kinetic profiles under consideration. Furthermore, target fluxes such as energy exchange ($\langle P_{ei}\rangle$), radiation ($\langle P_{rad}\rangle$) and alpha power ($\langle P_{\alpha,e}\rangle$, $\langle P_{\alpha,i}\rangle$) strongly depend on the profiles themselves.
Here it is assumed that the momentum flux, $\langle \mathbf{\Pi}\cdot\nabla r\rangle$, includes the residual stress, and that there are not intrinsic sources of the impurity of interest (hence the null-flux condition).
\referee{In the reminder of this paper, auxiliary heating power delivered to each species is assumed to be fixed throughout the simulations and therefore must be obtained by external means, such as interpretive simulations with \TRANSP or \ASTRA, and assumed constant with the variations of the kinetic profiles from initial to steady-state conditions. This is a current limitation of the framework, but it is expected to be addressed in future work.
The rest of target flux components are calculated self-consistently with the kinetic profiles. Details on these calculations are available in Ref.~\cite{rodriguez-fernandez_nonlinear_2022}.}

The process to bring an initial set of kinetic profiles to steady-state conditions when external heating, fueling and torque sources are imposed, requires a number of local, $\delta f$ turbulence simulations to provide the transport fluxes at each iteration.
Both in time dependent (with a criterion that accounts for the stationarity of the predicted quantities) and time independent (with a criterion based on the closeness between transport and target fluxes) solvers, the information of previous simulations is often not used, including the local Jacobian at each iteration.
The paradigm presented here proposes that a \textit{global} surrogate model for each of the transport fluxes is constructed and refined during the convergence process. All the simulations, including the ones far from convergence, are utilized to train the surrogate model and build the dependencies of transport fluxes with respect to the free parameters of the problem.

\section{Flux-matching as a surrogate-based optimization problem}
\label{sec:FM}

\begin{figure}
    \centering
    \includegraphics[width=1.0\columnwidth]{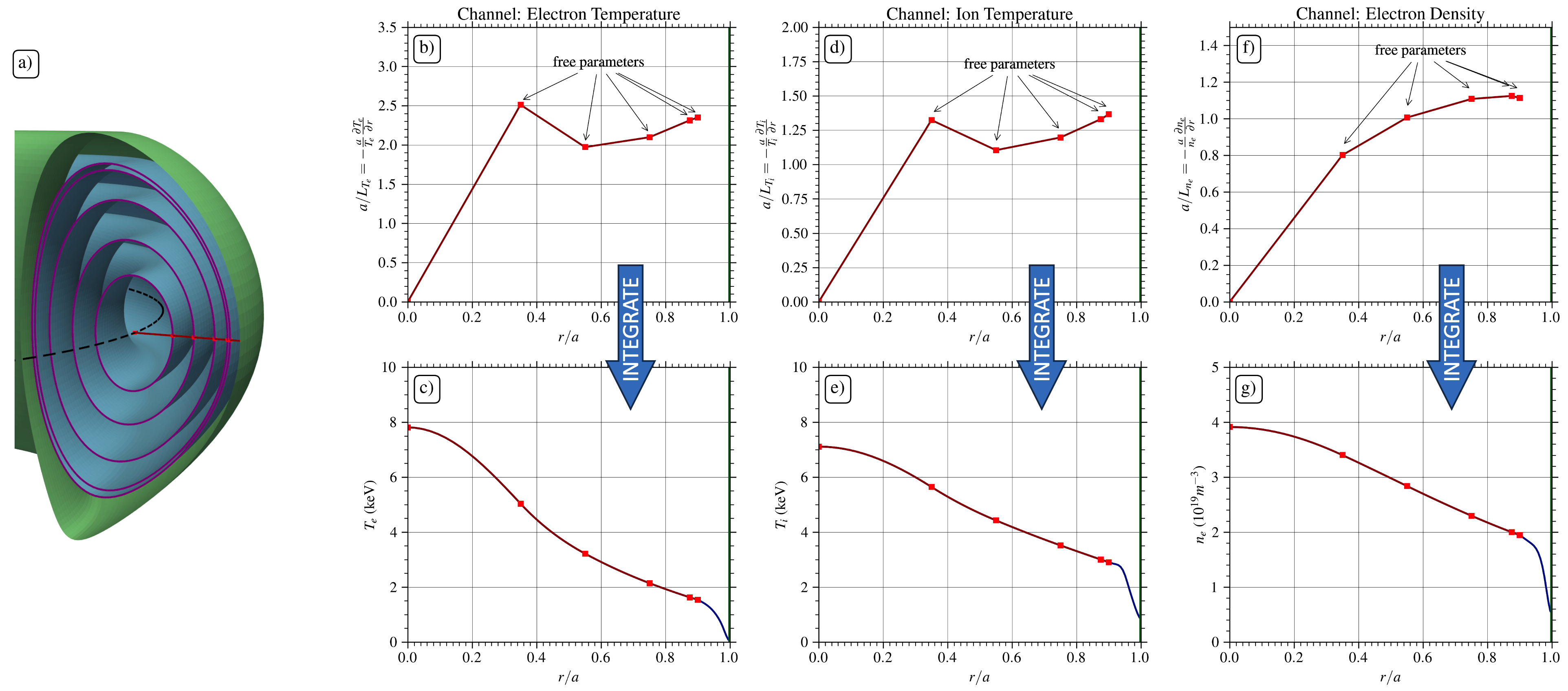}
    \caption{
    Visualization of problem geometry and free parameters of a three-channel prediction.
    a) Illustration of nested flux surfaces for a DIII-D ITER Similar Shape plasma \cite{howard_simultaneous_2024}, with magnetic axis in dashed black line, 5 selected flux surfaces in blue (cross section in purple) and last closed flux surface in green.
    Note that the lower x-point region is smoothed out due to Fourier-moments decomposition typical of transport solvers such as \TRANSP \cite{Breslau2018}.
    Subplots b), d) and f) depict the piece-wise linear function representations of the normalized logarithmic gradient of electron temperature, ion temperature and electron density with the value at 5 selected flux surfaces (values to be predicted by the transport solver).
    Subplots c), e) and g) depict the corresponding kinetic profiles of each channel, produced by radial integration of logarithmic gradient from an edge boundary condition (dark blue).
    }
    \label{Fig:rendering}
\end{figure}

The set of Equations~\ref{eq:FSA} can be generalized and written as a multi-residual minimization problem:
\begin{align}
    \label{eq:Minimization}
    \begin{split}
    &R_{j,c} = F^{\text{tr}}_{j,c}(y_{j,\forall c},\nabla y_{j,\forall c}) -
    \overbrace{\frac{1}{V'_{j}}
    \int_{0}^{r_j}
        f^{\text{target}}_{j,c}(y_{j,\forall c},\nabla y_{j,\forall c}) V'(r) dr}
    ^{ F^{\text{target}}_{j,c}(y_{r\leq r_j, \forall c},\nabla y_{r\leq r_j,\forall c})} \\
    &\xi=\frac{1}{N_c\cdot N_\rho}\sqrt{{\sum_{j,c}R_{j,c}^2}}\\
    &\textbf{y}^* = \argmin_{y_{j,c}\epsilon [y^L_{j,c},y^U_{j,c}]}
    \xi
    \end{split}
\end{align}
where $F^{\text{tr}}_{j,c}$ is representative of the transport flux of channel $c$ at radial location $r_j$, which is a function of the local values of all the channels, $y_{j,\forall c}=\{n_e(r_j),T_e(r_j),T_i(r_j),\omega_0(r_j),n_Z(r_j)\}$, and their local gradients, $\nabla y_{j,c}\equiv\frac{\partial y_c}{\partial r}\rvert_{r_j}$.
Similarly, $f^{\text{target}}_{j,c}$ represents the target flux density for channel $c$ at radial location $r_j$, which is a function of the local values of all the channels. For generalization purposes, we also assume that the target density can be a function of local gradients, for example in the case of turbulent energy exchange \cite{Candy2013}.
It is important to note that the target fluxes $F^{\text{target}}_{j,c}$ at flux surface $r_j$ are non-local due to volume integration, but the transport fluxes $F^{\text{tr}}_{j,c}$ are fully defined locally, a consequence of the local approximation of turbulence.
The number of radial points chosen to represent the logarithmic gradient profiles (piece-wise linear functions as illustrated in Fig.~\ref{Fig:rendering}) is $N_\rho$, the number of channels to simulate simultaneously and self-consistently is $N_c$, and the number of model evaluations is $N_m$.

Equations~\ref{eq:Minimization} represent a minimization problem of the scalarized, multi-channel residual, $\xi$, with each channel profile, $y_{j,c}$, free to evolve.
The set of radial profiles that minimize the residual, $y^*_c(r)$, defines the steady-state plasma with self-consistent transport and targets and represents the solution to the problem.

With a fixed boundary condition for each channel, $y_{b,c}=y_c(r=r_b)$, the gradients and the local values of the profiles are linked via radial integration.
Under the framework of tokamak FSA transport equations, it is convenient to define the normalized logarithmic gradients, $z_{c}=a/L_{y_c}=-\frac{a}{y_c}\frac{\partial y_c}{\partial r}$, as the free parameters of the problem.
This way, the logarithmic gradients (for each channel $c$ and radial location $j$) uniquely define the profiles \cite{Candy2009a} and the use of a normalized flux surface label, $\rho=r/a$, can be introduced readily:
\begin{subequations}
\label{eq:MinimizationFull}
\begin{align}
    &R_{j,c} = F^{\text{tr}}_{j,c}(z_{j,\forall c},y_{j,\forall c}) -
    \overbrace{
    \frac{1}{\frac{\partial V}{\partial r}\lvert_j}
    \int_{0}^{\rho_j} 
        f^{\text{target}}_{j,c}(z_{j,\forall c},y_{j,\forall c}) \frac{\partial V}{\partial\rho}(\rho)
    d\rho
    }
    ^{ F^{\text{target}}_{j,c}(z_{r\leq r_j, \forall c},y_{r\leq r_j,\forall c})} 
    \label{eq:MinimizationFull_R}\\
    &y_{j,c} = y_{c,b}\exp\left({\int_\rho^{\rho_b}z_{c}(\zeta)d\zeta}\right)\label{eq:MinimizationFull_x}\\
    &\xi=\frac{1}{N_c\cdot N_\rho}\sqrt{{\sum_{j,c}R_{j,c}^2}}\label{eq:MinimizationFull_xi}\\
    &\textbf{z}^*= \argmin_{z_{j,c}\epsilon [z^L_{j,c},z^U_{j,c}]} \xi
    \label{eq:MinimizationFull_z}
\end{align}
\end{subequations}

As it will be described in Section~\ref{sec:Bench}, standard numerical techniques such as Newton methods to solve this problem do not scale well for high $N_c$ (number of channels), as the local Jacobian must be estimated from finite differences methods and require several transport model evaluations per iteration, even in the case of assuming a block-diagonal Jacobian.
Leveraging advances in machine learning and statistics, here we adapt a standard Bayesian optimization workflow to solve for the flux-matching of multi-channel, steady-state transport as formulated in Equations~\ref{eq:MinimizationFull}.
The evaluation of the residual, $\xi$, requires the evaluation of turbulence and neoclassical transport fluxes and the target flux densities at each radial location.
When first-principles, nonlinear gyrokinetics is used, each residual evaluation (Eq.~\ref{eq:MinimizationFull_R}) requires $N_\rho$ simulations brought to saturation, which can quickly become computationally expensive if many evaluations $N_m$ are required to solve Eq.~\ref{eq:MinimizationFull_z}.
This is when Bayesian optimization techniques can help reduce the computational cost of the flux-matching problem.

\paragraph{Bayesian Optimization}

Instead of starting from a single initial condition, in Bayesian optimization, an initial set of training data (model evaluations) is used to fit a probabilistic \emph{surrogate model}\footnote{In this context, a surrogate model refers to an analytical expression or a reduced model that is much faster to evaluate but that captures key parametric dependencies} for the metric to minimize, usually a Gaussian process (GP) model. 
In the context of flux-matching solvers, this initial training phase consists of running a set of local, $\delta f$ turbulence simulations with variations in the free parameters (logarithmic gradients, $z_c$, but with consistent changes in the profile values, $y_c$). The information from these local simulations is used to fit GP models to reproduce the transport fluxes v.s. free parameters behavior, and to reconstruct a model for the residual, $\xi$.

The GP model, trained on all thus-far observed data points, provides a prediction for the posterior distribution of the residual, and is used to find the optimal points to evaluate next with the expensive turbulence simulations.
This phase requires the definition and optimization of an \emph{acquisition function}, i.e. a function that quantifies the value of evaluating the costly model at any given point in the domain for optimizing the outcome (in this context the residual). 
The work presented here makes use of the \GPyTorch \cite{Gardner} and  \BoTorch \cite{balandat_botorch_2020} packages for GP modeling and Bayesian optimization, respectively. 
The reader interested in the details on Bayesian optimization or GP theory and implementation is referred to Refs.~\cite{balandat_botorch_2020,Gardner,Frazier2018,Preuss2018,Brochu2010}.

The implementation of the flux-matching problem of $\delta f$ transport models as a surrogate-based optimization problem to leverage Bayesian optimization methods is done within the \PORTALS\footnote{``Performance Optimization of Reactors via Training of Active Learning Surrogates"} framework \cite{rodriguez-fernandez_nonlinear_2022} and the rest of this paper describes implementation details that were needed to achieve a significant speed-up with respect to standard methods, which enable some of the highest fidelity predictions of core kinetic profiles performed to date.

\referee{
For clarification, we must note that the results obtained with \PORTALS are not results of surrogate modeling of turbulence and transport, but of \textit{direct} turbulence modeling (either with quasilinear gyrofluid \TGLF or nonlinear gyrokinetic \CGYRO for the cases shown in this paper).
Surrogates are simply used to accelerate multi-channel convergence, but the steady-state predictions of profiles are a full-physics result and must be interpreted as such.
}

\subsection{Implementation of domain-information}
\label{sec:DI}

Instead of using the full residual, $\xi$, as both the objective function (Eq.~\ref{eq:MinimizationFull_z}) and the surrogate output, here we choose the divide $\xi$ in each component (channel and radius) as defined in Eq.~\ref{eq:MinimizationFull_R} to construct a composite surrogate model.
This ensures that each transport surrogate is only fit to the corresponding local input parameters, significantly reducing the dimensionality of the problem.
While $\xi$ is a nonlinear function of $N_c\cdot N_\rho$ variables, each transport flux $F^{\text{tr}}_{j,c}(z_{j,\forall c},y_{j,\forall c})$ is only a function of $2\cdot N_c$ variables (local gradients $z_{j,\forall c}$ and local profile values $y_{j,\forall c}$).
Therefore, constructing individual surrogates for each flux component is very advantageous to minimize the number of iterations to achieve accurate enough surrogate models within the Bayesian optimization workflow. 
This is essentially an application of \emph{composite Bayesian Optimization} \cite{Astudillo2019}.

\subsubsection{Input space transformation}
\label{sec:Input}

Each of the transport fluxes includes neoclassical and turbulent fluxes, $F^{\text{tr}}_{j,c} = F^{\text{neocl.}}_{j,c} + F^{\text{turb.}}_{j,c}$, which are evaluated in separated simulations, and also fitted by separate surrogate models.
By using index $m$ to represent the turbulent ($F^{\text{turb.}}$) or neoclassical ($F^{\text{neocl.}}$) fluxes, the aim of the surrogates is to reproduce the $2\cdot N_c$ dimensional functions (local in space, $j$, but with multi-channel dependencies, $\forall c$):
\begin{align}
\label{eq:2Nc}
    F^m_{j,c}(z_{j,\forall c},y_{j,\forall c})
\end{align}

While the normalized logarithmic gradients $z_{j,c}$ are suitable as direct input parameters to reproduce transport fluxes, the profile values $y_{j,\forall c}=\{n_e(r_j),T_e(r_j),T_i(r_j),\omega_0(r_j),n_Z(r_j)\}$ can be transformed into variables that have a more distinct effect on turbulent and neoclassical transport.
Defining the problem with $\widehat{y}_{j,\forall c}=\{\widehat{\nu}_{ei}(r_j),T_i/T_e(r_j),\beta_e(r_j),\omega_0/c_s(r_j),n_Z/n_e(r_j)\}$, while still retaining a complete set, allows the surrogates to be directly fitted to quantities of interest and of direct effect to the turbulence dynamics, which aids in the training of surrogate models.
Here, $c_s=f(T_e)$ is the ion sound speed,
$\widehat{\nu}_{ei}=f(T_e,n_e)$ represents the electron-ion collision frequency normalized to the ion sound speed, and $\beta_e=f(T_e,n_e)$ is the magnetic-field normalized electron pressure.

\referee{We note that in the current implementation, separate surrogate models for turbulent and neoclassical fluxes are used, but the framework is flexible enough to include a single model for both, if the user is interested in a more general approach.
In cases where the neoclassical fluxes are of high fidelity (e.g. with the \NEO code \cite{belli_eulerian_2009}), even if much cheaper than nonlinear gyrokinetic calculations, it is important to build surrogate models to their flux response. This is because their evaluation (unless analytical) will be much more expensive than a GP evaluation, and automatic differentiation would be, generally, not available; making the acquisition function optimization (the flux-matching process per se) significantly more expensive.
}

The rotation, which can go through zero throughout the radial profile, requires a special treatment, as its gradient cannot properly be defined as a normalized logarithmic gradient.
Here, we use the radial gradient normalized to the ion sound speed over minor radius: $z_{\omega_0,j}=-\frac{a}{c_s}\frac{d\omega_0}{dr}$.
This factor is proportional to the $E\times B$ shearing rate, $\gamma_{E\times B}$, and will affect all other channels via $E\times B$ shear stabilization of the background turbulence.

We note that the varying input parameters to the transport model are not limited to those in the basis, but any other variable can be derived from some combination of the fixed problem parameters and the basis parameters, such as the total pressure gradient.
However, some complex dependencies of fluxes with respect to input parameters can appear if one is not careful about synergies with fixed parameter and gyrokinetic model assumptions.
For instance, in a case where $a/L_{ne}$ is a free parameter in a plasma with two ions, if $a/L_{ni}$ is not varied self-consistently, each local gyrokinetic simulation will not be quasineutral. Similarly, if quasineutrality is achieved by compensating with one of the ions and not the other, the differing ion density gradients will affect background turbulence in a non-trivial way, making it challenging for the surrogates to describe the behavior with few training points.
In this work, we make the following choices:
1) the thermal ion species always have the same temperature as the main ions and the same logarithmic density gradient for non-trace ions,
and 2) variations of the electron density induce an equal change in the densities of the non-trace ions by keeping the same concentrations\footnote{This choice satisfies quasineutrality at all times if impurities are not evolved. In the case of using the impurity density as an additional evolving channel, there could be minor (of the order of the trace impurity concentration times its charge) deviations to quasineutrality}.

\subsubsection{Outcome space transformation}
\label{sec:Outcome}

Similarly to the input transformations, the outputs of the transport simulations are more properly fitted if transformed to gyro-Bohm units:
\begin{align}
\label{eq:GB}
    \widehat{F}^m_{j,c}(z_{j,\forall c},\widehat{y}_{j,\forall c}) = \frac{F^m_{j,c}}{G(T_{e,j},n_{e,j})}
\end{align}
where $G(T_{e,j},n_{e,j})$ is the gyro-Bohm normalization factor for each transport channel:
\begin{align}
 \begin{split}
\label{eq:GBs}
    \text{particle flux}\rightarrow    \Gamma_{GB} &= n_ec_s\left(\rho_s/a\right)^2 \\
    \text{energy flux}\rightarrow    Q_{GB} &= n_eT_ec_s\left(\rho_s/a\right)^2\\
    \text{momentum flux}\rightarrow    \Pi_{GB} &= an_eT_e\left(\rho_s/a\right)^2 \\
    \text{energy exchange}\rightarrow    S_{GB} &= n_eT_ec_s/a\left(\rho_s/a\right)^2\\
\end{split}
\end{align}
where $\rho_s$ is the ion gyroradius evaluated with the ion sound speed, $c_s$.
We note that this outcome transformation depends on the input parameters and therefore it is not fixed for all evaluations.

In this framework, the residual will be defined in real units and not gyro-Bohm normalized, which eases the convergence process when core and edge have wide variations in temperature or density (e.g. in the case of L-mode predictions from the edge to the core).
Therefore, the input and outcome transformations are both applied prior to model training and during the evaluation of the surrogate posterior, by leveraging \BoTorch's transformation methods, \texttt{botorch.models.transforms}.
After the transformation to the proper physics quantities for training (different set per surrogate model), we apply normalization (to the unit cube) and standardization (zero mean, unit variance) of inputs and outputs respectively, which eases the model hyperparameter training with \BoTorch standard priors.

Acknowledging the different units for each transport channel, for the definition of the scalar residual $\xi$ from Eq.~\ref{eq:MinimizationFull_xi} we make a units transformation for the particle transport residuals.
The electron particle flux residual enters in Eq.~\ref{eq:MinimizationFull_xi} as a convective flux component, $\widehat{\Gamma}_e=\frac{5}{2}T_e\Gamma_e$, where the factor $\frac{5}{2}$ is chosen following common approaches in transport modeling.
Impurity transport flux is also converted into a convective flux, but it is additionally divided by the original impurity density concentration to avoid dramatically different channel residuals.
These choices are ad-hoc, but provide residual definitions such that each local, channel components have similar magnitudes.

\referee{Note that since the total residual $\xi$ in~\eqref{eq:MinimizationFull_xi} is a nonlinear function of the individual residuals $R_{j,c}$, its posterior distribution under the surrogate model is non-Gaussian. We therefore} employ a Monte-Carlo based objective function to transform the posterior distributions of each local channel residual into the scalarized $\xi$ total residual. By leveraging BoTorch's sample-average approximation approach and PyTorch's auto-differentiation capabilities, we are able to back-propagate gradients through the acquisition function, objective, GP model, and transforms all the way back to the inputs, which allows us to effectively optimize the acquisition function to determine the next simulation parameters for the expensive model.

\subsubsection{Simple relaxation as initialization method}
\label{sec:SR}

Typically, Bayesian optimization workflows start with a random set of points.
Here, we leverage the fact that turbulence (dominant over neoclassical transport) is strongly affected by the normalized logarithmic gradients when written in normalized units (see Sec.~\ref{sec:Outcome}), and therefore it makes sense to ensure that gradients are properly sampled during training and that the profiles to evaluate are realistic.

Instead of random training (employed in the first \PORTALS publications \cite{rodriguez-fernandez_nonlinear_2022,howard_simultaneous_2024}), we utilize the derivative-free simple relaxation (SR) method also implemented in \TGYRO \cite{Candy2009a}.
The SR technique consists of neglecting cross-channel, cross-radius interactions and assuming that the local transport matrix is diagonal, with positive diffusion coefficients built from the relative difference between target and transport fluxes at each iteration.
Formally, the SR iteration scheme can be written as:
\begin{align}
\label{eq:SR}
    z_{j,c}^{(i+1)} = z_{j,c}^{(i)}+\eta_{j,c}\frac{F^{\text{target}}_{j,c}-F^{\text{tr}}_{j,c}}{\sqrt{\left(F^{\text{target}}_{j,c}\right)^2+\left(F^{\text{tr}}_{j,c}\right)^2}}\cdot \lvert z_{j,c}^{(i)}\rvert
\end{align}
where $\eta_{j,c}$ is an ad-hoc parameter that determines the relative step in normalized logarithmic gradients, $z_{j,c}$, to move the system towards reducing the difference between transport, $F^{\text{tr}}_{j,c}$, and target, $F^{\text{target}}_{j,c}$, fluxes.
When implementing Eq.~\ref{eq:SR} in a code such as \texttt{TGYRO}, it can also be helpful to impose a maximum value of $\delta z_{j,c}=z_{j,c}^{(i+1)} - z_{j,c}^{(i)}$ on a given iteration, which aids in smoothing the trajectory to convergence, particularly for early evaluations where there can be large mismatches in the target and model fluxes.

\begin{figure}
    \centering
    \includegraphics[width=0.9\columnwidth]{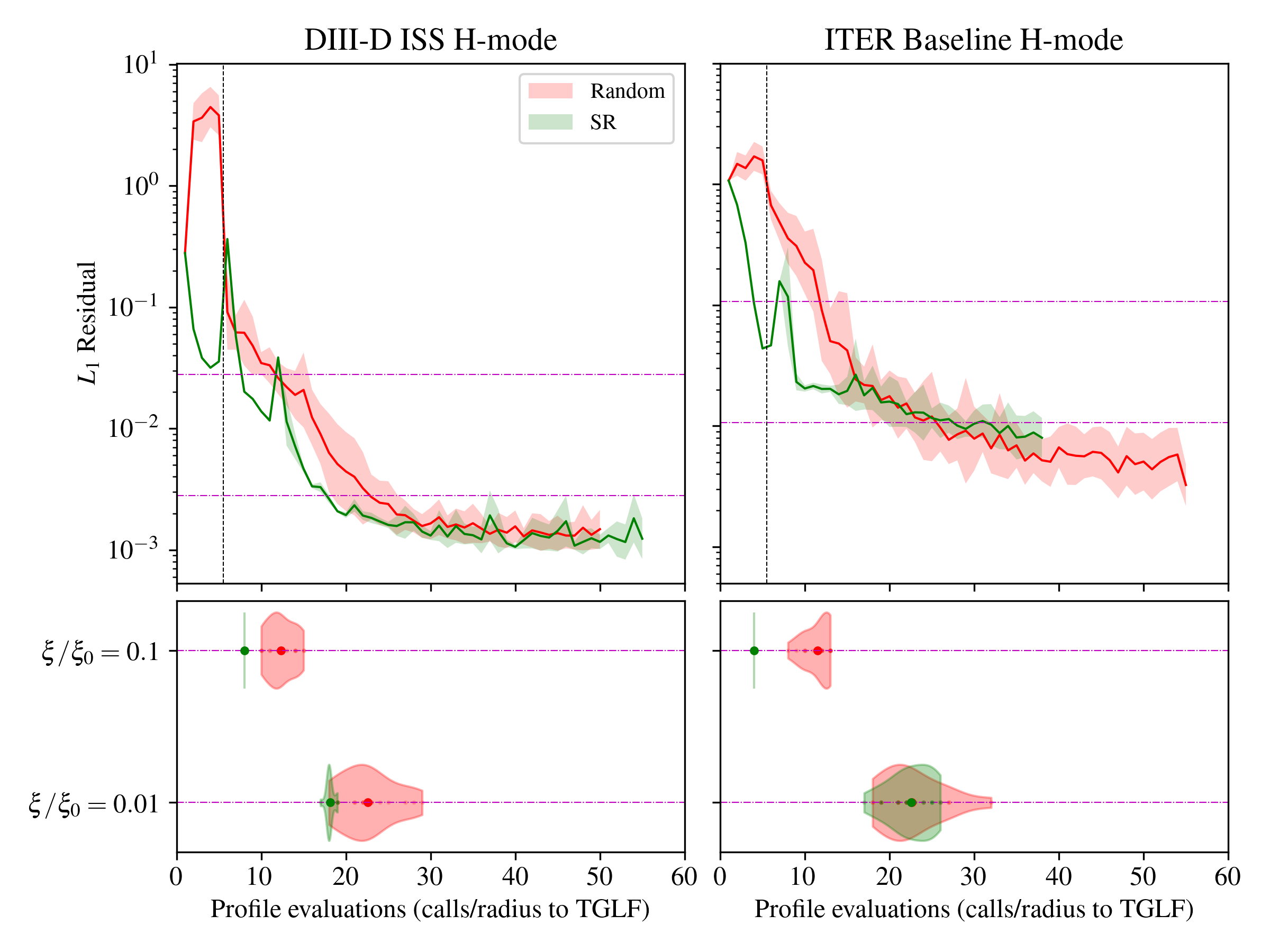}
    \caption{
    Performance of \PORTALS at achieving $10\times$ and $100\times$ residual reduction with \TGLF in two example H-mode plasmas: DIII-D ITER Similar Shape (ISS) and ITER Baseline \cite{howard_simultaneous_2024}. Random and SR initialization methods are compared, starting from the same initial condition. (top) Residual vs evaluations, with vertical line delimiting initialization phase and purple horizontal lines indicating the $10\times$ and $100\times$ residual reductions. (bottom) Violin plots representing distribution of number of evaluations that were required to achieve residual goals. Both methods were initialized with a random seed, and distributions were obtained for 16 seeds. As expected, SR is not strongly affected by the seed, particularly at the beginning of the flux matching process.
    }
    \label{Fig:violin}
\end{figure}

Performance tests were performed with the quasilinear Trapped Gyro-Landau Fluid (\TGLF) model \cite{Staebler2007}.
Our tests found that generating initial training samples with the SR method (with a constant value of $\eta_{j,c}=0.2$) was slightly more efficient in building accurate surrogates than random initialization (e.g. with Latin Hypercube sampling).
Two examples of SR vs random initialization performance are shown in Fig.~\ref{Fig:violin}.
SR is particularly efficient at reducing the residual for the first 15 evaluations, saving a few extra evaluations compared to random initialization. Generally, the benefits of SR vanish as more samples are added to the training database, as expected.

While the effect is not major in the number of total evaluations required, the SR method ensures realistic profiles, where big radial variations in logarithmic gradients or unrealistic de-coupling of the profiles (e.g., electrons and ions de-coupled in a way not consistent with collisional equilibration) are not allowed from a macroscopic transport point of view.
This eases the initial-value $\delta f$ simulations, which can otherwise be difficult and expensive to saturate if random choices of gradients are requested to be evaluated by \PORTALS, as some variations inevitably lead to unrealistically over-driven or completely stabilized turbulence. 

Apart from using SR for initialization of the first 5 profiles, we also make an adjustment to the dimensionality of the first surrogates fitted.
Although the transport models, $F^m_{j,c}(z_{j,\forall c},y_{j,\forall c})$, require both logarithmic gradients, $z_{j,\forall c}$, and profile values, $y_{j,\forall c}$, to be fully defined, for the first 5 Bayesian optimization iterations we only fit models to the logarithmic gradients, $F^m_{j,c}(z_{j,\forall c})$. 
Although it only has a small effect, this technique has also been observed to help convergence.

\subsubsection{Physics-informed GP and acquisition optimization}
\label{sec:PI}

In the current implementation of \PORTALS, we utilize a Radial Basis Function (RBF) covariance kernel for the Gaussian processes of turbulent and neoclassical fluxes.
We found that the use of a linear mean for gradient quantities, $z_{j,\forall c}$, and constant mean for profile quantities, $\widehat{y}_{j,\forall c}$, works well \referee{for a few test cases performed so far. However, a proper characterization of the benefit of this approach \textit{v.s.} the same polynomial degree for all input parameters has not been performed yet in a general manner and is subject of future work}.
This choice is motivated by the physics understanding that logarithmic gradients most strongly affect the background turbulence and profile quantities provide smaller corrections.
This \referee{is expected to} be particularly true as the plasma reaches steady-state, a situation where profile quantities will barely change and turbulence will be above critical-gradient, thus a higher-order polynomial makes sense for $z_{j,\forall c}$.
Future work will further utilize physics information to build better surrogates, to include constraints such as $F^m_{j,c}(z_{j,c}=0)=0$ (zero flux at zero gradient) or  $\frac{\partial F^m_{j,c}}{\partial z_{j,c}}>0$
(positive diagonal of transport matrix).
The addition of such constraints could help build better surrogate models that can accurately capture the system's behavior with lower number of evaluations.

In this work, we leverage the posterior mean of the predicted distribution of the residual $\xi$ to decide the next points to evaluate, which can be readily implemented with multi-dimensional root-finding methods (described below).
In practice, other acquisition functions such as Expected Improvement~(EI)~\citep{mockus_bayesian_1975, jones_efficient_1998}, Upper Confidence Bound (UCB), or information-based variants such as entropy search~\citep{wang_max-value_2018} are often used in the Bayesian optimization literature, as they balance exploration (sampling uncertain regions) and exploitation (sampling promising regions) more effectively.
We will investigate the effect of different acquisition function choices in \PORTALS, including any generalizations of the existing acquisition functions to the multi-residual structure of Eq.~\eqref{eq:MinimizationFull}, as part of our future work. \referee{However, as noted in Section~\ref{sec:Outcome}, as the total residual $\xi$ is a non-linear function of the individually modeled residuals, the full posterior distribution of these residuals is used in computing our target objective and thus the probabilistic nature of our surrogate models is crucial to our approach.}

To optimize the acquisition function (i.e. solving for flux-matching in the mean of the surrogate posterior distribution), we employ a sequential combination of three numerical methods.
Firstly, we take advantage of the decomposition of the total residual, $\xi$, into $\left(N_c\cdot N_\rho\right)$ individual local channel residuals.
For this, we use the multi-dimensional root-finding methods available in \texttt{SciPy} \cite{Virtanen2020}. Specifically, the Levenberg-Marquardt (LM) method \cite{levenberg_method_1944,marquardt_algorithm_1963} has shown superior performance for this particular problem.
The exact Jacobian of the surrogate models $F_{j, c}^m$ is provided to the LM method, as facilitated by automatic differentiation in \PyTorch \cite{Paszke2019}, encompassing all transformations outlined in Sections \ref{sec:Input} and \ref{sec:Outcome} as well as the residual reconstruction in Eqs.~\ref{eq:MinimizationFull}.
\referee{We must note that this first optimization method operates directly using the mean of the prediction of each surrogate flux evaluation, and therefore does not take advantage of the Monte-Carlo based objective transformation that would properly capture non-Gaussian properties of the nonlinear Equation~\eqref{eq:MinimizationFull_xi}. This would be required when uncertainty estimations from the Gaussian processes are taken into consideration for more advanced acquisition functions.
This method was still kept in \PORTALS as part of the historical familiarity of steady-state, flux-matching frameworks as multi-variable root-finding problems. 
Future work will be dedicated to understanding to what degree the decomposition of the total residual into each channel, radial flux components can be leveraged for acquisition optimization when moving away from the simplistic posterior-mean acquisition.
}

Secondly, in cases where the multi-variate LM method fails to achieve a satisfactory level of residual reduction, we resort to leveraging 
multi-start, scalarized-residual optimization in \BoTorch. 
A heuristic generation of initialization points is used from which to start the L-BFGS-B local optimization algorithm \cite{Zhu1997}.
Lastly, we utilize a genetic algorithm from the \texttt{DEAP} package \cite{Fortin2012}, to check for additional, possibly better optima that might have been missed by the local search algorithms of the previous steps.

\subsubsection{Other physics considerations}

\paragraph{Impurity particle transport}
As written in Eq.~\ref{eq:FSA}, the impurity density ($n_Z$) evolution is assumed to be source free, and therefore the workflow aims at solving for the null-flux condition, $\langle \mathbf{\Gamma}_Z\cdot\nabla r\rangle =0$. 
Impurity density peaking in such situations happens by means of a convective pinch that arises from neoclassical and turbulent transport and that results in a detrimental inward flux of impurities to the plasma core.
Formulating the impurity density flux with a diffusion-convection ansatz, summing over charge states and ignoring the effect of charge on core transport\footnote{We note that recent work \cite{howard_simultaneous_2024} suggests that this approximation to study impurity transport might need to be revisited and future work with \PORTALS could include a charge-state ionization recombination model.}:
\begin{align}
\label{eq:GZ}
    \Gamma_Z^m=-D^m_Z\nabla n_Z + V^m_z n_Z
\end{align}

Under the assumption of trace impurity, one can divide by $n_Z$ and arrive to an expression for a ``modified" impurity density flux that is only a function of the logarithmic gradient of the impurity density and not directly of its concentration:
\begin{align}
\label{eq:GZ1}
    \widehat{\Gamma}_Z^m=\frac{\Gamma_Z^m}{n_Z} = -\frac{D^m_Z}{a}a/L_{n_Z} + V^m_z \longrightarrow \widehat{\Gamma}_Z^m(a/L_{n_Z},\text{background})
\end{align}

As $n_Z$ is always non-zero and positive, the null flux condition is equivalent for the ``modified" impurity density flux:
\begin{align}
\label{eq:GZ2}
    \Gamma_Z^{\text{neocl.}}+\Gamma_Z^{\text{turb.}} = 0 
    \longrightarrow
    \widehat{\Gamma}_Z^{\text{neocl.}}+\widehat{\Gamma}_Z^{\text{turb.}}=0
\end{align}

As a consequence of Eq.~\ref{eq:GZ2}, we can reduce the dimensionality of the surrogate models for the impurity particle transport fluxes, as $n_Z$ is not required to fully describe $\widehat{\Gamma}_Z^m$ in null-flux conditions.
It is also important to note that under the trace impurity assumption, neither $n_Z$ nor $a/L_{n_Z}$ should affect the background turbulence state, and therefore the surrogate models for the rest of fluxes ($Q_e^m$, $Q_i^m$, $\Gamma_e^m$, $\Pi^m$) do not need to include those extra variables for training and prediction.

The previous approximation for trace impurities does not mean that the impurity density channel is decoupled from the rest. Even in trace quantities, high-Z impurities can radiate substantial amounts of power that will affect the evolution of electron energy, thus affecting the background turbulence in an indirect way.
However, the separation of $\delta f$ and background distribution functions allows for the coupling to only occur on a ``macroscopic transport" scale, and therefore handled by \PORTALS directly and not by each individual turbulence simulation, a feature that will be especially exploited in Section~\ref{sec:restart}.

\paragraph{Decoupling of radial grids}
\label{sec:decouplegrids}

\begin{figure}
    \centering
    \includegraphics[width=1.0\columnwidth]{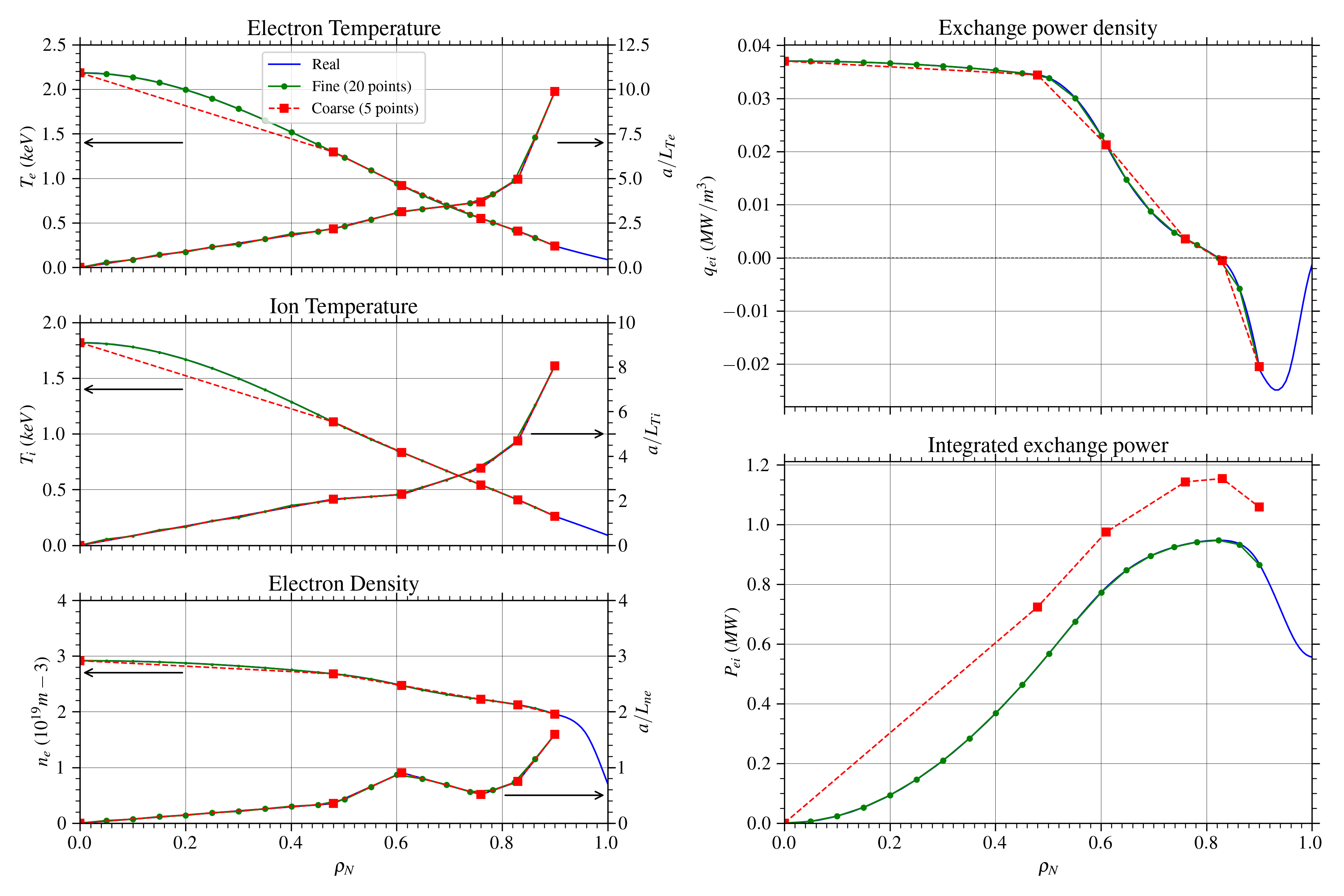}
    \caption{
    Summary of effect of radial grid on integrated exchange power calculation on an example L-mode plasma.
    Electron temperature, ion temperature and electron density are parameterized with piecewise linear logarithmic gradients.
    The effect of three grids on the calculation of targets are shown ([blue] full profile, [green] 20 grid points and [red] 5 grid points). While kinetic profiles and local power density overlap at grid points, as expected, the integrated exchange power can have significant deviations due to the volume integration.
    }
    \label{Fig:radial}
\end{figure}

As will be discussed in detail in Section~\ref{sec:knots}, the number of local transport model evaluations can be greatly reduced without loss of accuracy due to lack of normalized gradient profile structure in the plasmas of interest.
However, the calculation of target fluxes, $F^{\text{target}}$, requires a more careful treatment because it is constructed as the inner volume integral of the local target flux density, $f^{\text{target}}$.
Using a small set of radial points to perform the volume integrals can result in non-negligible errors, particularly in the calculation of the self-consistent energy exchange power, $\int_0^r\langle P_{ei}\rangle V'dr$, as shown in Fig.~\ref{Fig:radial}.

In the current implementation of \PORTALS, the target flux calculations are performed on a finer radial grid of 20 points (usually 4-5x more points than the transport calculations), which is sufficient for an accurate representation of volume integrals while at the same time does not introduce much overhead during acquisition optimization.

\paragraph{Turbulent exchange}

The framework of FSA transport equations described in Eq.~\ref{eq:FSA} also allows the generalization of the energy exchange power, $\langle P_{ei}\rangle$, to include contributions from both classical and turbulent exchange.
While the classical exchange can be calculated from standard net-exchange collisions formulas (see original \PORTALS \cite{rodriguez-fernandez_nonlinear_2022}), the turbulent exchange or anomalous heating is provided by the turbulence $\delta f$ simulations \cite{Candy2013}.
This means that the turbulent exchange becomes an additional quantity to include during the residual calculation and requires its own surrogate model. At flux-matching, however, the turbulent exchange enters as a target flux density in Eqs.~\ref{eq:MinimizationFull} and needs to be volume-integrated.
This is the only component in the target calculations that cannot be extended to the finer radial grid described previously, and hence its effect on the macroscopic profiles will be subject to volume integration errors.
\referee{
We must clarify that the turbulent exchange or any other target flux that requires surrogate modeling (when non-analytic) is incorporated into the surrogate modeling framework via the target flux density, $f^{\text{target}}_{j,c}$, and not via the integrated flux.
The volume integration is performed as part of the transformation from flux (surrogate) quantities to residual quantities during optimization.
}

Even if small, the inclusion of the turbulent exchange in \PORTALS allows to assess its effect on profile predictions directly with nonlinear gyrokinetic simulations instead of relying on quasilinear modeling.

\subsection{Benchmarking examples with the \TGLF model}
\label{sec:Bench}

The previously described workflow is general for any local, $\delta f$ transport model, and therefore the physics fidelity (and associated computational cost) of the kinetic profiles prediction depends on the chosen model fidelity.
Benchmarking with nonlinear gyrokinetic models would require significant computing resources for a proper study.
Therefore, here we approach this by using TGLF \cite{Staebler2007} ---a fast, quasilinear transport model--- to compare the performance of \PORTALS and standard Newton methods, from the perspective of how many profile evaluations are required to achieve flux-matching.

\begin{figure}
    \centering
    \includegraphics[width=1.0\columnwidth]{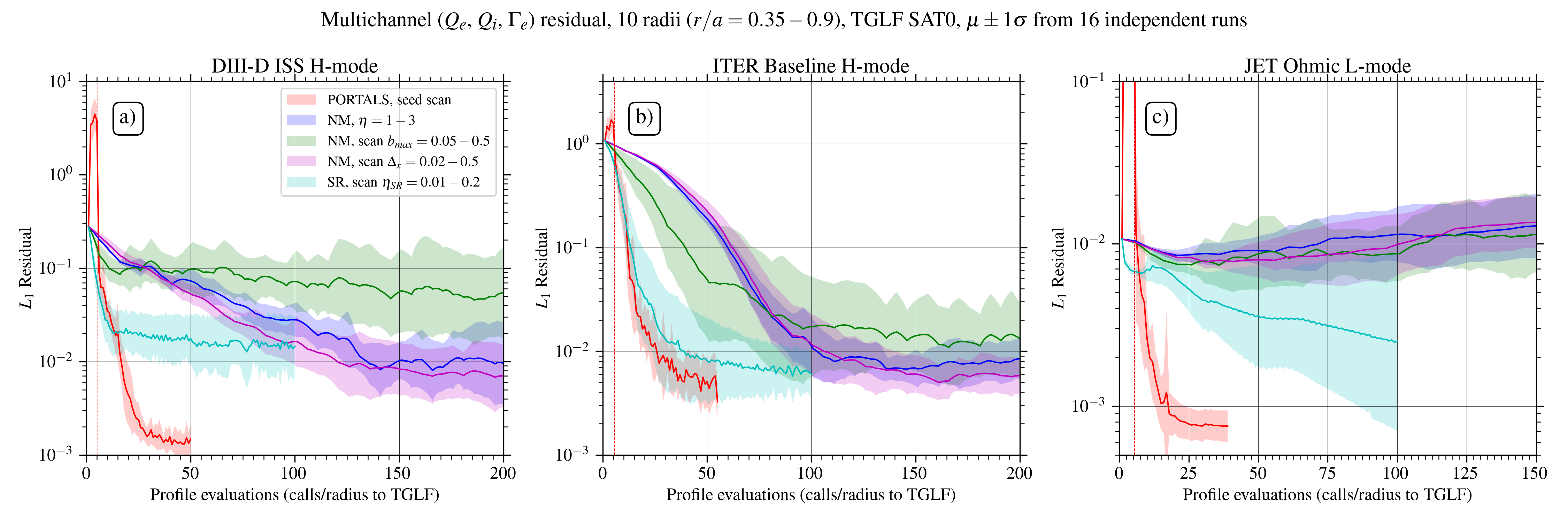}
    \caption{
    Comparison of residual evolution in \PORTALS with random initial training, a standard Newton method (NM) and simple relaxation (SR) schemes.
    For this exercise, we employed the quasilinear Trapped Gyro-Landau Fluid (\TGLF) SAT0 model \cite{Staebler2007}.
    Three example plasmas were used: (a) DIII-D ISS \cite{howard_simultaneous_2024}, (b) ITER Baseline \cite{howard_simultaneous_2024} and (c) JET Ohmic L-mode \cite{prf_2023_eps}.
    \PORTALS was initialized with different random seeds, which affect the initial training (5 profiles, vertical dashed line) and the subsequent training of Gaussian processes and acquisition optimization techniques.
    NM corresponds to \TGYRO method-$1$ and each of the three numerical parameters (relaxation parameter $\eta$, maximum step size $b_{max}$ and finite differences step $\Delta_x$) were varied within reasonable ranges.
    SR corresponds to \TGYRO method-$6$ with variations of the relaxation parameter $\eta_{SR}$.
    We note that, even though \TGYRO works with gyro-Bohm normalized residuals, here we plot them in real units, $MW/m^2$, for a direct comparison of performance.
    }
    \label{Fig:residuals}
\end{figure}

Figure~\ref{Fig:residuals} shows the evolution of the residual for the prediction of three kinetic profiles ($T_e$, $T_i$ and $n_e$) at 10 radial locations uniformly distributed between $r/a=0.35-0.9$ with evolving targets and with the \TGLF SAT0 model \cite{Staebler2007}.
\PORTALS has been run with 16 different seeds (with random initialization) and it is compared to a standard Newton method with finite differences Jacobian (approximated to be block-diagonal) \cite{Candy2009a} and a simple relaxation model.
Numerical input parameters to the Newton method such as the relaxation parameter $\eta$, maximum step size $b_{max}$ and finite differences step $\Delta_x$ are varied.
We note that these 1D scans of numerical parameters are not representative of the potentially ``best" performance of the Newton method, but reflects the often large variability of the results.

While only three specific cases are shown in Figure~\ref{Fig:residuals}, further numerical experiments reflect the same trend: the surrogate-based optimization in \PORTALS usually achieves a significant speedup when compared to the block-diagonal, finite-differences Newton method.
The simple relaxation method in some situations can get similar performance to \PORTALS, but it is very sensitive to the choice of relaxation parameter and can get stuck in local optima.
The increased cost of Newton methods for flux matching (i.e. more transport model evaluations) is mostly a consequence of the need to calculate the local Jacobian numerically at each iteration, which results in $(1+N_c)$ profile evaluations per Newton step.

It is important to note, however, that the methods as currently implemented in \PORTALS introduce a significant overhead in wall-time cost.
Standard numerical methods (as those implemented in \TGYRO) require close-to-negligible extra cost and the total computing time required for a full profile prediction can be approximated as the sum of each individual transport model evaluation: $t_{total} \sim N_m\cdot\left(N_\rho\cdot t_{m}\right)$, where $t_{m}$ is the cost of each local $\delta f$ simulation, including any internal parallelization technique\footnote{In this argument it was assumed that model evaluations are not parallelized. In principle, parallelization can occur in radial location and within each iteration for calculation of the finite-differences Jacobian.}.
When the flux-matching problem is formulated as a surrogate-based optimization one, the cost of fitting the surrogate models and of acquisition function optimization need to be taken into consideration: $t_{total} \sim N_m\cdot\left(N_\rho\cdot t_{m} + t_{fit}+ t_{acq}\right)$, where $t_{fit}$ and $t_{acq}$ represent the time spent in fitting the surrogates and in optimizing the acquisition function at each iteration.
In fact, if the transport model evaluation is cheap (as in the case of the \TGLF model), the cost of the flux-matching exercise can be dominated by the surrogate fitting and optimization overhead.
However, the improved convergence properties of the surrogate-based techniques combined with the comparatively (to nonlinear gyrokinetics) fast execution of quasilinear models may still make the adoption of these techniques more desirable than traditional Newton methods, but more work to characterize the numerical robustness of \PORTALS with quasilinear models is required to draw definite conclusions.

As it has been described throughout Sec.~\ref{sec:DI}, the high performance of \PORTALS techniques is achieved via the decomposition of the residual $\xi$ into smaller, self-contained components.
This also results in the need to fit and evaluate multiple surrogate models, usually $N_\rho\cdot\left(3\cdot N_c + 1\right)$, where it was assumed that each channel requires neoclassical, turbulent and target flux surrogate models, and that each radius also requires a model for the turbulent exchange.
The surrogate overhead $\left(t_{fit}+t_{acq}\right)$ takes on average $\sim10$ wall-time minutes utilizing an AMD EPYC 7543 32-Core Processor for a problem whose residual requires the evaluation of 90 surrogate models (10 radial locations, 3 channels with zero target particle flux) with 30 free parameters.
Under the conservative assumption of $5\times$ speed up in \PORTALS, we arrive at the rule of thumb that \PORTALS methods are appropriate if each transport model requires more than $3$ minutes to evaluate.
This condition is not satisfied with most quasilinear transport models like \TGLF \cite{Staebler2007} and \QuaLiKiz \cite{Bourdelle2005}, that take a few tens of seconds per evaluation.
On the other hand, nonlinear gyrokinetic modeling does benefit dramatically from the use of \PORTALS, as simulations of realistic plasma conditions range from a few tens of thousands to millions of CPU-hours (or in recent GPU architectures hundreds of GPU-hours) per evaluation.
Future work will devise methods to reduce the wall-time cost of each individual \PORTALS iteration (such as parallelizing the evaluation of the GP models and leveraging modern GPU hardware) to bring the benefits of surrogate modeling to the realm of quasilinear profile predictions.

\begin{figure}
    \centering
    \includegraphics[width=1.0\columnwidth]{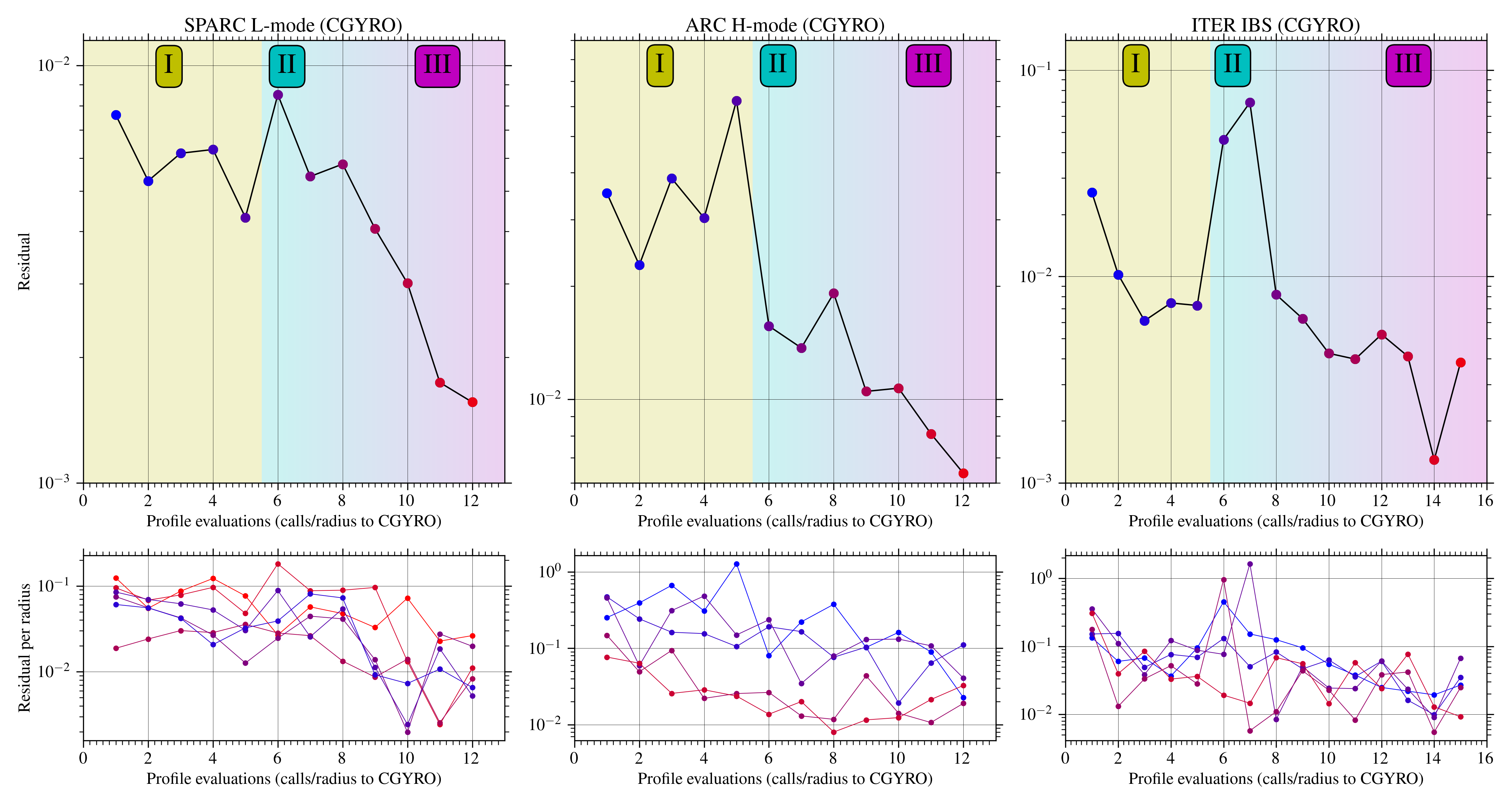}
    \caption{
    Characteristic \PORTALS phases during the prediction of three burning plasmas ([left] SPARC Primary Reference Discharge (PRD) \cite{Rodriguez-Fernandez2020a}, [center] ARC \cite{holland_2023_aps} and [right] ITER \cite{howard_2024_iter}) with nonlinear \CGYRO.
    Phase I is the initial surrogate training phase, Phase II represents the first \PORTALS iterations (often with oscillatory behavior as it learns key parametric dependencies) and Phase III is the final convergence phase. The transition from II to III is smooth and not characterized by a change in the \PORTALS iteration scheme but it represents a transition to a situation where the surrogates are capturing the turbulence dynamics with enough accuracy to drive the system towards steady state.
    \referee{Bottom subplots show the evolution of the residual per radial location separately, shaded from red (core) to blue (edge).}
    }
    \label{Fig:phases}
\end{figure}

Fig.~\ref{Fig:phases} shows characteristic phases for the decrease of multi-channel residual of \PORTALS-\CGYRO simulations, for three examples.
Phase I consists of the initial SR training, which often decreases residual if channels are not coupled or strongly dominated by one type of turbulence mode, although the simplicity of this scheme (as addressed in Sec.~\ref{sec:SR}) can also lead to oscillations or higher residual (as in the middle plot in Fig.~\ref{Fig:phases}).
However, the goal of this phase is not to observe a steady decrease of the residual, but of producing proper training points for subsequent phases of the \PORTALS workflow.
Phase II is when surrogates are fitted and acquisition function optimization informs next points to evaluate. During Phase II, oscillations are usually observed for 2-5 evaluations, a consequence of the surrogates exploring cases near the logarithmic gradients bounds, or combinations that were not explored during Phase I.
In Phase III, the residual drops dramatically and flux-matching occurs.
Sometimes, following Phase III, we observe oscillations due to critical gradient and stiff behaviors of turbulence.
Profiles and gradients are converged, but fluxes can jump up and down the targets, as was discussed in Ref.~\cite{rodriguez-fernandez_nonlinear_2022} and will be described in detail in Sec.~\ref{sec:5c}.
\referee{Generally, as depicted in the bottom subplots in Fig.~\ref{Fig:phases}, the residual evolution is not uniform across radial locations, but no specific trends are found. In some situations, the core reaches low residuals faster than the edge, and vice versa.}

\referee{Due to the need for expert knowledge to interpret if sufficient convergence has been achieved, together with the need for a careful nonlinear gyrokinetic setup (to avoid wasting computational resources), the current implementation of \PORTALS requires significant ``human in the loop'' component. Future work will focus on the development of a more automated approach that can make high-fidelity profile predictiosn available to a wider user base.}

\subsection{Restarting simulations for parameter scans}
\label{sec:restart}

During the prediction of profiles with the \PORTALS framework, it is currently assumed that the rest of plasma and geometry parameters are fixed.
While this assumption can be relaxed, the accuracy of the surrogate models would be compromised if such parameters are not included as input variables during training.
The inclusion of more variables, while possible, would require a higher number of local simulations with the transport model for surrogate training, which can quickly become intractable if expensive turbulence codes are used.

\begin{figure}
	\centering
	\includegraphics[width=1.0\columnwidth]{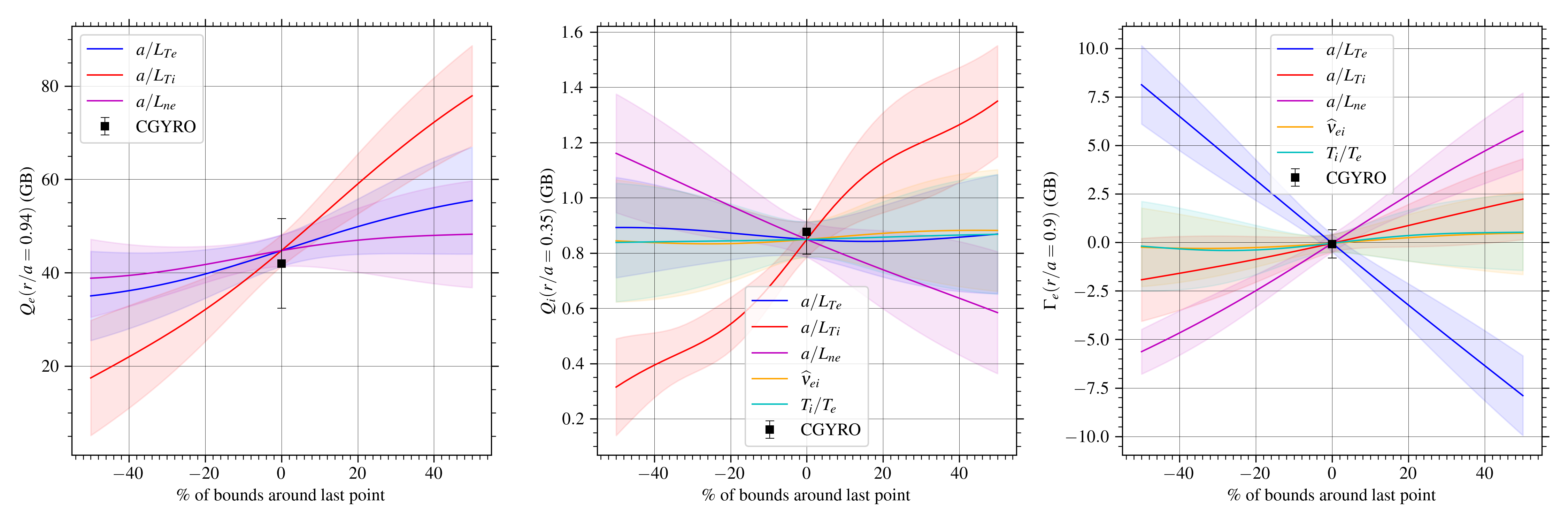}
	\caption{Example of GP model predictions (left: $Q_e$, center: $Q_i$ and right: $\Gamma_e$) around the last point evaluated after first converged profiles from Fig.~\ref{Fig:restarting} (point \#12) at three representative locations.
    The example $Q_e$ at the leftmost plot is only fitted to logarithmic gradients because the boundary condition for this profile prediction ($r/a=0.943$) is not allowed to vary, but the logarithmic gradients can.
    }.
	\label{Fig:surrogates}
\end{figure}

An advantage of using surrogate-based techniques with fixed geometry and background electromagnetics is that the surrogates trained during a profile prediction can be re-utilized to produce new scenarios with variations in certain input parameters.
Input parameters that do not affect the local $\delta f$ turbulence simulations but that affect the macroscopic evolution (e.g. auxiliary input power, fueling or torque) are suitable for surrogate-reutilization.
Input parameters whose effect is already captured by the set of profile quantities, $\widehat{y}_{j,c}$, that were used to train the surrogates (e.g. edge density or pressure boundary condition) are also suitable for this technique.
Fig.~\ref{Fig:surrogates} shows an example of how surrogate models look like for characteristic channels.
Leveraging this aspect of surrogate modeling results in a much reduced number of new evaluations required to reach new converged states, and allows for extensive study of reactor scenarios.

\begin{figure}
	\centering
	\includegraphics[width=1.0\columnwidth]{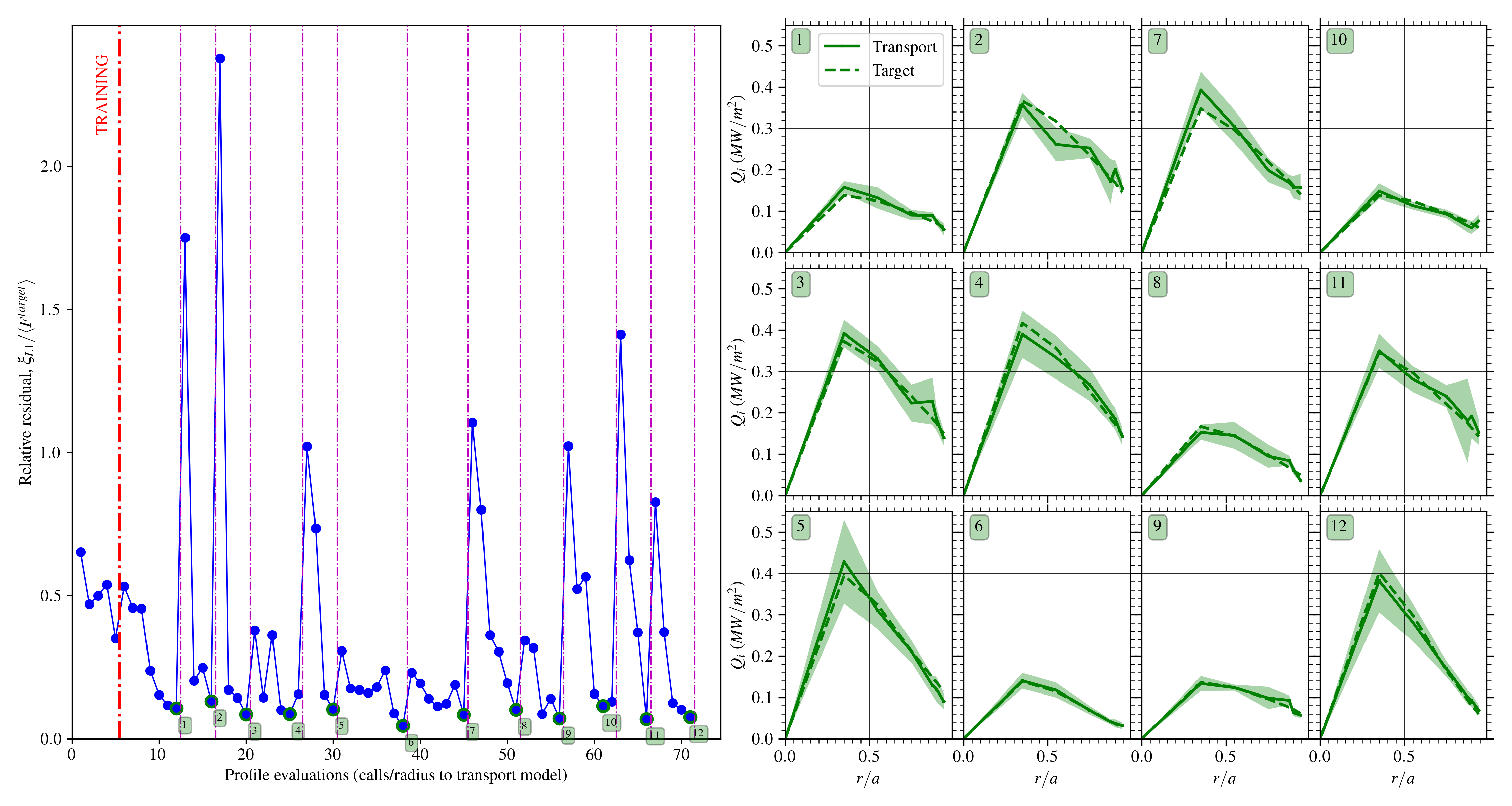}
	\caption{Summary of SPARC L-mode investigation using \PORTALS and nonlinear \CGYRO (6 radial locations, 3 channels). On the left, the $L_1$ residual normalized to the radial-average of the target flux is plotted for all iterations. Initial training had 5 profile evaluations and first converged profile was achieved with a total of 12 evaluations. The re-utilization of surrogates enabled the prediction of kinetic profiles ($T_e$, $T_i$, $n_e$) and fusion performance of 12 scenarios (variation of $P_{input}$, $n_{edge}$, $T_{edge}$) with only a total of 71 profile evaluations (426 local $\delta f$ \CGYRO simulations). On the right, ion heat flux matching between transport and targets for each of the scenarios. Similar flux-matching quality was achieved for electron energy and particle fluxes.}
	\label{Fig:restarting}
\end{figure}

This technique was exploited for the study of the parameter space of the SPARC tokamak \cite{Creely2020a,Rodriguez-Fernandez2022} when operating in L-mode-like conditions \referee{\cite{rodriguez-fernandez_core_2024}}.
We utilized nonlinear \CGYRO \cite{Candy2016} to evaluate core transport at 6 radial locations ($r/a=0.35, 0.55, 0.75, 0.875, 0.9,0.943$) and scanned auxiliary input power, edge density boundary condition and edge temperature boundary condition. 
As shown in Fig.~\ref{Fig:restarting}, first convergence took 12 profile evaluations, but subsequent scenarios required an average of 5 extra evaluations for convergence in the three predicted channels ($T_e$, $T_i$, $n_e$).
This unique study enabled the investigation of 12 scenarios at a total cost of 71 profile evaluations (426 local $\delta f$ \CGYRO simulations) \referee{\cite{rodriguez-fernandez_core_2024}}.
Physics results from this and other applications of \PORTALS will be part of upcoming publications and conference presentations.

It is important to note, however, that the re-utilization of surrogate models for parameter scans has the limitation of exploring scenarios with all other parameters fixed, such as the plasma geometry, safety factor profile, or effective charge, $Z_{eff}$.
If additional parameters change, and there is a non-negligible effect on turbulent or neoclassical transport, the re-utilization of results from previous optimization runs in a na\"ive fashion can potentially delay convergence, as the parameters that are used to train the surrogates would not capture all the changes in transport fluxes.
In such a case, it becomes more efficient to start the \PORTALS simulations from scratch.

In future work, we intend to explore leveraging multi-task GP modeling approaches \cite{bonilla_multi-task_2007} that do not assume identical dependence of the outcomes on input parameters from different simulation settings and instead learn correlations between the outcomes. Such ``transfer learning'' has the potential to substantially reduce the number of evaluations required even in cases where simulated behaviors are not identical. This also opens the door for ``multi-fidelity'' optimization, where different simulation setups with different cost/accuracy trade-offs can be leveraged together in a principled fashion via a joint surrogate model. 

\section{Guidelines for accurate profile predictions}
\label{sec:Guide}

At the time of publication, we have utilized this workflow for the nonlinear gyrokinetic profile prediction of \referee{over 30} different plasma conditions, many of which were for burning plasmas. 
\referee{
\PORTALS has been employed to perform direct nonlinear gyrokinetic simulations with \CGYRO \cite{Candy2016} to predict steady-state plasmas in SPARC PRD \cite{rodriguez-fernandez_nonlinear_2022}, ITER Baseline Scenario \cite{howard_2024_iter}, DIII-D ITER Similar Shape \cite{howard_simultaneous_2024}, ARC \cite{holland_2023_aps}, SPARC L-modes \cite{rodriguez-fernandez_core_2024} and JET H, D and T Ohmic plasmas \cite{prf_2023_eps}.
These studies are part of a broader effort to both validate nonlinear gyrokinetics in current experiments and to predict the performance of future devices.
}
This section is intended to outline ``best practices" that have been determined from these efforts.

\subsection{Selection of radial grid}
\label{sec:knots}

Thanks to the de-coupling of transport flux training and flux-matching solver, it is found that the cost of profile predictions with \PORTALS scales linearly with the number of local, flux-tube simulations used to represent the full profiles.
This is because the surrogate training occurs locally, so the nonlinearities introduced by the evolution of sources (e.g. energy exchange and alpha heating) and sinks (e.g. radiation) and the coupling of distant radial locations via gradient integration are handled directly by the trained surrogate models.
The use of automatic differentiation in \PyTorch presents clear advantages as access to the exact, local Jacobian matrix enables a more robust convergence process.

Therefore, the cost of the profile predictions directly depends on the choice of a radial grid that accurately represent the profiles.
The coarser the radial grid the cheaper the profile prediction, but misrepresentations of profile shape and fusion performance may occur if the profiles are too constrained by the chosen grid.
In this work, we focus on the prediction of inductive, on-axis heating plasma discharges, where the presence of internal transport barriers and a fine gradient structure are not expected.

\begin{figure}
	\centering
	\includegraphics[width=1.0\columnwidth]{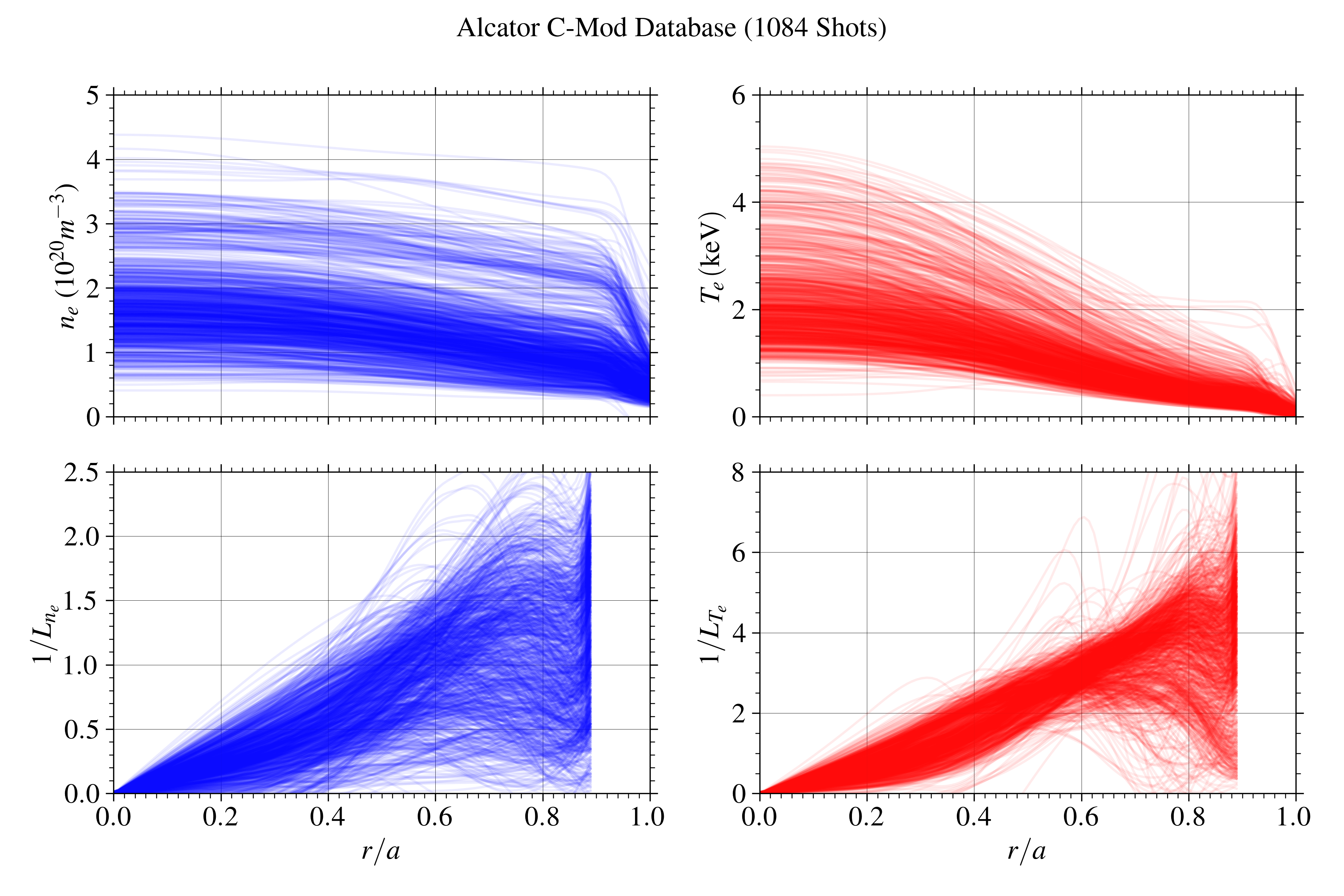}
	\caption{Electron temperature ($T_e$) and density ($n_e$) fits and their corresponding inverse gradient scale lengths ($1/L_{T_e}$, $1/L_{n_e}$) for all 1084 shots included in the database. In this work, we only explore gradients inside of $r/a<0.9$.}
	\label{Fig:DBcmod}
\end{figure}

To study the minimum requirements for the prediction of such scenarios we focus our attention to a database of 1084 discharges in the Alcator C-Mod tokamak \cite{Greenwald2013}, spanning various operational scenarios (including L-, I- and H-modes).
Discharges were required to have at least 150ms ($\sim5$ energy confinement times) period of steady plasma current, magnetic field, total input power and line averaged density.
Electron density and temperature data from Thomson Scattering \cite{hughes_thomson_2003} and Electron Cyclotron Emission \cite{oshea_notitle_1997} were used to produce a database of kinetic profiles that were fitted using sparse Gaussian processes techniques \cite{meneghini_integrated_2015}.
Filtering of bad fits (e.g. due to outliers) was used to ensure realistic profile shapes.
The full database of fitted profiles is depicted in Figure~\ref{Fig:DBcmod}. Core electron densities ranged from $\sim0.5-4\cdot$ $10^{20}m^{-3}$, and temperatures from $\sim0.5-5 keV$.

\begin{figure}
	\centering
	\includegraphics[width=1.0\columnwidth]{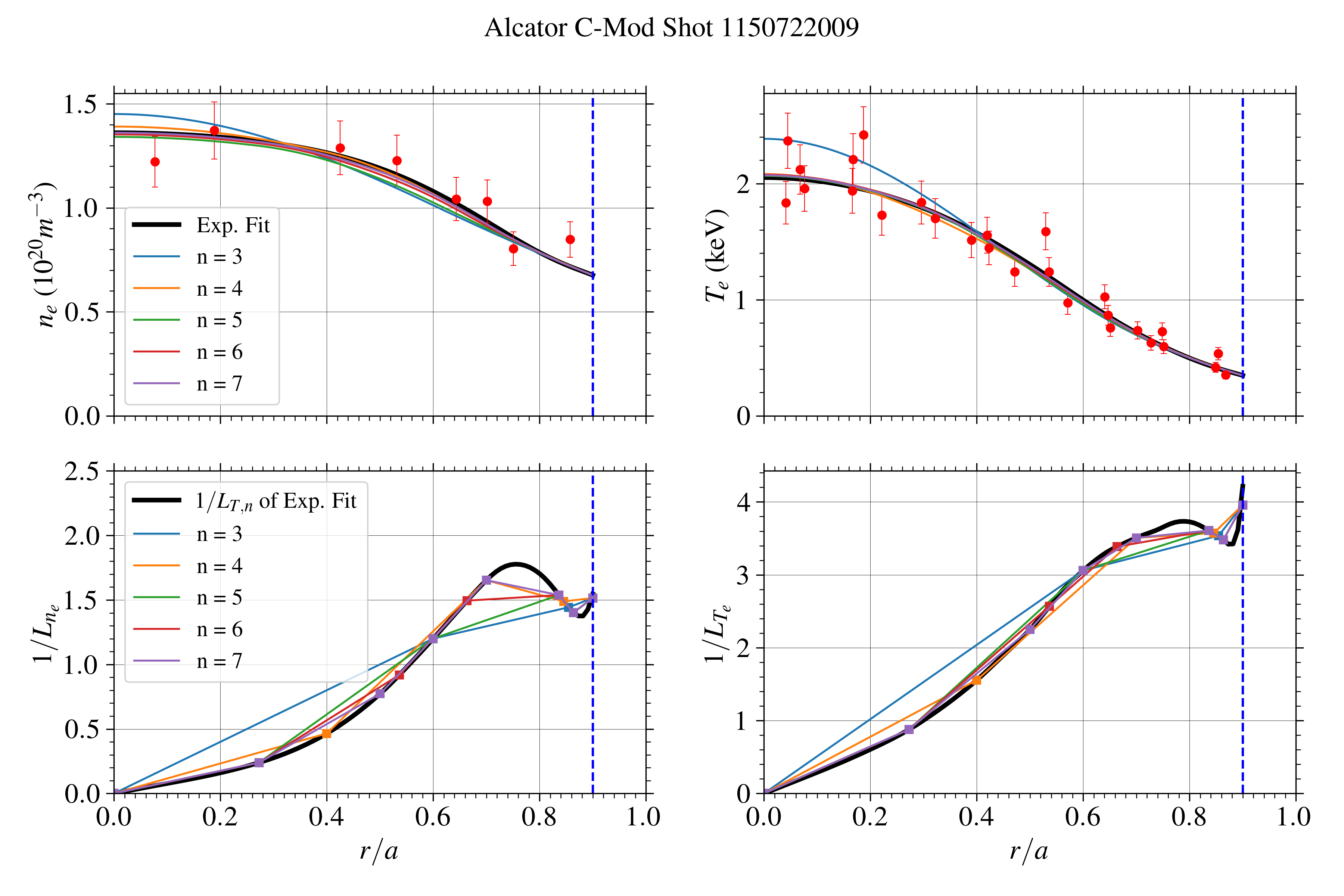}
	\caption{Example of profile parameterization using piece-wise linear $1/L_{T,n}$ interpolation for an Alcator C-Mod plasma in the database.
    3 to 7 radial knots evenly-spaced out between $r/a=0.2$ and $0.9$ are used to parameterize the gradient scale lengths, which are plotted in the bottom ($1/L_{n_e}$ and $1/L_{T_e}$) along with their integrated profiles on top ($n_e$ and $T_e$). Experimental data (red points) was assumed to have a $\sim10\%$ error.}
	\label{Fig:example_param}
\end{figure}

To assess the radial requirements for profile predictions, we parameterize the normalized inverse gradient scale lengths $a/L_{T,n}$ with piece-wise linear functions, with a finite number of knots (equal for both $a/L_{T}$ and $a/L_{n}$) that represent the locations that one would use to run $\delta f$ turbulence simulations.
Several potential radial knot locations were evaluated to determine the optimal number and location (in normalized minor radius, $r/a$) of such knots.
Potential knot positions from $r/a=0.2-0.9$ with $0.05$ spacing were considered, and $3-7$ radial knots were evaluated, using $r/a=0.9$ always as the anchor point and linearly interpolating to zero from the point that is nearest to the magnetic axis.
Note that the choice of $r/a=0.9$ as the anchor point ensures that pedestal-like structures are not covered by the prediction.
For any chosen number and location of knots, profiles of $T_e$ and $n_e$ can be obtained by radial integration of the gradient scale lengths.
For visualization purposes, Figure~\ref{Fig:example_param} depicts 5 possible knot grids (evenly-spaced points) and the changes in the integrated temperatures and densities.

To assess the goodness of any given knot grid, we determine a ``modified" fusion power ratio, $\widehat{P}_{f}/\widehat{P}_{f,fit}$.
We assume equilibrated and pure plasmas ($T_i=T_e$, $n_i=n_e$), and the temperatures are multiplied by a factor of $10$ to bring the average on-axis temperature of $\sim2keV$ in Alcator C-Mod to $\sim20keV$ expected in burning plasma experiments such as in SPARC \cite{Creely2020a,Rodriguez-Fernandez2020a}.
This, along with the assumption of a 50-50 D-T mixture, transforms the database to burning-plasma-like conditions and the deviations in fusion power due to the coarse radial grid are directly representative for what would be expected when predicting burning plasmas and reactor concepts.

\begin{figure}
	\centering
	\includegraphics[width=1.0\columnwidth]{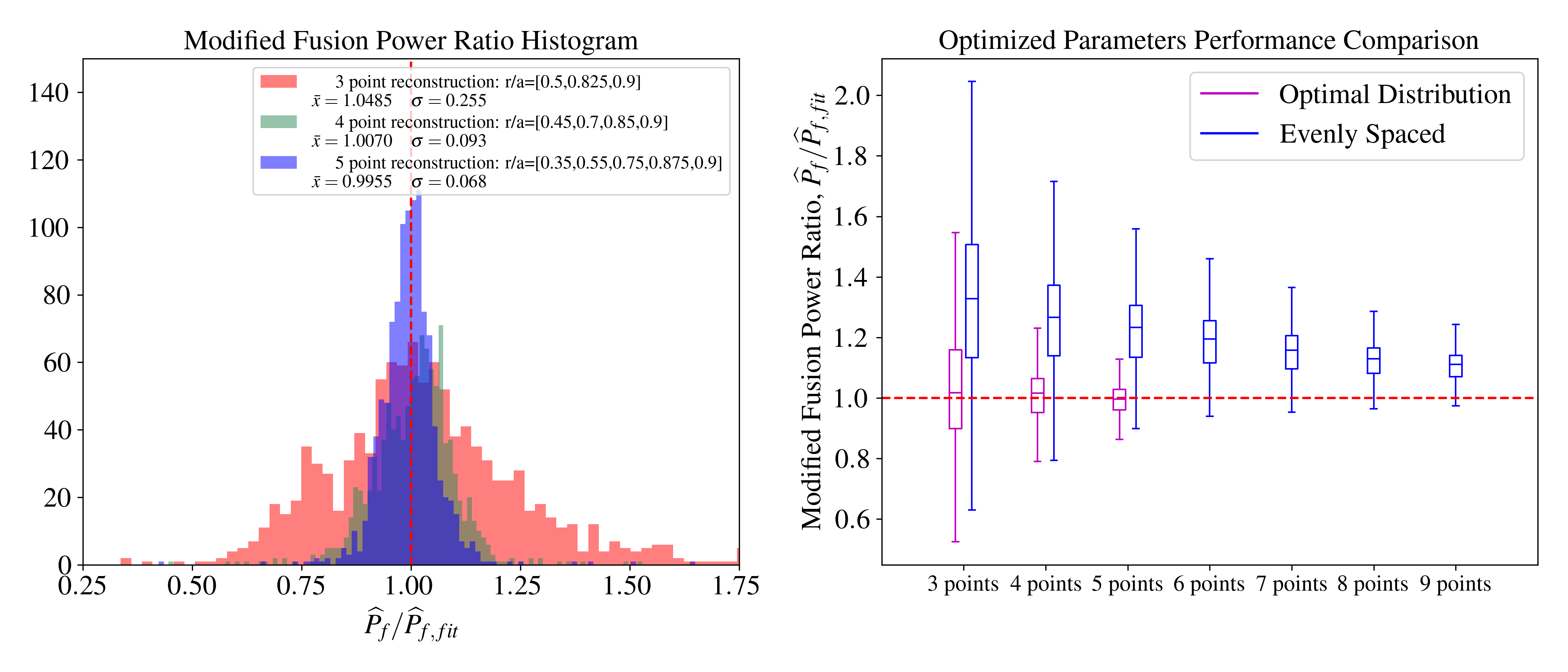}
	\caption{(left) Superposition of histograms of the modified fusion power ratio $\widehat{P}_{f}/\widehat{P}_{f,fit}$ for all shots in the database using the optimal set of parameters for each of the gradient reconstructions with piecewise linear functions. (right) Set of box-plots comparing $\widehat{P}_{f}/\widehat{P}_{f,fit}$ of using optimal parameters vs. evenly-spaced parameters.}
	\label{Fig:histogram}
\end{figure}

For any given number of knots, the optimal knot locations to minimize the error of the database is found by performing a grid search.
Figure~\ref{Fig:histogram} (left) shows the histogram of the ratio of the modified fusion power between the parameterized profiles and the actual profile fit for the optimized locations for each number of radial knots. 
Figure~\ref{Fig:histogram}(right) displays the very different parameterization performances between evenly-spaced and optimized knot locations. 

First, it is found that, for any number of chosen points, the optimization of the knot locations (non-uniform grid) produces significantly reduced parameterization error than if only evenly-spaced points are chosen.
Particularly, the edge region is found to require larger number of points, while inside mid-radius, it is found that one knot is sufficient to minimize the fusion power error. This occurs because the edge has always more structure than the core of the plasma, likely a consequence of potentially less-stiff transport in that region.
Furthermore, due to the consideration of realistic plasma geometry, most of the contribution to fusion power comes from $r/a\sim0.3$ (as shown for example in SPARC in Figure~\ref{Fig:scan_axis}c), requiring a more accurate representation of the edge and mid-radius gradients than those in the inner core.

This investigation also reveals that $5$ radial knots--- with radial locations at $r/a=[0.35,0.55,0.75,0.875,0.9]$--- are sufficient for an accurate representation of the fusion power, with a mean of the database centered at $\widehat{P}_{f}/\widehat{P}_{f,fit}=0.9955$ and standard deviation of $6.8\%$.
In the case of the stored energy, $W$, the representation with $5$ knots resulted in a distribution with mean centered at $W/W_{fit}=0.9990$ and standard deviation of $2.6\%$.
Note that the use of different fitting techniques (GPs to fit profiles and piecewise-linear-$a/L_{T,n}$ for finding optimal knots) ensures that we are not introducing bias in finding the best radial grid.

Therefore, with this investigation we recommend to perform profile predictions by using not more than 5 locations, as long as the plasmas to be predicted are not expected to exhibit features of advanced scenarios, with large off-axis current drive, internal transport barriers, reversed shear, or other phenomena that may introduce structure in the gradient profiles.
These results are reminiscent of the concept of profile resilience and consistency, an attribute of burning plasma profiles (usually with small gyro-Bohm normalized heat and particle fluxes) that results in the near invariability of $a/L_T$ profiles throughout the plasma core as a consequence of stiff turbulent transport driven by ion temperature gradient (ITG) mode turbulence \cite{Kotschenreuther1995}.
The edge, however, via the formation of a pedestal or via non-stiff transport \cite{Sauter2014} in colder, L-mode plasmas can significantly vary and is more sensitive to plasma parameters and input power.

\subsection{On the importance of near-axis simulations}

Section~\ref{sec:knots} has presented evidence that, for inductive tokamak scenarios, accurate enough ($6.8\%$ standard deviation) predictions of fusion power in burning-plasma regimes are attained even if it is assumed that the inverse gradient scale length is linearly interpolated from the value at $r/a=0.35$ to $a/L_{T,n}=0.0$ on axis.
In this section we further explore the reasons behind this perhaps surprising result. This finding can have important implications as running turbulence transport models near the magnetic axis is difficult due to the low magnetic shear and high gyro-Bohm units, and other physics (such as fast ions or sawtooth) may further complicate such simulations and interpretation of the results.

\begin{figure}
	\centering
	\includegraphics[width=1.0\columnwidth]{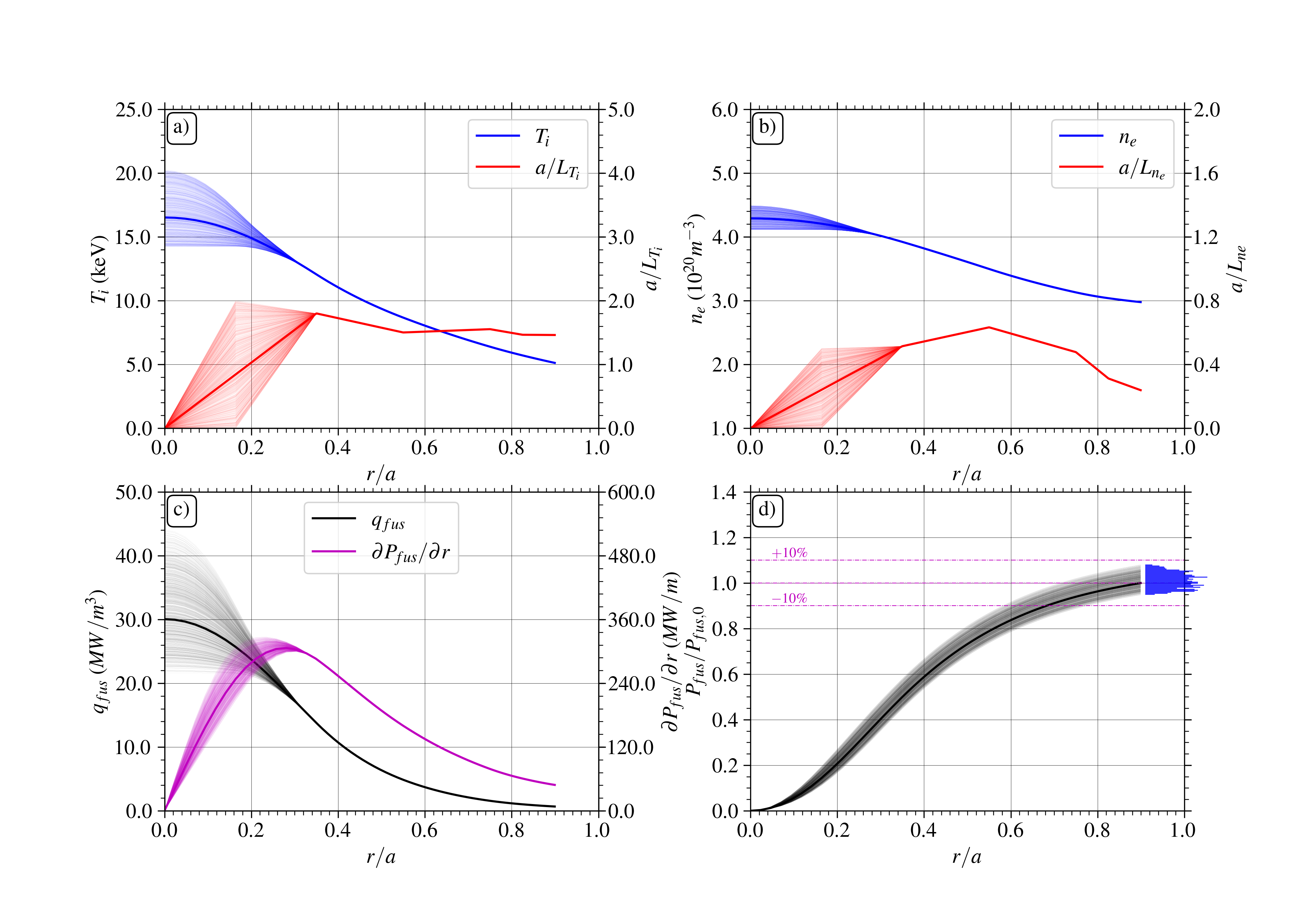}
	\caption{Evaluation of fusion power in SPARC PRD plasma using random combinations of gradient profiles in temperature and density inside of $r/a<0.35$. a) Ion temperature and inverse normalized gradient scale length, b) electron density and inverse normalized gradient scale length, c) fusion power density profile and radial derivative of its volume integral, and d) ratio of volume integrated fusion power between random profiles and original (linear interpolation from $r/a=0.35$ to $0.0$). Histogram of total fusion power is added on the right for visualization purposes.}
	\label{Fig:scan_axis}
\end{figure}

Setting aside the fact that there is experimental evidence that suggest that a linear interpolation to $a/L_{T,n}=0.0$ on axis is sufficient (as shown in Figure~\ref{Fig:DBcmod} for Alcator C-Mod plasmas), even if such feature was not observed, the small plasma volume near the plasma core results in only small variations in fusion power when the linear interpolation assumption is relaxed.
Figure~\ref{Fig:scan_axis} shows the small effect that the inner gradients (inside of $r/a<0.35$) have on the total fusion power, using the SPARC Primary Reference Discharge (PRD) scenario \cite{rodriguez-fernandez_nonlinear_2022} as the base case.
An additional point at $r/a=0.175$ was added to the $a/L_{T,n}$ parameterized profile and the values of the gradients were sampled 1000 times from uniform distributions ($0-2$ in $a/L_T$ and $0-0.5$ in $a/L_n$). Even with such large deviations from the original ``interpolation-to-zero" profile, the standard deviation of the distribution of fusion powers was $3.2\%$ (Figure~\ref{Fig:scan_axis}d).
Although the differences in the fusion power density (Figure~\ref{Fig:scan_axis}c) were significant, the small volume of that region results in only small differences to the integrated fusion power.

Consequently, from the perspective of predicting fusion performance and fusion power output from a burning plasma, the small gain in accuracy does not seem to be compensated by the additional cost of the turbulence simulations inside of $r/a<0.35$.

\subsection{Ion-scale simulations}

Although the \PORTALS techniques are insensitive to the choice of the $\delta f$ transport model ---allowing high-fidelity simulations to be brought to stationary conditions at reduced number of evaluations---, the choice of model setup affects directly the cost of the simulation exercise.
In the case of gyrokinetic modeling, the choice of the simulated normalized poloidal wavenumber range, $k_y\rho_s$, is an important parameter that affects strongly the simulation cost and the type of turbulence instabilities and interactions that are captured.
Although not a limitation, all the simulations performed so far with \PORTALS rely on the assumption that ion-scale turbulence is the dominant turbulent transport mechanism, and that high-wavenumber effects or interactions with long wavelength turbulence are sufficiently small.

Validation studies in current experiments, with widely varying heating schemes, collisionality regimes and plasma parameters, require additional work to understand the validity of ion-scale gyrokinetics to reproduce core transport.
Performing linear stability and numerical resolution scans around experimentally-measured conditions is an approach that is often taken in the community (e.g. Ref.~\cite{howard_simultaneous_2024} for predictions of DIII-D plasmas).

However, in the case of reactor and burning plasmas predictions, there is no access to experimental conditions to evaluate the turbulence spectrum.
A careful look at the requirements for fusion power production in conventional, inductive tokamaks can yield some insights to motivate the assumption of ion-scale turbulence modeling.
Ref.~\cite{holland_development_2023} showed that despite the low collisionality of reactor-relevant plasmas, their energy confinement times are sufficiently large that they remain well-coupled. Thus, although such plasmas are inherently electron heating dominated (due to the inherent dominance of alpha heating in a reactor-relevant plasma), the combination of strong coupling and radiation losses results in $Q_i>Q_e$ for all the reactor-relevant conditions studied with \PORTALS to date. 
Due to the dominance of the turbulent ion heat flux in the core plasma (where neoclassical transport is found to be negligible due to the low collisionality), it appears that gyrokinetic simulations which only resolve the long-wavelength ($k_y \rho_{s} \lesssim 1$) turbulence that drives the ion heat flux are sufficient for these conditions.
This conclusion is consistent with the ``fingerprint paradigm'' of Kotschenreuther \textit{et. al} \cite{Kotschenreuther2019} who note that only long-wavelength instabilities such as ITG, trapped-electron modes (TEMs), or kinetic ballooning modes (KBMs) are (at least linearly) capable of producing $Q_i/Q_e > 1$ ratios consistent with the observed reactor characteristics.
Whether multi-scale turbulence dynamics become important in other reactor scenarios (such as steady-state advanced tokamak regimes), or lead to significant nonlinear enhancements of the ion scale fluctuations remains to be resolved in future work.

\subsection{Ion bundling}

Similarly to the transport vs target radial grid decoupling presented in Sec.~\ref{sec:decouplegrids}, another technique that is possible in \PORTALS and desirable to reduce the computational cost of the predictive exercise is the de-coupling of the ion species.
While calculation of targets (particularly radiation) requires a realistic mix of impurities, the effect of the impurity mix in turbulence modeling is often captured by its effect on the effective charge $Z_{eff}$ and main ion dilution $f_{main}$. 
Therefore, nonlinear gyrokinetic simulations may benefit from the reduction of the number of ion species, with the lumping of impurity species being a common approach to reduce the computational expense.
We must note, however, that when the density of impurities become additional channels to simulate and bring to steady-state, this approach is not possible.

Similarly to the impurity lumping, it is observed \cite{howard_simultaneous_2024} that the effect of separating hydrogenic ions into deuterium and tritium is marginal on core turbulence modeling of burning plasmas and therefore it becomes advantageous to simulate a lumped $A=2.5$ hydrogenic main ion.
Studies of isotope effect in current machines \cite{Belli2021} show that using an effective main ion mass can reproduce many aspects of the turbulence.
However, the effect on alpha heating and fusion production is, of course, retained during the calculation of targets.

\section{5-channel nonlinear gyrokinetic prediction of DIII-D ITER Similar Shape Plasma}
\label{sec:5c}

To demonstrate the efficiency of \PORTALS, here we aim at simultaneously bringing 5 channels (electron temperature, ion temperature, electron density,  lithium impurity density and angular rotation) to steady-state of the DIII-D ITER Similar Shape (ISS) H-mode plasma from Ref.~\cite{howard_simultaneous_2024}, with ion-scale, nonlinear \CGYRO \cite{Candy2016} and \NEO \cite{Belli2008} simulations providing the turbulent and neoclassical components of cross-field transport.
We allow radiation (Bremsstrahlung, line and synchrotron) and energy exchange (classical and turbulent) to evolve.
We employ the 5 radial locations described in Sec.~\ref{sec:knots} ($r/a=[0.35,0.55,0.75,0.875,0.9]$, as illustrated in Fig.~\ref{Fig:rendering} for this specific DIII-D ISS case) but calculate targets in a higher resolution grid of 20 points from $r/a=0.0$ to $0.9$.

Nonlinear gyrokinetic simulations using the \CGYRO code were performed at each of the radial locations. The general simulation setup is similar to that reported in Ref.~\cite{howard_simultaneous_2024} with some modest changes that are meant to lead to improve the turbulence statistics and therefore lead to more reliable time averaged heat and particle fluxes. 
Although the exact box sizes varied slightly across the radius, the nominal simulation setup was targeting box sizes in the radial and bi-normal direction of $\sim 120 \times 120 \rho_s$ for $L_x$ and $L_y$.
This simulation domain was represented with 24 toroidal modes ($n_n$) spanning from $0.053$ to $1.219$ in $k_y\rho_s$. Nominally 512 radial modes ($n_r$), 24 toroidal modes ($n_n$), 24 theta points ($n_\theta$), 24 pitch angles $(n_\xi)$ and 8-12 energies ($n_{energy}$) were used, with the exact number of radial modes varying somewhat depending on the radial location simulated.
Simulations utilized high accuracy collisions implemented with the Sugama collision operator \cite{Sugama1997}, realistic geometry implemented with the Miller Extended Harmonic formulation \cite{Arbon2020}, electromagnetic turbulence ($\delta \phi$, $\delta A_{\|}$, $\delta B_{\|}$), and included 5 gyrokinetic species (D, T, Lumped impurity species, trace Li).
The addition of the trace Li species ($n_{Li}/n_e=1\cdot10^{-6}$) is to allow for the prediction of the lithium density profiles in the simulation.

Each individual local $\delta f$ \CGYRO simulation was performed on the GPU partition of the NERSC Perlmutter supercomputer with each simulation utilizing 12 nodes, with each node comprised of 4 NVIDIA A100 (40GB) GPUs and requiring approximately 3-4 hours to achieve saturated conditions.
Therefore each ``black-box" evaluation (i.e. profile evaluation with 5 radial points) results in 720 GPU-hours (180 node-hours), making the use of surrogate-modeling techniques in \PORTALS more efficient than standard numerical methods, as described in Sec.~\ref{sec:Bench}.
As will be discussed in the following, 30 profile evaluations were required to achieve steady state, amounting to 21,600 GPU-hours for a 5-channel, self-consistent nonlinear \CGYRO prediction of the core plasma in the DIII-D ISS.

\begin{figure}
	\centering
	\includegraphics[width=1.0\columnwidth]{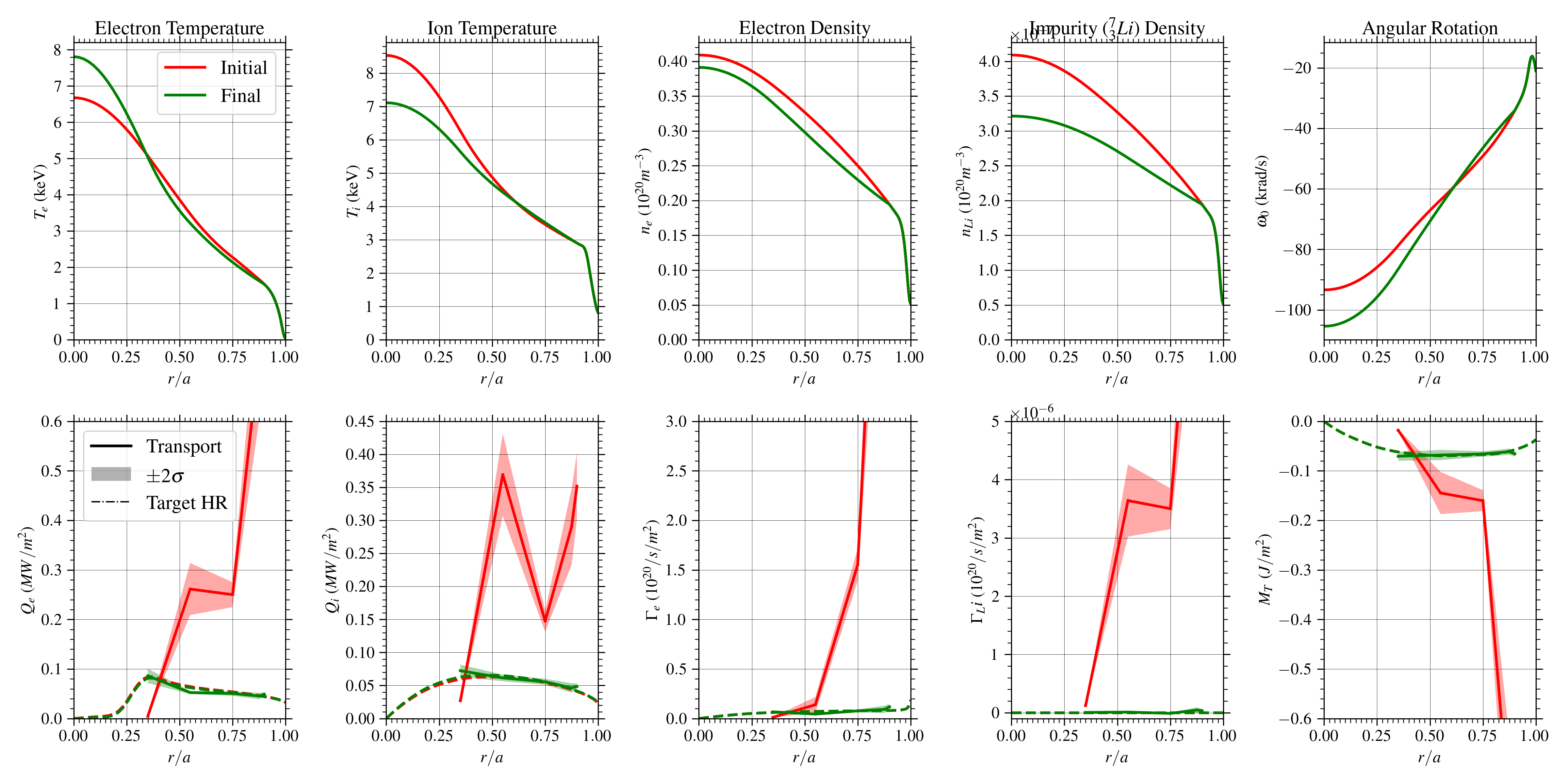}
	\caption{Summary of 5-channel prediction in DIII-D ISS with nonlinear \CGYRO and \NEO simulations. Kinetic profiles ($T_e$, $T_i$, $n_e$, $n_{Li}$, $\omega_0$) and corresponding transport flux (turbulent and neoclassical) and high-resolution (HR) target flux profiles. Two iterations are plotted: original (\#0) and best residual (\#26).}
	\label{Fig:multi}
\end{figure}

\begin{figure}
	\centering
	\includegraphics[width=1.0\columnwidth]{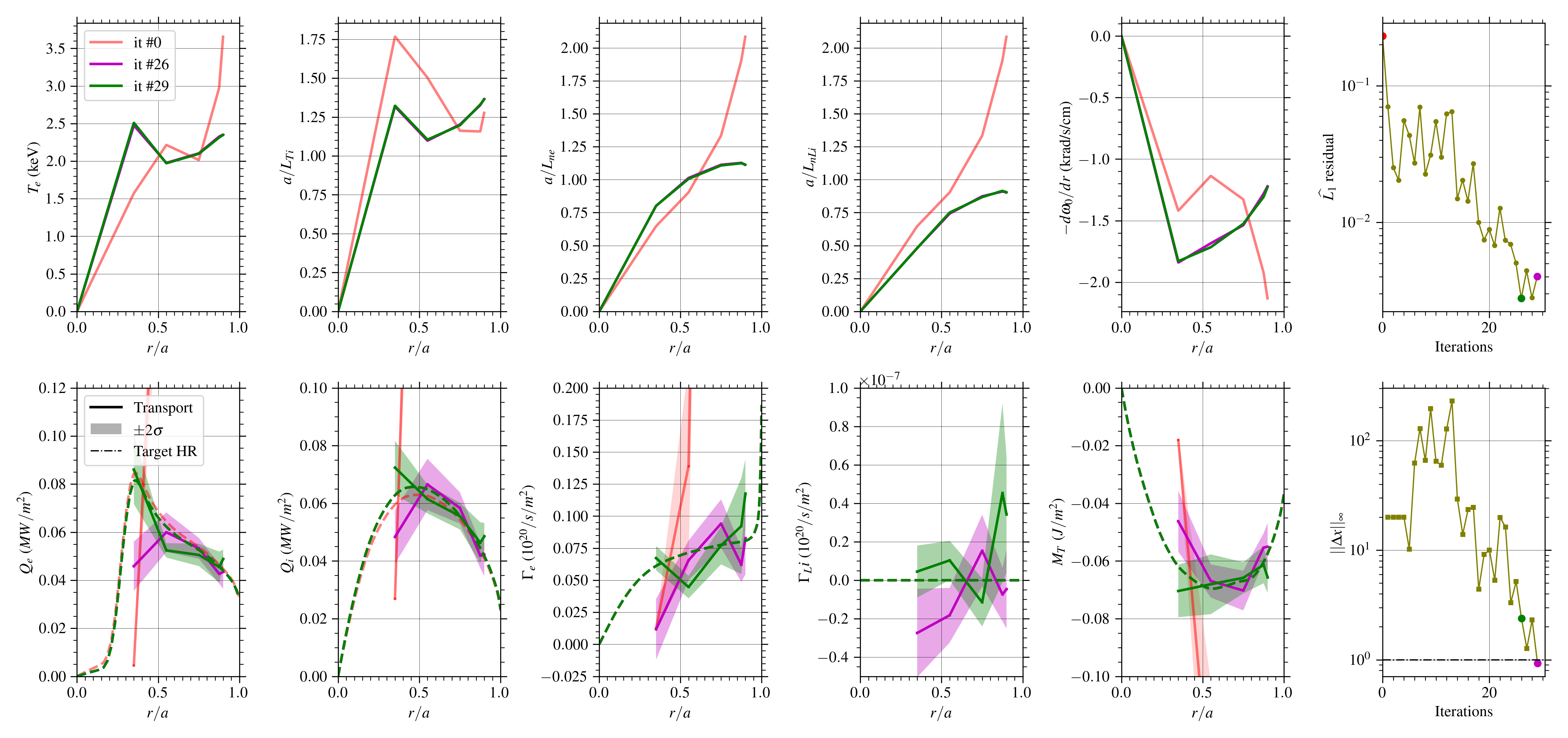}
	\caption{Investigation of stiff behavior. The gradient profiles of each kinetic profile are plotted for three iterations (\#0, \#26, \#29), along with the flux profiles (zoomed-in version of Fig.~\ref{Fig:multi}). Right most column depicts the $L_1$ residual (best residual achieved in iteration \#26, approximately 2 orders of magnitude lower than original) and the convergence metric $\|\Delta \mathbf{y}\|_\infty$ that represents the $L_\infty$ norm of the difference between the input parameters of the current iteration and the closest point evaluated. This metric reaches the stopping criterion of $1\%$, which indicates that the maximum change in any of the logarithmic gradient is below 1\%, demonstrating sufficient stationarity.
    }
	\label{Fig:multi2}
\end{figure}

Fig.~\ref{Fig:multi} shows the evolution of the 5 kinetic profiles from the original parameterized profiles (red) \cite{howard_simultaneous_2024} to the multi-channel flux-matched ones (green).
Starting transport fluxes (red lines in bottom subplots) are very far from targets, but after 26 transport evaluations per radius (130 nonlinear \CGYRO simulations), the plasma is brought to steady-state in all coupled channels simultaneously.
For visualization purposes, transport vs target fluxes are plotted in both Fig.~\ref{Fig:multi} and Fig.~\ref{Fig:multi2} with different y-axis range. 

From a naive inspection of the flux-matching profiles of transport and target fluxes for the best residual plasma, one could argue that the plasma is not truly in steady-state, as target fluxes are not within confidence bounds of the turbulent fluxes evaluations.
However, this is a consequence of stiff transport behavior and extreme sensitivity with respect to input parameters, as it was also reported for the SPARC PRD in Ref.~\cite{rodriguez-fernandez_nonlinear_2022}.
Fig.~\ref{Fig:multi2} plots two cases towards the end of the simulation: the best residual case and the last one evaluated.
The gradient profiles are nearly unchanged, yet the fluxes jump in some locations from below to above targets.
This somewhat oscillatory behavior could be resolved by running extremely long $\delta f$ simulations, where any error introduced by the limited time averaging is suppressed. However, such simulations will not give any more accuracy to the profile predictions, which are already converged and no significant changes to gradients are expected to happen.

\begin{figure}
	\centering
	\includegraphics[width=1.0\columnwidth]{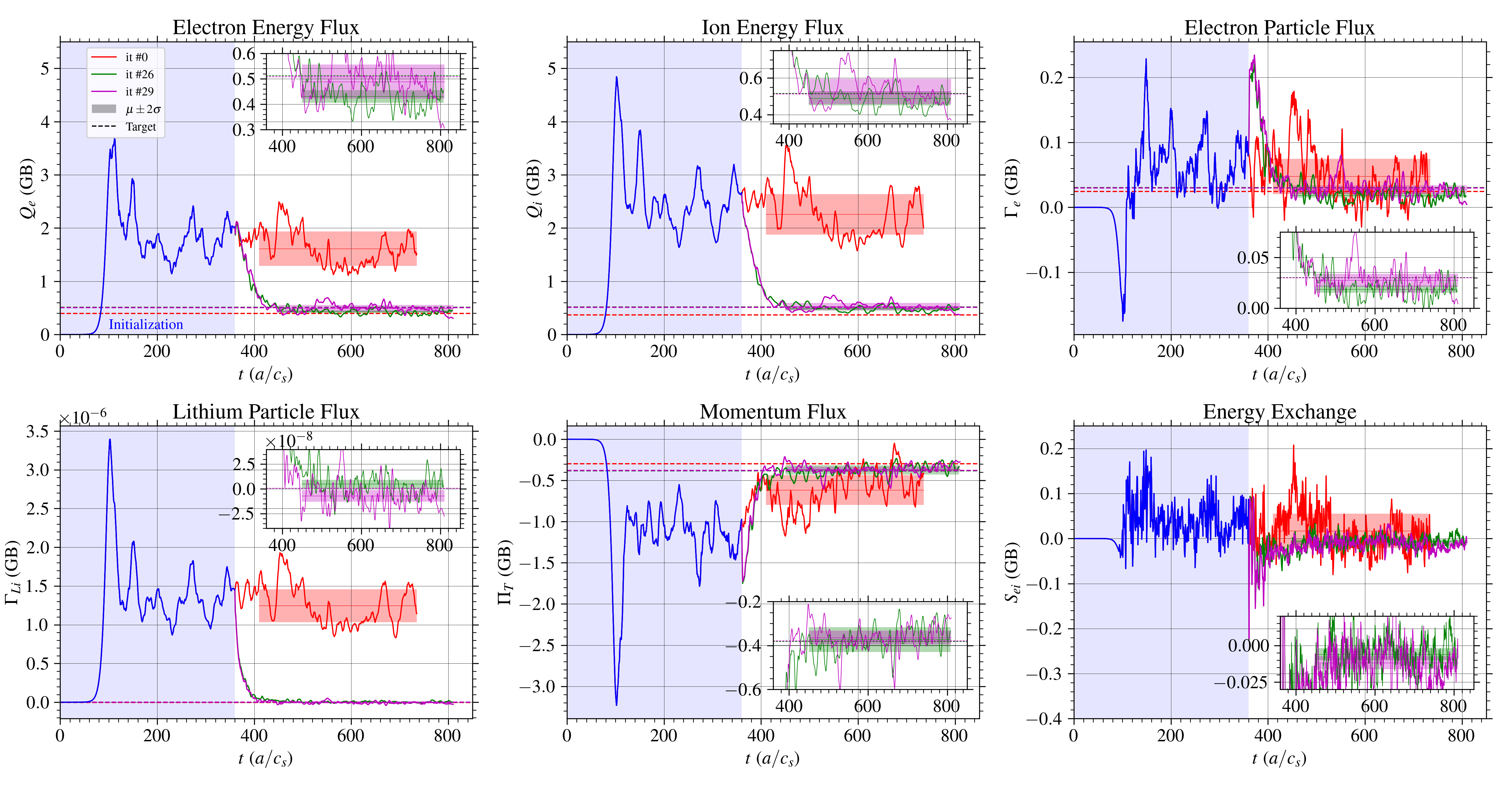}
	\caption{Time traces of turbulent transport fluxes from \CGYRO simulations at $r/a=0.55$.
 Six moments (electron energy, ion energy, electron particle, lithium particle, momentum and energy exchange fluxes) are plotted for the same three iterations in Fig.~\ref{Fig:multi2}.
 Each plot contains the value of the target turbulent flux (targets with neoclassical transport subtracted) as dashed horizontal lines and the turbulent transport flux as solid horizontal line, represented as the mean of the time traces once they saturate ($\Delta t\sim350 a/c_s$) with estimated confidence bounds around them.
 Each plot contains also a zoomed-in subplot around targets for the near-convergence cases.
    }
	\label{Fig:turbulence}
\end{figure}

Fig.~\ref{Fig:turbulence} shows the time traces of the \CGYRO $\delta f$ simulations at $r/a=0.55$ as representative location.
Simulations are restarted from a baseline case (blue shaded region) that was run for $\sim360 a/c_s$ and each of them is let to evolve to their corresponding saturated state.
At this location it was assumed that running for an additional $\sim400 a/c_s$ was sufficient to evaluate the saturated state, which was determined by visual inspection, but metrics for automation of \PORTALS are under investigation.
\referee{We must note that this approach ---given the limited time averaging caused by the otherwise prohibitely cost of running for longer time periods--- would not capture changes in the transport fluxes from turbulence and zonal flow patterns that may developed on long time scales, as observed in some gyrokinetic studies \cite{weikl_ion_2017,rath_transport_2022}.
This is generally a limitation of flux-matching frameworks of $\delta f$ simulations that aim to achieve steady-state with high-fidelity gyrokinetic simulations, under the constraint of computational cost.
However, most long-time patterns, such as those in the aforementioned papers, typically show early signs (e.g., intermittency or bursts) that might trigger a collapse. These are features that are actively looked for in the simulations used to bring the plasma to steady-state in our workflow, and have not been observed in the specific regimes that \PORTALS has been run on so far, including the predictions showed in this paper.
}

The mean flux in the saturated state is determined by the mean of the time trace in the saturated region, and the confidence bounds are $2\sigma_S$, where $\sigma_S=\frac{\sigma}{\sqrt{N}}$. Here, $\sigma$ is the standard deviation of the time trace and $N$ is the number of independent samples, found by decomposing the full time window into periods of $3\tau_C$, where $\tau_C$ is the autocorrelation time.
This assumes that time points separated by more than 3 autocorrelation periods can be considered to be independent from each other and contribute to the reduction of the total error of the time signal.
Here we also make the assumption that target and neoclassical fluxes have no error associated to their calculation, which could be included in future work when constructing better metrics to determine convergence in \PORTALS simulations.
\referee{This utilization of the estimation of uncertainties in observed data to build more robust surrogate models enabled by the built-in capabilities of the Gaussian processes is one of the key advantages of \PORTALS, as discussed in Section~\ref{sec:Discussion}.
The details of the implementation of fixed Gaussian noise (coming from this estimation of the error in the time evolution of the saturated state in the initial value simulations) is further discussed in Ref.~\cite{rodriguez-fernandez_nonlinear_2022}.}

\section{Discussion}
\label{sec:Discussion}

This paper has presented the \PORTALS framework that formulates the steady-state multi-channel transport equations in the plasma core as a surrogate-based optimization problem.
This formulation has allowed arguably the highest fidelity predictions of kinetic profiles and performance in tokamaks, by the direct use of the \CGYRO code for steady-state flux matching.
A unique, 5-channel nonlinear gyrokinetic prediction of core profiles was presented, which only required 26 evaluations to attain enough convergence of the profiles.
The difficulties of dealing with stiff transport and its effect on the development of robust convergence metrics were discussed.
In this section we share some thoughts on why the field of magnetic-confinement fusion, as it enters the era of burning plasma experiments and power plant design, may need high-fidelity transport physics and simulations.

\subsection{On the need for self-consistent, multi-channel predictions}
\label{sec:need}

\paragraph{For prediction of burning plasmas and reactor scoping}

At a time when the fusion community is focusing on designing potential fusion power plants and when the planning of the experimental operational campaigns of ITER \cite{Doyle2007} and SPARC \cite{Creely2020a} is requiring careful, predict-first simulations, the development of techniques that can accurately describe plasma profiles and performance is critical.
Historically, tokamak devices have been designed using simplified models of transport and performance.
In fact, the use of global energy confinement scaling laws, along with empirical formulas for peaking factors and functional forms for profile shapes, is widespread in the community.
While such approaches are useful to have a quick view of potential operational scenarios in new machines and as a check for simulation results, the uncertainties in the predictions are too high to trust fusion endeavors solely to such simplified formulas.

Generally speaking, there are two types of uncertainties introduced by the use of empirical scaling laws to project burning-plasma performance.
First, capturing the complex, high-dimensional dynamics of tokamak plasmas with just a few global metrics is, by nature, an approach subject to inaccuracies. Some of these errors are captured by the fit error bars of scaling laws, but higher errors may be present if one accounts for measurement uncertainties as well.
The extrapolation of current scaling laws to burning-plasma regimes, with their high degree of non-linearity magnifies this error.
Predictions made with a physics-based understanding of the underlying dynamics \cite{angioni_dependence_2023,Angioni2022} are a powerful tool for reducing this uncertainty.
Extra attention must be taken if power plant designs are aimed at exploiting unexplored regimes that were not fully accounted for when building the scalings laws or saturation rules for quasilinear models, such as negative triangularity plasmas \cite{frank_radiative_2022} or non-conventional aspect ratios \cite{wilson_steppathway_2020}.
Operational regimes with high plasma rotation, high positive triangularity and high core radiation fractions are also examples where standard empirical scalings laws are more prone to inaccuracies, although most recent updates to the ITPA scaling laws are promising \cite{Verdoolaege2021}.
Secondly, even if confinement and peaking were predicted exactly, the choice of parameterized functional forms for temperature and density profiles introduces a non-negligible uncertainty. This explains why the performance of the SPARC PRD plasma varied from $Q\approx11.0$ \cite{Creely2020a} (empirical) to $Q\approx9.2$ \cite{Rodriguez-Fernandez2020a} (physics-based) as published in its physics basis, even though the predicted peaking and confinement with physics models were the same as those given by the empirical formulas.

\begin{figure}
    \centering
    \includegraphics[width=1.0\columnwidth]{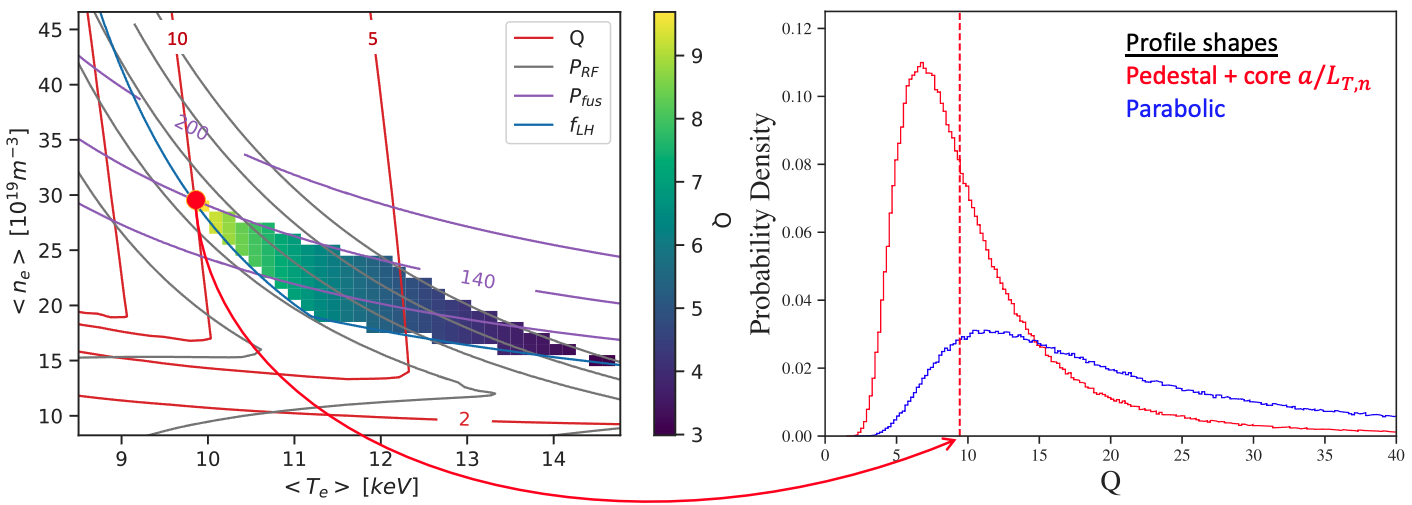}
    \caption{Monte-Carlo analysis of fusion gain in SPARC PRD plasma. 
    (a) POPCON for the SPARC PRD plasma, with operational point at the intersection of L-H power transition curve and the $140MW$ nominal fusion power limit.
    (b) Histogram of fusion gain for $10^6$ randomly drawn samples from independent Gaussian distributions (with $\sigma=5\%$) of transport-related assumptions into POPCONs: $H_{98,y2}$ ($\mu=1.0$), $\nu_{n_e}$ ($\mu=\nu_{scaling}$ \cite{Angioni2007}), $T_i/T_e$ ($\mu=0.85$), $\nu_T$ ($\mu=2.5$) and $a/L_{T}$ ($\mu=2.5$).
    Two different profile shapes (parabolic and pedestal-$a/L_T$) are analyzed.}
    \label{Fig:Q_MC}
\end{figure}

To better visualize the effect of both of these uncertainties, Figure~\ref{Fig:Q_MC} shows the variation of fusion gain ($Q=P_{fus}/P_{in}$) for a potential SPARC PRD operational point when transport-related input parameters to the Plasma OPeration CONtours (POPCON) analysis \cite{battaglia_cfs-energycfspopcon_2023} are drawn from Gaussian distributions.
Chosen distributions (mean and standard deviation) for this analysis are specified in the figure footnote.
We note that the $5\%$ standard deviation assumed here, although ad-hoc, is on the conservative end.
POPCON assumptions were similar to those in Refs~\cite{Creely2020a,Rodriguez-Fernandez2022}.
In the same analysis, we also included the uncertainty propagation analysis if a physics-informed profile shape is used instead of parabolic. For this physics-informed profile shape, it is assumed that the core has constant gradient scale lengths ($a/L_T$ and $a/L_n$) and an edge pedestal is varied to match the choice of $H_{98,y2}$ and volume average density.

Fig.~\ref{Fig:Q_MC} unequivocally demonstrates that even assuming rather small uncertainties in transport assumptions in empirical modeling can lead to drastically different fusion gain results.
As changing design parameters in burning-plasma experiments is costly and time consuming, it becomes clear that incorporating high physics fidelity simulations early in the design of new devices is key to the success of fusion as an energy source.
We note that even though we took SPARC PRD as an example, the same results apply to any burning-plasma and reactor-relevant scenario.
It is expected that reactors near ignition conditions or with high fusion gains may potentially be more sensitive to transport assumptions.
Furthermore, in the example of Fig.~\ref{Fig:Q_MC} it was assumed that the plasma remains in H-mode (i.e., energy confinement scaling $\tau_E^{98y2}$ is applicable) throughout the parameter scans. L-H power threshold considerations and its associated uncertainties will certainly further widen the probably density of fusion gain and fusion power.
In the case of the SPARC PRD plasma, the risk from relying on empirical scalings and POPCON analysis has been mitigated by comprehensive transport modeling via quasilinear simulations in \TRANSP-\TGLF \cite{Rodriguez-Fernandez2020a}, standalone gyrokinetics with \CGYRO \cite{Howard2021} and self-consistent \PORTALS-\CGYRO simulations \cite{rodriguez-fernandez_nonlinear_2022}.

\paragraph{For validation of gyrokinetics in current experiments}

Prior to predicting new devices, validating our understanding and simulation codes of core turbulence is key to build confidence in the results.
Validation in core turbulence research consists of comparing transport simulations to experimental results \cite[and references therein]{White2019}.
Due to the high computational cost of gyrokinetic simulations, validation studies in this area have historically focused on using experimental gradients as inputs to simulation codes or, at most, 1- or 2-dimensional scans of driving gradients (e.g. $a/L_{T_i}$ and $a/L_{T_e}$).
The concept of ``$Q_i$-matched" simulation has been used in the literature to refer to simulations that recover the experimentally-inferred ion heat flux by varying experimental inputs within error bars.
Given the stiff-transport nature of turbulence and the often large uncertainties in experimental gradients, it is common that a $Q_i$-matched simulation exists within error bars \cite{Rodriguez-Fernandez2018a}.
However, such simulation may not necessarily reproduce transport in other channels, such as electron energy and particle transport. Additionally, flux-matching in this context often refers only to local results, and propagation of gradient variations (with the associated changes in plasma parameters and power flows) from edge to core are rarely accounted for.

\begin{figure}
    \centering
    \includegraphics[width=0.5\columnwidth]{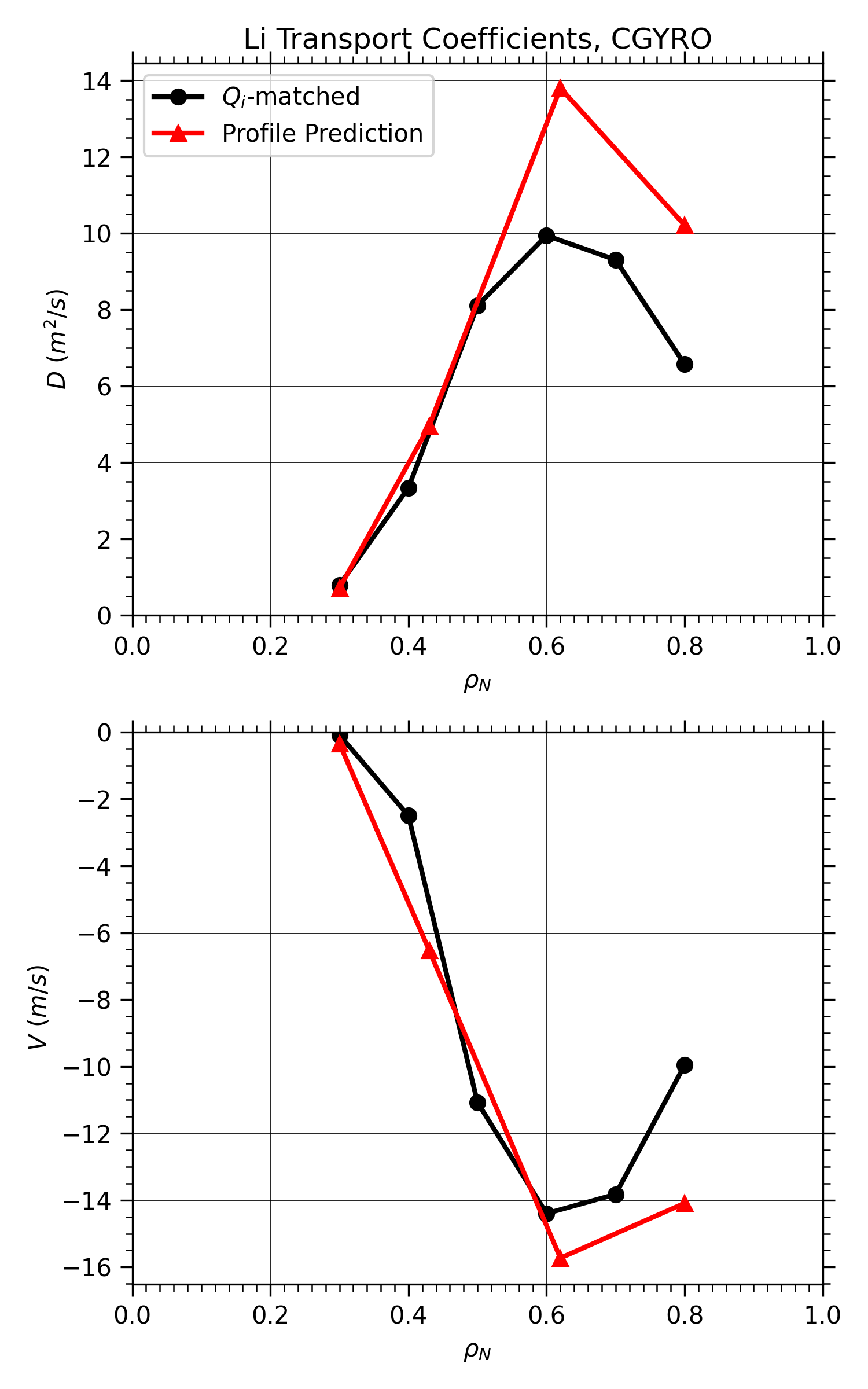}
    \caption{
    Example of predicted lithium particle transport coefficients with \CGYRO for the DIII-D simulation presented in Ref.~\cite{howard_simultaneous_2024}. Coefficients calculated with background profiles from (black) $Q_i$-matched, standalone simulations and (red) 3-channel profile predictions.
    }
    \label{Fig:validation}
\end{figure}

To illustrate this, Fig.~\ref{Fig:validation} displays the differences of the calculated impurity particle transport coefficients with \CGYRO between using the self-consistent, 3-channel predicted profiles or only using the local, standalone, $Q_i$-matched gradients.
This example correspond to the DIII-D ISS plasma \cite{howard_simultaneous_2024}.
Differences are such that can lead incorrect validation conclusions if not self-consistent results are compared to experimentally-inferred transport coefficients.

\referee{This work, particularly the capability of \PORTALS to bring to steady-state multi-channel profiles, including rotation, opens avenues for more comprehensive validation studies in current experiments.
This can be important to, for example, complement validation studies of turbulent transport models to capture intrinsic rotation and momentum transport dynamics \cite{grierson_main-ion_2017,hornsby_global_2018,hornsby_effect_2017}.
}

\subsection{Advantages and challenges of \PORTALS methods}
\label{sec:Advantages}

This paper has presented a newly developed framework to enable flux matching and has presented benchmarks and comparisons with standard numerical methods.
The \PORTALS framework has generally delivered convergence at lower number of profile evaluations than standard Newton and simple relaxation methods.
With the risk of oversimplification, this can be explained from the perspective of \PORTALS ---or any Bayesian optimization framework with only-exploitation acquisition functions\footnote{We would like to note that, usually, the concept of Bayesian optimization is intimately tied to trading exploration and exploitation and therefore the workflow presented here and implemented so far in \PORTALS could be referred to as surrogate-based optimization.}--- being a generalization of local Jacobian methods. In the latter, first order derivatives are constructed usually by assuming a linear model in the vicinity of the current point. This derivative information is then used to inform the next point to evaluate, that is predicted by the linear model to move towards convergence (e.g., reducing residual). In such methods, however, the information of previous iterations is lost, and new linear models are constructed locally at each iteration.
When formulating the problem with surrogate-based optimization, a global model is constructed by utilizing all evaluations that were previously performed, maximizing the information leveraged for the optimization.
In the context of purely-exploitation acquisition functions (as the one used here, mean of the posterior distribution), the next point is chosen such that it is predicted to optimize the underlying function the most (in expectation).
In core transport modeling particularly with first-principles simulations, there is certain smoothness to be expected (under the assumption of infinite simulation time) between transport fluxes and input gradients that drive turbulence. Critical gradient behavior is usually well defined within $\delta f$ codes and therefore the use of global surrogates provide important advantages.

The seemingly high dimensionality of the problem (e.g., 6 input variables in a 3-channel prediction) for the low number of samples ($<20$ in most cases studied, usually $12$-$16$ iterations) can be explained from different perspectives.
First, Gaussian processes are well known to be highly sample-efficient surrogate models. 
Second, the \textit{effective dimensionality} of turbulence during the flux-matching problem is usually lower due to the higher influence of logarithmic gradients on the outcome and somewhat ``residual effect'' of additional terms such as $T_i/T_e$ and $\widehat{\nu}_{ei}$.
This is quickly picked up by the surrogates, mostly as a consequence of the specialized initial training that scans gradients and quickly drives the system towards the region around marginal stability.
While the critical gradient may be in turn affected by these ``residual" terms (such as the effect of $T_i/T_e$), most of the parameter sensitivity is captured by the logarithmic gradients.
Once the plasma searches around the region of marginal stability (with large variation in fluxes), the ``residual" parameters constructed from $y_{j,\forall c}=\{n_e,T_e,T_i,\omega_0,n_Z\}$ do not vary significantly and their effect in the background turbulence is small, an effect that is achieved by the outcome transformation described in Sec.~\ref{sec:Outcome}.
The formulation of the transport fluxes in gyro-Bohm units ensures that most turbulent transport sensitivities are captured by the logarithmic gradients.
Future work will explore the possibility to encompass the ``residual" parameters into a correction factor to the GPs, which can further aid convergence.

The use of Gaussian process regression models as surrogates works well with limited time-averaging when turbulence saturation is defined, as estimated error bars can be used for model training.
This provides important advantages to model pressure gradient driven turbulent transport, as it is often the case that in near-marginal stability conditions large oscillations, predator-prey behavior and the need for long time averages are required. By the proper definition of evaluation error and its use to train the surrogate models, we prevent the solver from getting stuck below or at critical gradients.
Furthermore, it is generally the case that \PORTALS does not ``waste time'' in stable regions, with otherwise small local derivatives with respect to gradients, as the global nature of the surrogates can quickly bring the plasma to turbulent conditions, by making larger steps than would otherwise be requested by local Jacobian methods with fixed step size.

An additional important advantage of \PORTALS is that by further de-coupling the macroscopic flux-matching problem from the local $\delta f$ simulations we ensure robustness against nonlinear transport-target instabilities.
In other words, because the macroscopic flux-matching is performed with the surrogate models, large number of evaluations can be made, and heuristic, global methods such as genetic algorithms, can be leveraged to find flux-matching conditions (e.g., as in Ref~\cite{Honda2018}).
Cases with targets that are a strong function of profiles (e.g., burning plasmas via alpha heating, Ohmic plasmas via energy exchange), do not require more iterations than plasmas with de-coupled sources.

On the other hand, the formulation in \PORTALS presents also challenges compared to standard methods.
As discussed in Sec.~\ref{sec:Bench}, \PORTALS is a tool that is more efficient with high-fidelity modeling (e.g., nonlinear gyrokinetics), as the overhead introduced by both surrogate fitting and acquisition optimization can be substantial.
This makes the use of these methods with low (e.g., analytical) and medium fidelity (e.g., quasilinear) transport models more expensive.
While this is generally a limitation of Bayesian optimization ---developed explicitly for expensive black-box function optimization--- future work will be devoted to make the surrogate training and acquisition optimization more efficient.

As with most probabilistic-modeling optimization methods, the techniques developed here are subject to some vulnerabilities:
\\
1) \textbf{Sensitivity to bounds}: allowable values for the choice of free logarithmic gradients may limit the search space. This challenge could be mitigated by allowing extrapolations of the GP outside of its training bounds, or to perform a pre-optimization step that characterizes the magnitude of the fluxes vs reasonable targets and selects the appropriate bounds.
In the profile predictions performed so far, defining bounds $\pm50\%$ or $\pm100\%$ the experimental (or predicted by quasilinear models) gradients was enough to ensure the existence of a steady-state solution for the high fidelity nonlinear gyrokinetic system, but this remains a choice of the expert user, as a trade-off with surrogate accuracy may be considered. If bounds are too large, very stiff behavior or transitions between unstable micro-instabilities can result in the requirement to observe more data points for accurate surrogate representation of transport fluxes.
\\
2) \textbf{Overfitting}: small training sets could limit \PORTALS' optimization potential. This challenge could be mitigated by the introduction of some exploration in the acquisition function, either through the acquisition function definition (e.g., expected improvement or others) or via the inclusion of random samples recurrently.
\\
3) \textbf{Scale disparity}: difficulties arise when training GP models that aim to capture outputs of very dissimilar magnitude (e.g., order of magnitude differences between near-stable and unstable turbulence), with heteroskedastic noise estimations (e.g., differences in turbulence autocorrelation periods depending on gradient value), and with covariance length-scales strongly non-stationary (e.g., pre and post critical gradient).
These issues can be approached by the development of proper pre-training transformation, better priors and kernel definitions, which will be subject of future work.
\\
4) \textbf{Oscillatory convergence}: defining proper convergence metrics becomes challenging. As discussed in Sec.~\ref{sec:5c}, the stiff behavior of tokamak core transport results in large variations in fluxes that may oscillate around targets with minimal change in free parameters. When using fast transport models, convergence metrics can be developed that account for how much variation in input space has happened for a number of iterations, but the application of such techniques with expensive gyrokinetic codes remains challenging.

\subsubsection{Potential for future work}
\label{sec:Future}

While the \PORTALS techniques have shown promising results and great performance in the predictions using nonlinear \CGYRO, there are potential avenues for improvement.
As part of future work, further development of the surrogate-based optimization techniques will involve a more careful selection and investigation of Bayesian optimization aspects to further increase the efficiency of the formulation.
These may include physics-informed GP surrogate models (e.g. that capture critical gradient behavior \referee{and that considers the expected smoothness of the flux responses}), other acquisition functions (accounting for the exploration / exploitation tradeoff) and better metrics to determine if converged-gradients solutions have been achieved.

In the context of optimization and reactor scenario design, the techniques developed here are also suitable for the high-fidelity optimization with direct numerical simulations.
During the construction of surrogate models and as the simulation approaches converged solutions, it is possible to leverage the current information for decision making and modification of input parameters that were not part of the original free parameters.
For example, as kinetic profiles are predicted with fixed geometry, if the plasma is predicted to not reach desirable performance goals (e.g. fusion power or L-H thresholds), it is possible to modify the geometry and drive the system to optimal solutions before converging the profiles in the early stages.
This sort of ``early stopping" or ``sequential decision making" can be vital to maximize the high-fidelity physics information during the design of new fusion devices and will be explored as part of future work.

\referee{An interesting potential avenue for future work is the exploitation of the uncertainty quantification properties of the GP surrogate models for sensitivity studies of modeling assumptions. The propagation of uncertainties in the assumed models or boundary conditions could provide valuable insights into the robustness of the predictions and their validation against experimental data.
Once trained, the GP models can be used to perform sensitivity studies of the input parameters, by finding flux-matched conditions not only for the mean of the posterior distribution, but also for the upper and lower bounds of the confidence intervals.
This capability has not been explored yet but will be subject of future investigation.
}

\section{Conclusions and Prospects}
\label{sec:Conclusions}

This paper has presented the approach of formulating the multi-channel steady-state transport equations in tokamak geometry as a surrogate-based optimization problem. 
The implementation of this approach in \PORTALS is demonstrated to reduce the number of $\delta f$ transport simulations required to bring the plasma to flux-matching conditions, as compared to standard methods.
Advantages of the framework were discussed, particularly those relevant to power plant design and burning plasma simulations.
\referee{This paper has not delved into any particular transport model validation study nor has provided practical burning plasma predictions ---which are done as part of separate publications (e.g. \cite{rodriguez-fernandez_nonlinear_2022,howard_2024_iter,howard_simultaneous_2024,holland_2023_aps,rodriguez-fernandez_core_2024,prf_2023_eps})---.
In describing this unique formulation and solution to the steady-state transport problem, we aimed to offer a comprehensive perspective on \PORTALS' potential and its implications for future fusion research.
}

After outlining the general guidelines for the accurate prediction of burning plasmas and demonstrating the efficiency of \PORTALS, it is essential to look towards the future landscape of fusion research.
As the field progresses within the burning plasma era, with new experiments coming online soon, the dynamics of turbulence and core plasma transport remain central to designing efficient reactors that can adapt to energy market needs while remaining economically attractive.
The behaviors and intricacies of turbulence and transport are pivotal for optimizing the operational point and performance of fusion devices, as discussed and demonstrated in previous sections.

While the foundational theories of nonlinear gyrokinetics and new workflows that leverage advances in computer and data science have provided robust frameworks to simulate and predict core performance (such as \PORTALS-\CGYRO presented here but also \TANGO-\GENE \cite{siena_global_2022,siena_predictions_2023} and \TRINITY-\GX \cite{mandell_gx_2022} targeted for high-fidelity transport modeling), there remains an ongoing endeavor to refine and accelerate plasma models.
Furthermore, this era, marked by novel experiments coming online and advancements in diagnostics in current facilities, offers a unique vantage point.
Validation of transport models, physics assumptions and numerical techniques will be crucial going forward.
Real-world experimental data from D-D and D-T facilities will serve as benchmarks against which turbulence models and simulation techniques can be tested, ensuring accurate representation of the underlying physics and reliable prediction of fusion energy systems behavior. This is particularly important for alpha heating physics and the study of power plant relevant conditions at the plasma edge of D-T plasmas.
It will also be an exciting frontier to consider using surrogate-based models to accelerate multi-machine and multi-physics validation.

As new experimental findings help us shape the understanding of burning-plasma dynamics, the importance of collaboration between experimentalists and theorists and between private and public efforts will be paramount and the only way to unlock the full potential of fusion.
This synergy will drive improvements in prediction accuracy and computational efficiency, ultimately enabling more accurate control and optimization of core performance.
It is, in fact, interesting to look at the history of the fission industry and analyze the potential parallels for anticipating the evolution of core transport in fusion research. With the establishment and maturation of commercial fission reactors, the emphasis in fundamental nuclear reactor physics and neutron transport research shifted towards optimization, reducing reactor cost, increasing reliability and addressing emerging challenges.
We expect, and already observe, a similar trend in our community, with shifted focus on optimizing operational points, accelerating transport models and starting to use simulations during the design of fusion power plants.

Building on our understanding and new approaches to plasma dynamics, our work with accelerated core transport solvers and GPU-accelerated transport models opens doors to faster advancements in fusion reactor design and operation, along with the development of better models with more physics included.
Our strategy, enriched by real-world experiments and innovative computational tools, is geared up to tackle the challenges and unlock new potentials in core research, which was only possible thanks to decades of pioneering work by theorists, modelers and experimentalists in many scientific fields.
This collaborative and integrative effort signifies a progression from foundational theories to practical application, aimed at optimizing the potential of fusion energy in meeting contemporary needs.

\section*{Acknowledgments}

\referee{
The \PORTALS framework has recently been published open source as part of the MIT Integrated Modeling (MITIM) repository \cite{mitim}.}
The authors would like to thank W. Dorland, N.R. Mandell and A. di Siena for fruitful conversations on high-fidelity transport solvers.
We also appreciate discussions with colleagues during the 31st and 32nd US Transport Taskforce meetings (2022-2023) and the 2nd Meta Adaptive Experimentation workshop (2023).
We thank E. Belli and G. Staebler for the development of \CGYRO and \TGLF, respectively.
This work was funded by Commonwealth Fusion Systems under RPP020 and US DoE under grants DE-SC0017992, DE-SC0014264, DE-SC0024399 and DE-SC0018287.
This research used resources of the National Energy Research Scientific Computing Center (NERSC), a US Department of Energy Office of Science User Facility located at Lawrence Berkeley National Laboratory, operated under Contract No. DE-AC02–05CH11231, for the \PORTALS-\CGYRO simulations.
Clusters hosted at the Massachusetts Green High Performance Computing Center (MGHPCC) were used to perform the \PORTALS-\TGLF benchmark simulations of SPARC and ARC (MIT-NSE and MIT-RPP partitions) and of DIII-D, ITER and JET (MIT-PSFC partition, funded by DoE grant number DE-FG02-91-ER54109).
GPT-4 was used to enhance parts of the manuscript for clarity and coherence, and to assist in the creation of Fig.~\ref{Fig:rendering}a.

\bibliography{References/references_fixed.bib,References/references_extra.bib}


\begin{thebibliography}{86}
\ifx \bisbn   \undefined \def \bisbn  #1{ISBN #1}\fi
\ifx \binits  \undefined \def \binits#1{#1}\fi
\ifx \bauthor  \undefined \def \bauthor#1{#1}\fi
\ifx \batitle  \undefined \def \batitle#1{#1}\fi
\ifx \bjtitle  \undefined \def \bjtitle#1{#1}\fi
\ifx \bvolume  \undefined \def \bvolume#1{\textbf{#1}}\fi
\ifx \byear  \undefined \def \byear#1{#1}\fi
\ifx \bissue  \undefined \def \bissue#1{#1}\fi
\ifx \bfpage  \undefined \def \bfpage#1{#1}\fi
\ifx \blpage  \undefined \def \blpage #1{#1}\fi
\ifx \burl  \undefined \def \burl#1{\textsf{#1}}\fi
\ifx \doiurl  \undefined \def \doiurl#1{\url{https://doi.org/#1}}\fi
\ifx \betal  \undefined \def \betal{\textit{et al.}}\fi
\ifx \binstitute  \undefined \def \binstitute#1{#1}\fi
\ifx \binstitutionaled  \undefined \def \binstitutionaled#1{#1}\fi
\ifx \bctitle  \undefined \def \bctitle#1{#1}\fi
\ifx \beditor  \undefined \def \beditor#1{#1}\fi
\ifx \bpublisher  \undefined \def \bpublisher#1{#1}\fi
\ifx \bbtitle  \undefined \def \bbtitle#1{#1}\fi
\ifx \bedition  \undefined \def \bedition#1{#1}\fi
\ifx \bseriesno  \undefined \def \bseriesno#1{#1}\fi
\ifx \blocation  \undefined \def \blocation#1{#1}\fi
\ifx \bsertitle  \undefined \def \bsertitle#1{#1}\fi
\ifx \bsnm \undefined \def \bsnm#1{#1}\fi
\ifx \bsuffix \undefined \def \bsuffix#1{#1}\fi
\ifx \bparticle \undefined \def \bparticle#1{#1}\fi
\ifx \barticle \undefined \def \barticle#1{#1}\fi
\bibcommenthead
\ifx \bconfdate \undefined \def \bconfdate #1{#1}\fi
\ifx \botherref \undefined \def \botherref #1{#1}\fi
\ifx \url \undefined \def \url#1{\textsf{#1}}\fi
\ifx \bchapter \undefined \def \bchapter#1{#1}\fi
\ifx \bbook \undefined \def \bbook#1{#1}\fi
\ifx \bcomment \undefined \def \bcomment#1{#1}\fi
\ifx \oauthor \undefined \def \oauthor#1{#1}\fi
\ifx \citeauthoryear \undefined \def \citeauthoryear#1{#1}\fi
\ifx \endbibitem  \undefined \def \endbibitem {}\fi
\ifx \bconflocation  \undefined \def \bconflocation#1{#1}\fi
\ifx \arxivurl  \undefined \def \arxivurl#1{\textsf{#1}}\fi
\csname PreBibitemsHook\endcsname

\bibitem{rodriguez-fernandez_nonlinear_2022}
\begin{barticle}
\bauthor{\bsnm{Rodriguez-Fernandez}, \binits{P.}},
\bauthor{\bsnm{Howard}, \binits{N.T.}},
\bauthor{\bsnm{Candy}, \binits{J.}}:
\batitle{Nonlinear gyrokinetic predictions of {SPARC} burning plasma profiles
  enabled by surrogate modeling}.
\bjtitle{Nuclear Fusion}
\bvolume{62}(\bissue{7}),
\bfpage{076036}
(\byear{2022}).
\doiurl{10.1088/1741-4326/AC64B2}.
\bcomment{Publisher: IOP Publishing}.
2022-05-16
\end{barticle}
\endbibitem

\bibitem{Candy2016}
\begin{barticle}
\bauthor{\bsnm{Candy}, \binits{J.}},
\bauthor{\bsnm{Belli}, \binits{E.A.}},
\bauthor{\bsnm{Bravenec}, \binits{R.V.}}:
\batitle{A high-accuracy {Eulerian} gyrokinetic solver for collisional
  plasmas}.
\bjtitle{Journal of Computational Physics}
\bvolume{324},
\bfpage{73}--\blpage{93}
(\byear{2016}).
\doiurl{10.1016/j.jcp.2016.07.039}.
\bcomment{Publisher: Elsevier Inc.}
\end{barticle}
\endbibitem

\bibitem{Doyle2007}
\begin{barticle}
\bauthor{\bsnm{Doyle}, \binits{E.J.}},
\bauthor{\bsnm{Houlberg}, \binits{W.A.}},
\bauthor{\bsnm{Kamada}, \binits{Y.}},
\bauthor{\bsnm{Mukhovatov}, \binits{V.}},
\bauthor{\bsnm{Osborne}, \binits{T.H.}},
\bauthor{\bsnm{Polevoi}, \binits{A.}},
\bauthor{\bsnm{Bateman}, \binits{G.}},
\bauthor{\bsnm{Connor}, \binits{J.W.}},
\bauthor{\bsnm{Cordey}, \binits{J.G.}},
\bauthor{\bsnm{Fujita}, \binits{T.}},
\bauthor{\bsnm{Garbet}, \binits{X.}},
\bauthor{\bsnm{Hahm}, \binits{T.S.}},
\bauthor{\bsnm{Horton}, \binits{L.D.}},
\bauthor{\bsnm{Hubbard}, \binits{A.E.}},
\bauthor{\bsnm{Imbeaux}, \binits{F.}}, \betal:
\batitle{Progress in the {ITER} {Physics} {Basis} {Chapter} 2: {Plasma}
  confinement and transport}.
\bjtitle{Nuclear Fusion}
\bvolume{47},
\bfpage{18}
(\byear{2007}).
\doiurl{10.1088/0029-5515/47/6/S02}
\end{barticle}
\endbibitem

\bibitem{Rodriguez-Fernandez2022}
\begin{barticle}
\bauthor{\bsnm{Rodriguez-Fernandez}, \binits{P.}},
\bauthor{\bsnm{Creely}, \binits{A.J.}},
\bauthor{\bsnm{Greenwald}, \binits{M.J.}},
\bauthor{\bsnm{Brunner}, \binits{D.}},
\bauthor{\bsnm{Ballinger}, \binits{S.B.}},
\bauthor{\bsnm{Chrobak}, \binits{C.P.}},
\bauthor{\bsnm{Garnier}, \binits{D.T.}},
\bauthor{\bsnm{Granetz}, \binits{R.}},
\bauthor{\bsnm{Hartwig}, \binits{Z.S.}},
\bauthor{\bsnm{Howard}, \binits{N.T.}},
\bauthor{\bsnm{Hughes}, \binits{J.W.}},
\bauthor{\bsnm{Irby}, \binits{J.H.}},
\bauthor{\bsnm{Izzo}, \binits{V.A.}},
\bauthor{\bsnm{Kuang}, \binits{A.Q.}},
\bauthor{\bsnm{Lin}, \binits{Y.}}, \betal:
\batitle{Overview of the {SPARC} physics basis towards the exploration of
  burning-plasma regimes in high-field, compact tokamaks}.
\bjtitle{Nuclear Fusion}
\bvolume{62}(\bissue{4}),
\bfpage{042003}
(\byear{2022}).
\doiurl{10.1088/1741-4326/AC1654}.
\bcomment{Publisher: IOP Publishing}.
2022-03-01
\end{barticle}
\endbibitem

\bibitem{frank_radiative_2022}
\begin{barticle}
\bauthor{\bsnm{Frank}, \binits{S.J.}},
\bauthor{\bsnm{Perks}, \binits{C.J.}},
\bauthor{\bsnm{Nelson}, \binits{A.O.}},
\bauthor{\bsnm{Qian}, \binits{T.}},
\bauthor{\bsnm{Jin}, \binits{S.}},
\bauthor{\bsnm{Cavallaro}, \binits{A.}},
\bauthor{\bsnm{Rutkowski}, \binits{A.}},
\bauthor{\bsnm{Reiman}, \binits{A.}},
\bauthor{\bsnm{Freidberg}, \binits{J.P.}},
\bauthor{\bsnm{Rodriguez-Fernandez}, \binits{P.}},
\bauthor{\bsnm{Whyte}, \binits{D.}}:
\batitle{Radiative pulsed {L}-mode operation in {ARC}-class reactors}.
\bjtitle{Nuclear Fusion}
\bvolume{62}(\bissue{12}),
\bfpage{126036}
(\byear{2022}).
\doiurl{10.1088/1741-4326/ac95ac}.
\bcomment{Publisher: IOP Publishing}.
2022-11-07
\end{barticle}
\endbibitem

\bibitem{meschini_review_2023}
\begin{botherref}
\oauthor{\bsnm{Meschini}, \binits{S.}},
\oauthor{\bsnm{Laviano}, \binits{F.}},
\oauthor{\bsnm{Ledda}, \binits{F.}},
\oauthor{\bsnm{Pettinari}, \binits{D.}},
\oauthor{\bsnm{Testoni}, \binits{R.}},
\oauthor{\bsnm{Torsello}, \binits{D.}},
\oauthor{\bsnm{Panella}, \binits{B.}}:
Review of commercial nuclear fusion projects.
Frontiers in Energy Research
\textbf{11}
(2023).
\doiurl{10.3389/fenrg.2023.1157394}.
2023-06-20
\end{botherref}
\endbibitem

\bibitem{Menard2022}
\begin{barticle}
\bauthor{\bsnm{Menard}, \binits{J.E.}},
\bauthor{\bsnm{Grierson}, \binits{B.A.}},
\bauthor{\bsnm{Brown}, \binits{T.G.}},
\bauthor{\bsnm{Rana}, \binits{C.}},
\bauthor{\bsnm{Zhai}, \binits{Y.}},
\bauthor{\bsnm{Poli}, \binits{F.M.}},
\bauthor{\bsnm{Maingi}, \binits{R.}},
\bauthor{\bsnm{Guttenfelder}, \binits{W.}},
\bauthor{\bsnm{Snyder}, \binits{P.B.}}:
\batitle{Fusion pilot plant performance and the role of a sustained high power
  density tokamak}.
\bjtitle{Nuclear Fusion}
\bvolume{62}(\bissue{3}),
\bfpage{036026}
(\byear{2022}).
\doiurl{10.1088/1741-4326/AC49AA}.
\bcomment{Publisher: IOP Publishing}.
2022-02-10
\end{barticle}
\endbibitem

\bibitem{doe_milestone}
\begin{botherref}
\oauthor{\bsnm{{U.S. Department of Energy}}}:
Milestone-Based Fusion Development Program, DE-FOA-0002809.
Founding Opportunity Announcements, DE-FOA-0002809
(2022).
\url{https://science.osti.gov/grants/FOAs/FOAs/2022/DE-FOA-0002809}
\end{botherref}
\endbibitem

\bibitem{Brizard2007}
\begin{botherref}
\oauthor{\bsnm{Brizard}, \binits{A.J.}},
\oauthor{\bsnm{Hahm}, \binits{T.S.}}:
Foundations of nonlinear gyrokinetic theory.
Reviews of Modern Physics
\textbf{79}
(2007).
\doiurl{10.1103/RevModPhys.79.421}
\end{botherref}
\endbibitem

\bibitem{krommes_nonlinear_2010}
\begin{barticle}
\bauthor{\bsnm{Krommes}, \binits{J.A.}}:
\batitle{Nonlinear gyrokinetics: a powerful tool for the description of
  microturbulence in magnetized plasmas}.
\bjtitle{Physica Scripta}
\bvolume{2010}(\bissue{T142}),
\bfpage{014035}
(\byear{2010}).
\doiurl{10.1088/0031-8949/2010/T142/014035}.
2023-02-12
\end{barticle}
\endbibitem

\bibitem{sugama_nonlinear_1998}
\begin{barticle}
\bauthor{\bsnm{Sugama}, \binits{H.}},
\bauthor{\bsnm{Horton}, \binits{W.}}:
\batitle{Nonlinear electromagnetic gyrokinetic equation for plasmas with large
  mean flows}.
\bjtitle{Physics of Plasmas}
\bvolume{5}(\bissue{7}),
\bfpage{2560}--\blpage{2573}
(\byear{1998}).
\doiurl{10.1063/1.872941}.
2023-11-30
\end{barticle}
\endbibitem

\bibitem{sugama_transport_1996}
\begin{barticle}
\bauthor{\bsnm{Sugama}, \binits{H.}},
\bauthor{\bsnm{Okamoto}, \binits{M.}},
\bauthor{\bsnm{Horton}, \binits{W.}},
\bauthor{\bsnm{Wakatani}, \binits{M.}}:
\batitle{Transport processes and entropy production in toroidal plasmas with
  gyrokinetic electromagnetic turbulence}.
\bjtitle{Physics of Plasmas}
\bvolume{3}(\bissue{6}),
\bfpage{2379}--\blpage{2394}
(\byear{1996}).
\doiurl{10.1063/1.871922}.
2023-11-30
\end{barticle}
\endbibitem

\bibitem{Jenko2000}
\begin{barticle}
\bauthor{\bsnm{Jenko}, \binits{F.}},
\bauthor{\bsnm{Dorland}, \binits{W.}},
\bauthor{\bsnm{Kotschenreuther}, \binits{M.}},
\bauthor{\bsnm{Kotschenreuther}, \binits{.M.}},
\bauthor{\bsnm{Rogers}, \binits{B.N.}}:
\batitle{Electron temperature gradient driven turbulence}.
\bjtitle{Physics of Plasmas}
\bvolume{7}(\bissue{5}),
\bfpage{1904}
(\byear{2000}).
\doiurl{10.1063/1.874014}.
\bcomment{Publisher: American Institute of PhysicsAIP}.
2021-12-11
\end{barticle}
\endbibitem

\bibitem{peeters_nonlinear_2009}
\begin{barticle}
\bauthor{\bsnm{Peeters}, \binits{A.G.}},
\bauthor{\bsnm{Camenen}, \binits{Y.}},
\bauthor{\bsnm{Casson}, \binits{F.J.}},
\bauthor{\bsnm{Hornsby}, \binits{W.A.}},
\bauthor{\bsnm{Snodin}, \binits{A.P.}},
\bauthor{\bsnm{Strintzi}, \binits{D.}},
\bauthor{\bsnm{Szepesi}, \binits{G.}}:
\batitle{The nonlinear gyro-kinetic flux tube code {GKW}}.
\bjtitle{Computer Physics Communications}
\bvolume{180}(\bissue{12}),
\bfpage{2650}--\blpage{2672}
(\byear{2009}).
\doiurl{10.1016/j.cpc.2009.07.001}.
2023-11-22
\end{barticle}
\endbibitem

\bibitem{mandell_gx_2022}
\begin{botherref}
\oauthor{\bsnm{Mandell}, \binits{N.R.}},
\oauthor{\bsnm{Dorland}, \binits{W.}},
\oauthor{\bsnm{Abel}, \binits{I.}},
\oauthor{\bsnm{Gaur}, \binits{R.}},
\oauthor{\bsnm{Kim}, \binits{P.}},
\oauthor{\bsnm{Martin}, \binits{M.}},
\oauthor{\bsnm{Qian}, \binits{T.}}:
{GX}: a {GPU}-native gyrokinetic turbulence code for tokamak and stellarator
  design
(2022).
\doiurl{10.48550/arXiv.2209.06731}.
\url{https://arxiv.org/abs/2209.06731v3}
2023-02-28
\end{botherref}
\endbibitem

\bibitem{Staebler2007}
\begin{barticle}
\bauthor{\bsnm{Staebler}, \binits{G.M.}},
\bauthor{\bsnm{Kinsey}, \binits{J.E.}},
\bauthor{\bsnm{Waltz}, \binits{R.E.}}:
\batitle{A theory-based transport model with comprehensive physics}.
\bjtitle{Physics of Plasmas}
\bvolume{14}(\bissue{5}),
\bfpage{055909}
(\byear{2007}).
\doiurl{10.1063/1.2436852}
\end{barticle}
\endbibitem

\bibitem{Bourdelle2015a}
\begin{barticle}
\bauthor{\bsnm{Bourdelle}, \binits{C.}},
\bauthor{\bsnm{Citrin}, \binits{J.}},
\bauthor{\bsnm{Baiocchi}, \binits{B.}},
\bauthor{\bsnm{Casati}, \binits{A.}},
\bauthor{\bsnm{Cottier}, \binits{P.}},
\bauthor{\bsnm{Garbet}, \binits{X.}},
\bauthor{\bsnm{Imbeaux}, \binits{F.}}:
\batitle{Core turbulent transport in tokamak plasmas: bridging theory and
  experiment with {QuaLiKiz}}.
\bjtitle{Plasma Physics and Controlled Fusion}
\bvolume{58}(\bissue{1}),
\bfpage{014036}
(\byear{2015}).
\doiurl{10.1088/0741-3335/58/1/014036}.
\bcomment{Publisher: IOP Publishing}.
2021-12-06
\end{barticle}
\endbibitem

\bibitem{Belli2008}
\begin{barticle}
\bauthor{\bsnm{Belli}, \binits{E.A.}},
\bauthor{\bsnm{Candy}, \binits{J.}}:
\batitle{Kinetic calculation of neoclassical transport including
  self-consistent electron and impurity dynamics}.
\bjtitle{Plasma Physics and Controlled Fusion}
\bvolume{50}(\bissue{9}),
\bfpage{095010}
(\byear{2008}).
\doiurl{10.1088/0741-3335/50/9/095010}
\end{barticle}
\endbibitem

\bibitem{Houlberg1997}
\begin{barticle}
\bauthor{\bsnm{Houlberg}, \binits{W.A.}},
\bauthor{\bsnm{Shaing}, \binits{K.C.}},
\bauthor{\bsnm{Hirshman}, \binits{S.P.}},
\bauthor{\bsnm{Zarnstorff}, \binits{M.C.}}:
\batitle{Bootstrap current and neoclassical transport in tokamaks of arbitrary
  collisionality and aspect ratio}.
\bjtitle{Physics of Plasmas}
\bvolume{4}(\bissue{9}),
\bfpage{3230}--\blpage{3242}
(\byear{1997}).
\doiurl{10.1063/1.872465}
\end{barticle}
\endbibitem

\bibitem{Candy2009a}
\begin{barticle}
\bauthor{\bsnm{Candy}, \binits{J.}},
\bauthor{\bsnm{Holland}, \binits{C.}},
\bauthor{\bsnm{Waltz}, \binits{R.E.}},
\bauthor{\bsnm{Fahey}, \binits{M.R.}},
\bauthor{\bsnm{Belli}, \binits{E.}}:
\batitle{Tokamak profile prediction using direct gyrokinetic and neoclassical
  simulation}.
\bjtitle{Physics of Plasmas}
\bvolume{16}(\bissue{6}),
\bfpage{060704}
(\byear{2009}).
\doiurl{10.1063/1.3167820}
\end{barticle}
\endbibitem

\bibitem{Barnes2010}
\begin{barticle}
\bauthor{\bsnm{Barnes}, \binits{M.}},
\bauthor{\bsnm{Abel}, \binits{I.G.}},
\bauthor{\bsnm{Dorland}, \binits{W.}},
\bauthor{\bsnm{Grler}, \binits{T.}},
\bauthor{\bsnm{Hammett}, \binits{G.W.}},
\bauthor{\bsnm{Jenko}, \binits{F.}}:
\batitle{Direct multiscale coupling of a transport code to gyrokinetic
  turbulence codes}.
\bjtitle{Physics of Plasmas}
\bvolume{17}(\bissue{5}),
\bfpage{056109}
(\byear{2010}).
\doiurl{10.1063/1.3323082}
\end{barticle}
\endbibitem

\bibitem{siena_global_2022}
\begin{barticle}
\bauthor{\bsnm{Siena}, \binits{A.D.}},
\bauthor{\bsnm{Navarro}, \binits{A.B.}},
\bauthor{\bsnm{Luda}, \binits{T.}},
\bauthor{\bsnm{Merlo}, \binits{G.}},
\bauthor{\bsnm{Bergmann}, \binits{M.}},
\bauthor{\bsnm{Leppin}, \binits{L.}},
\bauthor{\bsnm{Grler}, \binits{T.}},
\bauthor{\bsnm{Parker}, \binits{J.B.}},
\bauthor{\bsnm{LoDestro}, \binits{L.}},
\bauthor{\bsnm{Dannert}, \binits{T.}},
\bauthor{\bsnm{Germaschewski}, \binits{K.}},
\bauthor{\bsnm{Allen}, \binits{B.}},
\bauthor{\bsnm{Hittinger}, \binits{J.}},
\bauthor{\bsnm{Dorland}, \binits{B.W.}},
\bauthor{\bsnm{Hammett}, \binits{G.}}, \betal:
\batitle{Global gyrokinetic simulations of {ASDEX} {Upgrade} up to the
  transport timescale with {GENE}{Tango}}.
\bjtitle{Nuclear Fusion}
\bvolume{62}(\bissue{10}),
\bfpage{106025}
(\byear{2022}).
\doiurl{10.1088/1741-4326/ac8941}.
\bcomment{Publisher: IOP Publishing}.
2023-02-28
\end{barticle}
\endbibitem

\bibitem{Meneghini2016}
\begin{barticle}
\bauthor{\bsnm{Meneghini}, \binits{O.}},
\bauthor{\bsnm{Snyder}, \binits{P.B.}},
\bauthor{\bsnm{Smith}, \binits{S.P.}},
\bauthor{\bsnm{Candy}, \binits{J.}},
\bauthor{\bsnm{Staebler}, \binits{G.M.}},
\bauthor{\bsnm{Belli}, \binits{E.A.}},
\bauthor{\bsnm{Lao}, \binits{L.L.}},
\bauthor{\bsnm{Park}, \binits{J.M.}},
\bauthor{\bsnm{Green}, \binits{D.L.}},
\bauthor{\bsnm{Elwasif}, \binits{W.}},
\bauthor{\bsnm{Grierson}, \binits{B.A.}},
\bauthor{\bsnm{Holland}, \binits{C.}}:
\batitle{Integrated fusion simulation with self-consistent core-pedestal
  coupling}.
\bjtitle{Physics of Plasmas}
\bvolume{23},
\bfpage{42507}
(\byear{2016}).
\doiurl{10.1063/1.4947204}.
2018-11-28
\end{barticle}
\endbibitem

\bibitem{VanDePlassche2020}
\begin{barticle}
\bauthor{\bsnm{Van De~Plassche}, \binits{K.L.}},
\bauthor{\bsnm{Citrin}, \binits{J.}},
\bauthor{\bsnm{Bourdelle}, \binits{C.}},
\bauthor{\bsnm{Camenen}, \binits{Y.}},
\bauthor{\bsnm{Casson}, \binits{F.J.}},
\bauthor{\bsnm{Dagnelie}, \binits{V.I.}},
\bauthor{\bsnm{Felici}, \binits{F.}},
\bauthor{\bsnm{Ho}, \binits{A.}},
\bauthor{\bsnm{Mulders}, \binits{S.V.}},
\bauthor{\bsnm{Contributors}, \binits{J.}},
\bauthor{\bsnm{Van~Mulders}, \binits{S.}},
\bauthor{\bsnm{Affiliations}, \binits{.}}:
\batitle{Fast modeling of turbulent transport in fusion plasmas using neural
  networks}.
\bjtitle{Phys. Plasmas}
\bvolume{27},
\bfpage{22310}
(\byear{2020}).
\doiurl{10.1063/1.5134126}.
2020-02-21
\end{barticle}
\endbibitem

\bibitem{Ho2021}
\begin{barticle}
\bauthor{\bsnm{Ho}, \binits{A.}},
\bauthor{\bsnm{Citrin}, \binits{J.}},
\bauthor{\bsnm{Bourdelle}, \binits{C.}},
\bauthor{\bsnm{Camenen}, \binits{Y.}},
\bauthor{\bsnm{Casson}, \binits{F.J.}},
\bauthor{\bparticle{van~de} \bsnm{Plassche}, \binits{K.L.}},
\bauthor{\bsnm{Weisen}, \binits{H.}}:
\batitle{Neural network surrogate of {QuaLiKiz} using {JET} experimental data
  to populate training space}.
\bjtitle{Physics of Plasmas}
\bvolume{28}(\bissue{3}),
\bfpage{032305}
(\byear{2021}).
\doiurl{10.1063/5.0038290}
\end{barticle}
\endbibitem

\bibitem{zanisi_efficient_2024}
\begin{barticle}
\bauthor{\bsnm{Zanisi}, \binits{L.}},
\bauthor{\bsnm{Ho}, \binits{A.}},
\bauthor{\bsnm{Barr}, \binits{J.}},
\bauthor{\bsnm{Madula}, \binits{T.}},
\bauthor{\bsnm{Citrin}, \binits{J.}},
\bauthor{\bsnm{Pamela}, \binits{S.}},
\bauthor{\bsnm{Buchanan}, \binits{J.}},
\bauthor{\bsnm{Casson}, \binits{F.J.}},
\bauthor{\bsnm{Gopakumar}, \binits{V.}},
\bauthor{\bsnm{Contributors}, \binits{J.E.T.}}:
\batitle{Efficient training sets for surrogate models of tokamak turbulence
  with {Active} {Deep} {Ensembles}}.
\bjtitle{Nuclear Fusion}
\bvolume{64}(\bissue{3}),
\bfpage{036022}
(\byear{2024}).
\doiurl{10.1088/1741-4326/ad240d}.
\bcomment{Publisher: IOP Publishing}.
2024-03-07
\end{barticle}
\endbibitem

\bibitem{hornsby_gaussian_2024}
\begin{barticle}
\bauthor{\bsnm{Hornsby}, \binits{W.A.}},
\bauthor{\bsnm{Gray}, \binits{A.}},
\bauthor{\bsnm{Buchanan}, \binits{J.}},
\bauthor{\bsnm{Patel}, \binits{B.S.}},
\bauthor{\bsnm{Kennedy}, \binits{D.}},
\bauthor{\bsnm{Casson}, \binits{F.J.}},
\bauthor{\bsnm{Roach}, \binits{C.M.}},
\bauthor{\bsnm{Lykkegaard}, \binits{M.B.}},
\bauthor{\bsnm{Nguyen}, \binits{H.}},
\bauthor{\bsnm{Papadimas}, \binits{N.}},
\bauthor{\bsnm{Fourcin}, \binits{B.}},
\bauthor{\bsnm{Hart}, \binits{J.}}:
\batitle{Gaussian process regression models for the properties of micro-tearing
  modes in spherical tokamaks}.
\bjtitle{Physics of Plasmas}
\bvolume{31}(\bissue{1}),
\bfpage{012303}
(\byear{2024}).
\doiurl{10.1063/5.0174478}.
2024-03-21
\end{barticle}
\endbibitem

\bibitem{Rodriguez-Fernandez2019a}
\begin{botherref}
\oauthor{\bsnm{Rodriguez-Fernandez}, \binits{P.}}:
Perturbative transport experiments and time-dependent modeling in {Alcator}
  {C}-{Mod} and {DIII}-{D}.
PhD thesis,
Massachusetts Institute of Technology
(2019).
\url{https://dspace.mit.edu/handle/1721.1/123371}
\end{botherref}
\endbibitem

\bibitem{Jardin2008}
\begin{barticle}
\bauthor{\bsnm{Jardin}, \binits{S.C.}},
\bauthor{\bsnm{Bateman}, \binits{G.}},
\bauthor{\bsnm{Hammett}, \binits{G.W.}},
\bauthor{\bsnm{Ku}, \binits{L.P.}}:
\batitle{On {1D} diffusion problems with a gradient-dependent diffusion
  coefficient}.
\bjtitle{Journal of Computational Physics}
(\byear{2008}).
\doiurl{10.1016/j.jcp.2008.06.032}
\end{barticle}
\endbibitem

\bibitem{pereverzev_stable_2008}
\begin{barticle}
\bauthor{\bsnm{Pereverzev}, \binits{G.V.}},
\bauthor{\bsnm{Corrigan}, \binits{G.}}:
\batitle{Stable numeric scheme for diffusion equation with a stiff transport}.
\bjtitle{Computer Physics Communications}
\bvolume{179}(\bissue{8}),
\bfpage{579}--\blpage{585}
(\byear{2008}).
\doiurl{10.1016/j.cpc.2008.05.006}.
2022-09-15
\end{barticle}
\endbibitem

\bibitem{Pereverzev2002}
\begin{botherref}
\oauthor{\bsnm{Pereverzev}, \binits{G.V.}},
\oauthor{\bsnm{Yushmanov}, \binits{P.N.}}:
{ASTRA}.
IPP-Report
\textbf{IPP 5/98}(February)
(2002)
\end{botherref}
\endbibitem

\bibitem{Breslau2018}
\begin{barticle}
\bauthor{\bsnm{Breslau}, \binits{J.}},
\bauthor{\bsnm{Gorelenkova}, \binits{M.}},
\bauthor{\bsnm{Poli}, \binits{F.}},
\bauthor{\bsnm{Sachdev}, \binits{J.}},
\bauthor{\bsnm{Pankin}, \binits{A.}},
\bauthor{\bsnm{Perumpilly}, \binits{G.}},
\bauthor{\bsnm{Yuan}, \binits{X.}},
\bauthor{\bsnm{Glant}, \binits{L.}}:
\batitle{{TRANSP}}.
\bjtitle{Computer software. USDOE Office of Science (SC), Fusion Energy
  Sciences (FES)}
(\byear{2018}).
\doiurl{10.11578/DC.20180627.4}
\end{barticle}
\endbibitem

\bibitem{Candy2013}
\begin{barticle}
\bauthor{\bsnm{Candy}, \binits{J.}}:
\batitle{Turbulent energy exchange: {Calculation} and relevance for profile
  prediction}.
\bjtitle{Physics of Plasmas}
\bvolume{20}(\bissue{8}),
\bfpage{082503}
(\byear{2013}).
\doiurl{10.1063/1.4817820}.
\bcomment{Publisher: American Institute of PhysicsAIP}.
2021-12-24
\end{barticle}
\endbibitem

\bibitem{howard_simultaneous_2024}
\begin{barticle}
\bauthor{\bsnm{Howard}, \binits{N.T.}},
\bauthor{\bsnm{Rodriguez-Fernandez}, \binits{P.}},
\bauthor{\bsnm{Holland}, \binits{C.}},
\bauthor{\bsnm{Odstrcil}, \binits{T.}},
\bauthor{\bsnm{Grierson}, \binits{B.}},
\bauthor{\bsnm{Sciortino}, \binits{F.}},
\bauthor{\bsnm{McKee}, \binits{G.}},
\bauthor{\bsnm{Yan}, \binits{Z.}},
\bauthor{\bsnm{Wang}, \binits{G.}},
\bauthor{\bsnm{Rhodes}, \binits{T.L.}},
\bauthor{\bsnm{White}, \binits{A.E.}},
\bauthor{\bsnm{Candy}, \binits{J.}},
\bauthor{\bsnm{Chrystal}, \binits{C.}}:
\batitle{Simultaneous reproduction of experimental profiles, fluxes, transport
  coefficients, and turbulence characteristics via nonlinear gyrokinetic
  profile predictions in a {DIII}-{D} {ITER} similar shape plasma}.
\bjtitle{Physics of Plasmas}
\bvolume{31}(\bissue{3}),
\bfpage{032501}
(\byear{2024}).
\doiurl{10.1063/5.0175792}.
2024-03-01
\end{barticle}
\endbibitem

\bibitem{Gardner}
\begin{botherref}
\oauthor{\bsnm{Gardner}, \binits{J.R.}},
\oauthor{\bsnm{Pleiss}, \binits{G.}},
\oauthor{\bsnm{Bindel}, \binits{D.}},
\oauthor{\bsnm{Weinberger}, \binits{K.Q.}},
\oauthor{\bsnm{Wilson}, \binits{A.G.}}:
{GPyTorch}: {Blackbox} {Matrix}-{Matrix} {Gaussian} {Process} {Inference} with
  {GPU} {Acceleration}.
arXiv
(2019).
2019-12-12
\end{botherref}
\endbibitem

\bibitem{balandat_botorch_2020}
\begin{botherref}
\oauthor{\bsnm{Balandat}, \binits{M.}},
\oauthor{\bsnm{Karrer}, \binits{B.}},
\oauthor{\bsnm{Jiang}, \binits{D.R.}},
\oauthor{\bsnm{Daulton}, \binits{S.}},
\oauthor{\bsnm{Letham}, \binits{B.}},
\oauthor{\bsnm{Wilson}, \binits{A.G.}},
\oauthor{\bsnm{Bakshy}, \binits{E.}}:
{BoTorch}: {A} {Framework} for {Efficient} {Monte}-{Carlo} {Bayesian}
  {Optimization}.
arXiv.
arXiv:1910.06403 [cs, math, stat]
(2020).
\doiurl{10.48550/arXiv.1910.06403}.
\url{http://arxiv.org/abs/1910.06403}
2023-12-19
\end{botherref}
\endbibitem

\bibitem{Frazier2018}
\begin{botherref}
\oauthor{\bsnm{Frazier}, \binits{P.I.}}:
A {Tutorial} on {Bayesian} {Optimization}.
arXiv
(2018)
\end{botherref}
\endbibitem

\bibitem{Preuss2018}
\begin{barticle}
\bauthor{\bsnm{Preuss}, \binits{R.}},
\bauthor{\bparticle{von} \bsnm{Toussaint}, \binits{U.}}:
\batitle{Optimization {Employing} {Gaussian} {Process}-{Based} {Surrogates}}.
\bjtitle{Springer Proceedings in Mathematics and Statistics}
\bvolume{239},
\bfpage{275}--\blpage{284}
(\byear{2018}).
\doiurl{10.1007/978-3-319-91143-4_26}.
\bcomment{ISBN: 9783319911427}
\end{barticle}
\endbibitem

\bibitem{Brochu2010}
\begin{botherref}
\oauthor{\bsnm{Brochu}, \binits{E.}},
\oauthor{\bsnm{Cora}, \binits{V.M.}},
\oauthor{\bparticle{de} \bsnm{Freitas}, \binits{N.}}:
A {Tutorial} on {Bayesian} {Optimization} of {Expensive} {Cost} {Functions},
  with {Application} to {Active} {User} {Modeling} and {Hierarchical}
  {Reinforcement} {Learning}
(2010).
\doiurl{10.48550/arXiv.1012.2599}
\end{botherref}
\endbibitem

\bibitem{Astudillo2019}
\begin{barticle}
\bauthor{\bsnm{Astudillo}, \binits{R.}},
\bauthor{\bsnm{Frazier}, \binits{P.I.}}:
\batitle{Bayesian optimization of composite functions}.
\bjtitle{36th International Conference on Machine Learning, ICML 2019}
\bvolume{2019-June}(\bissue{2006}),
\bfpage{547}--\blpage{556}
(\byear{2019}).
\bcomment{ISBN: 9781510886988}
\end{barticle}
\endbibitem

\bibitem{belli_eulerian_2009}
\begin{barticle}
\bauthor{\bsnm{Belli}, \binits{E.A.}},
\bauthor{\bsnm{Candy}, \binits{J.}}:
\batitle{An {Eulerian} method for the solution of the multi-species
  drift-kinetic equation}.
\bjtitle{Plasma Physics and Controlled Fusion}
\bvolume{51}(\bissue{7}),
\bfpage{075018}
(\byear{2009}).
\doiurl{10.1088/0741-3335/51/7/075018}.
2022-12-09
\end{barticle}
\endbibitem

\bibitem{mockus_bayesian_1975}
\begin{bchapter}
\bauthor{\bsnm{Mokus}, \binits{J.}}:
\bctitle{On bayesian methods for seeking the extremum}.
In: \beditor{\bsnm{Marchuk}, \binits{G.I.}} (ed.)
\bbtitle{Optimization {Techniques} {IFIP} {Technical} {Conference}
  {Novosibirsk}, {July} 17, 1974}.
\bsertitle{Lecture {Notes} in {Computer} {Science}},
pp. \bfpage{400}--\blpage{404}.
\bpublisher{Springer},
\blocation{Berlin, Heidelberg}
(\byear{1975}).
\doiurl{10.1007/3-540-07165-2_55}
\end{bchapter}
\endbibitem

\bibitem{jones_efficient_1998}
\begin{barticle}
\bauthor{\bsnm{Jones}, \binits{D.R.}},
\bauthor{\bsnm{Schonlau}, \binits{M.}},
\bauthor{\bsnm{Welch}, \binits{W.J.}}:
\batitle{Efficient {Global} {Optimization} of {Expensive} {Black}-{Box}
  {Functions}}.
\bjtitle{Journal of Global Optimization}
\bvolume{13}(\bissue{4}),
\bfpage{455}--\blpage{492}
(\byear{1998}).
\doiurl{10.1023/A:1008306431147}.
2023-11-27
\end{barticle}
\endbibitem

\bibitem{wang_max-value_2018}
\begin{botherref}
\oauthor{\bsnm{Wang}, \binits{Z.}},
\oauthor{\bsnm{Jegelka}, \binits{S.}}:
Max-value {Entropy} {Search} for {Efficient} {Bayesian} {Optimization}.
arXiv.
arXiv:1703.01968 [cs, math, stat]
(2018).
\doiurl{10.48550/arXiv.1703.01968}.
\url{http://arxiv.org/abs/1703.01968}
2023-11-27
\end{botherref}
\endbibitem

\bibitem{Virtanen2020}
\begin{barticle}
\bauthor{\bsnm{Virtanen}, \binits{P.}},
\bauthor{\bsnm{Gommers}, \binits{R.}},
\bauthor{\bsnm{Oliphant}, \binits{T.E.}},
\bauthor{\bsnm{Haberland}, \binits{M.}},
\bauthor{\bsnm{Reddy}, \binits{T.}},
\bauthor{\bsnm{Cournapeau}, \binits{D.}},
\bauthor{\bsnm{Burovski}, \binits{E.}},
\bauthor{\bsnm{Peterson}, \binits{P.}},
\bauthor{\bsnm{Weckesser}, \binits{W.}},
\bauthor{\bsnm{Bright}, \binits{J.}},
\bauthor{\bparticle{van~der} \bsnm{Walt}, \binits{S.J.}},
\bauthor{\bsnm{Brett}, \binits{M.}},
\bauthor{\bsnm{Wilson}, \binits{J.}},
\bauthor{\bsnm{Millman}, \binits{K.J.}},
\bauthor{\bsnm{Mayorov}, \binits{N.}}, \betal:
\batitle{{SciPy} 1.0: fundamental algorithms for scientific computing in
  {Python}}.
\bjtitle{Nature Methods 2020 17:3}
\bvolume{17}(\bissue{3}),
\bfpage{261}--\blpage{272}
(\byear{2020}).
\doiurl{10.1038/s41592-019-0686-2}.
\bcomment{Publisher: Nature Publishing Group}.
2021-12-07
\end{barticle}
\endbibitem

\bibitem{levenberg_method_1944}
\begin{barticle}
\bauthor{\bsnm{Levenberg}, \binits{K.}}:
\batitle{A method for the solution of certain non-linear problems in least
  squares}.
\bjtitle{Quarterly of Applied Mathematics}
\bvolume{2}(\bissue{2}),
\bfpage{164}--\blpage{168}
(\byear{1944}).
\doiurl{10.1090/qam/10666}.
2023-10-19
\end{barticle}
\endbibitem

\bibitem{marquardt_algorithm_1963}
\begin{barticle}
\bauthor{\bsnm{Marquardt}, \binits{D.W.}}:
\batitle{An {Algorithm} for {Least}-{Squares} {Estimation} of {Nonlinear}
  {Parameters}}.
\bjtitle{Journal of the Society for Industrial and Applied Mathematics}
\bvolume{11}(\bissue{2}),
\bfpage{431}--\blpage{441}
(\byear{1963}).
\doiurl{10.1137/0111030}.
\bcomment{Publisher: Society for Industrial and Applied Mathematics}.
2023-10-19
\end{barticle}
\endbibitem

\bibitem{Paszke2019}
\begin{botherref}
\oauthor{\bsnm{Paszke}, \binits{A.}},
\oauthor{\bsnm{Gross}, \binits{S.}},
\oauthor{\bsnm{Massa}, \binits{F.}},
\oauthor{\bsnm{Lerer}, \binits{A.}},
\oauthor{\bsnm{Bradbury}, \binits{J.}},
\oauthor{\bsnm{Chanan}, \binits{G.}},
\oauthor{\bsnm{Killeen}, \binits{T.}},
\oauthor{\bsnm{Lin}, \binits{Z.}},
\oauthor{\bsnm{Gimelshein}, \binits{N.}},
\oauthor{\bsnm{Antiga}, \binits{L.}},
\oauthor{\bsnm{Desmaison}, \binits{A.}},
\oauthor{\bsnm{Kpf}, \binits{A.}},
\oauthor{\bsnm{Yang}, \binits{E.}},
\oauthor{\bsnm{DeVito}, \binits{Z.}},
\oauthor{\bsnm{Raison}, \binits{M.}}, et al.:
{PyTorch}: {An} {Imperative} {Style}, {High}-{Performance} {Deep} {Learning}
  {Library}.
arXiv
(2019).
Publisher: Neural information processing systems foundation.
2021-12-07
\end{botherref}
\endbibitem

\bibitem{Zhu1997}
\begin{barticle}
\bauthor{\bsnm{Zhu}, \binits{C.}},
\bauthor{\bsnm{Byrd}, \binits{R.H.}},
\bauthor{\bsnm{Lu}, \binits{P.}},
\bauthor{\bsnm{Nocedal}, \binits{J.}}:
\batitle{Algorithm 778: {L}-{BFGS}-{B}}.
\bjtitle{ACM Transactions on Mathematical Software (TOMS)}
\bvolume{23}(\bissue{4}),
\bfpage{550}--\blpage{560}
(\byear{1997}).
\doiurl{10.1145/279232.279236}.
\bcomment{Publisher: ACM PUB27 New York, NY, USA}.
2021-12-12
\end{barticle}
\endbibitem

\bibitem{Fortin2012}
\begin{barticle}
\bauthor{\bsnm{Fortin}, \binits{F.A.}},
\bauthor{\bsnm{De~Rainville}, \binits{F.M.}},
\bauthor{\bsnm{Gardner}, \binits{M.A.}},
\bauthor{\bsnm{Parizeau}, \binits{M.}},
\bauthor{\bsnm{Gage}, \binits{C.}}:
\batitle{{DEAP}: {Evolutionary} algorithms made easy}.
\bjtitle{Journal of Machine Learning Research}
\bvolume{13},
\bfpage{2171}--\blpage{2175}
(\byear{2012})
\end{barticle}
\endbibitem

\bibitem{prf_2023_eps}
\begin{botherref}
\oauthor{\bsnm{Rodriguez-Fernandez}, \binits{P.}},
\oauthor{\bsnm{Howard}, \binits{N.T.}},
\oauthor{\bsnm{Delabie}, \binits{E.}},
\oauthor{\bsnm{Lomanowski}, \binits{B.}},
\oauthor{\bsnm{Saltzman}, \binits{A.}},
\oauthor{\bsnm{Kantamneni}, \binits{S.}},
\oauthor{\bsnm{Candy}, \binits{J.}},
\oauthor{\bsnm{Holland}, \binits{C.}},
\oauthor{\bsnm{Nave}, \binits{M.F.F.}},
\oauthor{\bsnm{Biewer}, \binits{T.M.}},
\oauthor{\bsnm{Garcia}, \binits{J.}},
\oauthor{\bsnm{Lennholm}, \binits{M.}},
\oauthor{\bsnm{White}, \binits{A.E.}},
\oauthor{\bsnm{{JET Contributors}}}:
Prediction of core kinetic profiles and burning plasma performance with
  high-fidelity gyrokinetic simulations in tokamaks
(2023).
Bordeaux, France, July 3-7
\end{botherref}
\endbibitem

\bibitem{Bourdelle2005}
\begin{botherref}
\oauthor{\bsnm{Bourdelle}, \binits{C.}}:
Turbulent particle transport in magnetized fusion plasma.
Plasma Physics and Controlled Fusion
\textbf{47}(5 A)
(2005).
\doiurl{10.1088/0741-3335/47/5A/023}
\end{botherref}
\endbibitem

\bibitem{Rodriguez-Fernandez2020a}
\begin{barticle}
\bauthor{\bsnm{Rodriguez-Fernandez}, \binits{P.}},
\bauthor{\bsnm{Howard}, \binits{N.T.}},
\bauthor{\bsnm{Greenwald}, \binits{M.J.}},
\bauthor{\bsnm{Creely}, \binits{A.J.}},
\bauthor{\bsnm{Hughes}, \binits{J.W.}},
\bauthor{\bsnm{Wright}, \binits{J.C.}},
\bauthor{\bsnm{Holland}, \binits{C.}},
\bauthor{\bsnm{Lin}, \binits{Y.}},
\bauthor{\bsnm{Sciortino}, \binits{F.}},
\bauthor{\bsnm{Team}, \binits{t.S.}}:
\batitle{Predictions of core plasma performance for the {SPARC} tokamak}.
\bjtitle{Journal of Plasma Physics}
\bvolume{86}(\bissue{5}),
\bfpage{865860503}
(\byear{2020}).
\doiurl{10.1017/S0022377820001075}.
\bcomment{Publisher: Cambridge University Press}.
2021-07-12
\end{barticle}
\endbibitem

\bibitem{holland_2023_aps}
\begin{botherref}
\oauthor{\bsnm{Holland}, \binits{C.G.}}:
Characterization of predicted confinement and transport in an arc-class tokamak
  power plant
(2023).
Denver (CO), October 31
\end{botherref}
\endbibitem

\bibitem{howard_2024_iter}
\begin{botherref}
\oauthor{\bsnm{Howard}, \binits{N.T.}},
\oauthor{\bsnm{Rodriguez-Fernandez}, \binits{P.}},
\oauthor{\bsnm{Holland}, \binits{C.}},
\oauthor{\bsnm{Candy}, \binits{J.}}:
Prediction of performance and turbulence in iter burning plasmas via nonlinear
  gyrokinetic profile prediction.
(submitted)
(2024)
\end{botherref}
\endbibitem

\bibitem{Creely2020a}
\begin{botherref}
\oauthor{\bsnm{Creely}, \binits{A.J.}},
\oauthor{\bsnm{Greenwald}, \binits{M.J.}},
\oauthor{\bsnm{Ballinger}, \binits{S.B.}},
\oauthor{\bsnm{Brunner}, \binits{D.}},
\oauthor{\bsnm{Canik}, \binits{J.}},
\oauthor{\bsnm{Doody}, \binits{J.}},
\oauthor{\bsnm{Flp}, \binits{T.}},
\oauthor{\bsnm{Garnier}, \binits{D.T.}},
\oauthor{\bsnm{Granetz}, \binits{R.}},
\oauthor{\bsnm{Gray}, \binits{T.K.}},
\oauthor{\bsnm{Holland}, \binits{C.}},
\oauthor{\bsnm{Howard}, \binits{N.T.}},
\oauthor{\bsnm{Hughes}, \binits{J.W.}},
\oauthor{\bsnm{Irby}, \binits{J.H.}},
\oauthor{\bsnm{Izzo}, \binits{V.A.}}, et al.:
Overview of the {SPARC} tokamak.
Journal of Plasma Physics
\textbf{86}(5).
\doiurl{10.1017/S0022377820001257}.
2021-07-12
\end{botherref}
\endbibitem

\bibitem{rodriguez-fernandez_core_2024}
\begin{botherref}
\oauthor{\bsnm{Rodriguez-Fernandez}, \binits{P.}},
\oauthor{\bsnm{Howard}, \binits{N.T.}},
\oauthor{\bsnm{Saltzman}, \binits{A.}},
\oauthor{\bsnm{Shoji}, \binits{L.}},
\oauthor{\bsnm{Body}, \binits{T.}},
\oauthor{\bsnm{Battaglia}, \binits{D.J.}},
\oauthor{\bsnm{Hughes}, \binits{J.W.}},
\oauthor{\bsnm{Candy}, \binits{J.}},
\oauthor{\bsnm{Staebler}, \binits{G.M.}},
\oauthor{\bsnm{Creely}, \binits{A.J.}}:
Core performance predictions in projected {SPARC} first-campaign plasmas with
  nonlinear {CGYRO}.
arXiv.
arXiv:2403.15633 [physics]
(2024).
\doiurl{10.48550/arXiv.2403.15633}.
\url{http://arxiv.org/abs/2403.15633}
2024-04-04
\end{botherref}
\endbibitem

\bibitem{bonilla_multi-task_2007}
\begin{botherref}
\oauthor{\bsnm{Bonilla}, \binits{E.V.}},
\oauthor{\bsnm{Chai}, \binits{K.}},
\oauthor{\bsnm{Williams}, \binits{C.}}:
Multi-task {Gaussian} {Process} {Prediction}.
Advances in Neural Information Processing Systems
\textbf{20}
(2007).
2023-11-22
\end{botherref}
\endbibitem

\bibitem{Greenwald2013}
\begin{barticle}
\bauthor{\bsnm{Greenwald}, \binits{M.}}:
\batitle{Overview of experimental results and code validation activities at
  {Alcator} {C}-{Mod}}.
\bjtitle{Nuclear Fusion}
\bvolume{53},
\bfpage{104004}
(\byear{2013}).
\doiurl{10.1088/0029-5515/53/10/104004}
\end{barticle}
\endbibitem

\bibitem{hughes_thomson_2003}
\begin{barticle}
\bauthor{\bsnm{Hughes}, \binits{J.W.}},
\bauthor{\bsnm{Mossessian}, \binits{D.}},
\bauthor{\bsnm{Zhurovich}, \binits{K.}},
\bauthor{\bsnm{DeMaria}, \binits{M.}},
\bauthor{\bsnm{Jensen}, \binits{K.}},
\bauthor{\bsnm{Hubbard}, \binits{A.}}:
\batitle{Thomson scattering upgrades on {Alcator} {C}-{Mod}}.
\bjtitle{Review of Scientific Instruments}
\bvolume{74}(\bissue{3}),
\bfpage{1667}--\blpage{1670}
(\byear{2003}).
\doiurl{10.1063/1.1532764}.
2023-10-18
\end{barticle}
\endbibitem

\bibitem{oshea_notitle_1997}
\begin{botherref}
\oauthor{\bsnm{O'Shea}, \binits{P.J.}},
\oauthor{\bsnm{Bonoli}, \binits{P.}},
\oauthor{\bsnm{Hubbard}, \binits{A.}},
\oauthor{\bsnm{Porkolab}, \binits{M.}},
\oauthor{\bsnm{Takase}, \binits{Y.}}:
In: Proceedings of the 11th {Conference} on {Radio} {Frequency} {Power} In
  {Plasmas},
vol. 403,
p. 89.
AIP
(1997)
\end{botherref}
\endbibitem

\bibitem{meneghini_integrated_2015}
\begin{barticle}
\bauthor{\bsnm{Meneghini}, \binits{O.}},
\bauthor{\bsnm{Smith}, \binits{S.P.}},
\bauthor{\bsnm{Lao}, \binits{L.L.}},
\bauthor{\bsnm{Izacard}, \binits{O.}},
\bauthor{\bsnm{Ren}, \binits{Q.}},
\bauthor{\bsnm{Park}, \binits{J.M.}},
\bauthor{\bsnm{Candy}, \binits{J.}},
\bauthor{\bsnm{Wang}, \binits{Z.}},
\bauthor{\bsnm{Luna}, \binits{C.J.}},
\bauthor{\bsnm{Izzo}, \binits{V.A.}},
\bauthor{\bsnm{Grierson}, \binits{B.A.}},
\bauthor{\bsnm{Snyder}, \binits{P.B.}},
\bauthor{\bsnm{Holland}, \binits{C.}},
\bauthor{\bsnm{Penna}, \binits{J.}},
\bauthor{\bsnm{Lu}, \binits{G.}}, \betal:
\batitle{Integrated modeling applications for tokamak experiments with
  {OMFIT}}.
\bjtitle{Nuclear Fusion}
\bvolume{55}(\bissue{8}),
\bfpage{083008}
(\byear{2015}).
\doiurl{10.1088/0029-5515/55/8/083008}.
\bcomment{Publisher: IOP Publishing}.
2023-11-10
\end{barticle}
\endbibitem

\bibitem{Kotschenreuther1995}
\begin{barticle}
\bauthor{\bsnm{Kotschenreuther}, \binits{M.}},
\bauthor{\bsnm{Dorland}, \binits{W.}},
\bauthor{\bsnm{Beer}, \binits{M.A.}},
\bauthor{\bsnm{Hammett}, \binits{G.W.}}:
\batitle{Quantitative predictions of tokamak energy confinement from
  first-principles simulations with kinetic effects}.
\bjtitle{Physics of Plasmas}
\bvolume{2}(\bissue{6}),
\bfpage{2381}--\blpage{2389}
(\byear{1995}).
\doiurl{10.1063/1.871261}
\end{barticle}
\endbibitem

\bibitem{Sauter2014}
\begin{botherref}
\oauthor{\bsnm{Sauter}, \binits{O.}},
\oauthor{\bsnm{Brunner}, \binits{S.}},
\oauthor{\bsnm{Kim}, \binits{D.}},
\oauthor{\bsnm{Merlo}, \binits{G.}},
\oauthor{\bsnm{Behn}, \binits{R.}},
\oauthor{\bsnm{Camenen}, \binits{Y.}},
\oauthor{\bsnm{Coda}, \binits{S.}},
\oauthor{\bsnm{Duval}, \binits{B.P.}},
\oauthor{\bsnm{Federspiel}, \binits{L.}},
\oauthor{\bsnm{Goodman}, \binits{T.P.}},
\oauthor{\bsnm{Karpushov}, \binits{A.}},
\oauthor{\bsnm{Merle}, \binits{A.}},
\oauthor{\bsnm{Team}, \binits{T.}}:
On the non-stiffness of edge transport in {L}-mode tokamak plasmas.
Physics of Plasmas
\textbf{21}(5)
(2014).
\doiurl{10.1063/1.4876612}
\end{botherref}
\endbibitem

\bibitem{holland_development_2023}
\begin{barticle}
\bauthor{\bsnm{Holland}, \binits{C.}},
\bauthor{\bsnm{Bass}, \binits{E.M.}},
\bauthor{\bsnm{Orlov}, \binits{D.M.}},
\bauthor{\bsnm{McClenaghan}, \binits{J.}},
\bauthor{\bsnm{Lyons}, \binits{B.C.}},
\bauthor{\bsnm{Grierson}, \binits{B.A.}},
\bauthor{\bsnm{Jian}, \binits{X.}},
\bauthor{\bsnm{Howard}, \binits{N.T.}},
\bauthor{\bsnm{Rodriguez-Fernandez}, \binits{P.}}:
\batitle{Development of compact tokamak fusion reactor use cases to inform
  future transport studies}.
\bjtitle{Journal of Plasma Physics}
\bvolume{89}(\bissue{4}),
\bfpage{905890418}
(\byear{2023}).
\doiurl{10.1017/S0022377823000843}.
2023-08-30
\end{barticle}
\endbibitem

\bibitem{Kotschenreuther2019}
\begin{barticle}
\bauthor{\bsnm{Kotschenreuther}, \binits{M.}},
\bauthor{\bsnm{Liu}, \binits{X.}},
\bauthor{\bsnm{Hatch}, \binits{D.R.}},
\bauthor{\bsnm{Mahajan}, \binits{S.}},
\bauthor{\bsnm{Zheng}, \binits{L.}},
\bauthor{\bsnm{Diallo}, \binits{A.}},
\bauthor{\bsnm{Groebner}, \binits{R.}},
\bauthor{\bsnm{Hillesheim}, \binits{J.C.}},
\bauthor{\bsnm{Maggi}, \binits{C.F.}},
\bauthor{\bsnm{Giroud}, \binits{C.}},
\bauthor{\bsnm{Koechl}, \binits{F.}},
\bauthor{\bsnm{Parail}, \binits{V.}},
\bauthor{\bsnm{Saarelma}, \binits{S.}},
\bauthor{\bsnm{Solano}, \binits{E.}},
\bauthor{\bsnm{Chankin}, \binits{A.}}:
\batitle{Gyrokinetic analysis and simulation of pedestals to identify the
  culprits for energy losses using fingerprints}.
\bjtitle{Nuclear Fusion}
\bvolume{59}(\bissue{9}),
\bfpage{096001}
(\byear{2019}).
\doiurl{10.1088/1741-4326/AB1FA2}.
\bcomment{Publisher: IOP Publishing}.
2022-02-02
\end{barticle}
\endbibitem

\bibitem{Belli2021}
\begin{barticle}
\bauthor{\bsnm{Belli}, \binits{E.A.}},
\bauthor{\bsnm{Candy}, \binits{J.}}:
\batitle{Asymmetry between deuterium and tritium turbulent particle flows}.
\bjtitle{Physics of Plasmas}
\bvolume{28}(\bissue{6}),
\bfpage{062301}
(\byear{2021}).
\doiurl{10.1063/5.0048620}.
\bcomment{Publisher: AIP Publishing LLC AIP Publishing}.
2022-01-12
\end{barticle}
\endbibitem

\bibitem{Sugama1997}
\begin{barticle}
\bauthor{\bsnm{Sugama}, \binits{H.}},
\bauthor{\bsnm{Norton}, \binits{W.}}:
\batitle{Transport processes and entropy production in toroidally rotating
  plasmas with electrostatic turbulence}.
\bjtitle{Physics of Plasmas}
\bvolume{4}(\bissue{2}),
\bfpage{405}--\blpage{418}
(\byear{1997}).
\doiurl{10.1063/1.872099}
\end{barticle}
\endbibitem

\bibitem{Arbon2020}
\begin{barticle}
\bauthor{\bsnm{Arbon}, \binits{R.}},
\bauthor{\bsnm{Candy}, \binits{J.}},
\bauthor{\bsnm{Belli}, \binits{E.A.}}:
\batitle{Rapidly-convergent flux-surface shape parameterization}.
\bjtitle{Plasma Physics and Controlled Fusion}
\bvolume{63}(\bissue{1}),
\bfpage{012001}
(\byear{2020}).
\doiurl{10.1088/1361-6587/abc63b}
\end{barticle}
\endbibitem

\bibitem{weikl_ion_2017}
\begin{barticle}
\bauthor{\bsnm{Weikl}, \binits{A.}},
\bauthor{\bsnm{Peeters}, \binits{A.G.}},
\bauthor{\bsnm{Rath}, \binits{F.}},
\bauthor{\bsnm{Grosshauser}, \binits{S.R.}},
\bauthor{\bsnm{Buchholz}, \binits{R.}},
\bauthor{\bsnm{Hornsby}, \binits{W.A.}},
\bauthor{\bsnm{Seiferling}, \binits{F.}},
\bauthor{\bsnm{Strintzi}, \binits{D.}}:
\batitle{Ion temperature gradient turbulence close to the finite heat flux
  threshold}.
\bjtitle{Physics of Plasmas}
\bvolume{24}(\bissue{10}),
\bfpage{102317}
(\byear{2017}).
\doiurl{10.1063/1.4986035}.
2024-03-21
\end{barticle}
\endbibitem

\bibitem{rath_transport_2022}
\begin{barticle}
\bauthor{\bsnm{Rath}, \binits{F.}},
\bauthor{\bsnm{Peeters}, \binits{A.G.}}:
\batitle{Transport hysteresis in electromagnetic microturbulence caused by
  mesoscale zonal flow pattern-induced mitigation of high turbulence runaways}.
\bjtitle{Physics of Plasmas}
\bvolume{29}(\bissue{4}),
\bfpage{042305}
(\byear{2022}).
\doiurl{10.1063/5.0081846}.
2024-03-21
\end{barticle}
\endbibitem

\bibitem{angioni_dependence_2023}
\begin{barticle}
\bauthor{\bsnm{Angioni}, \binits{C.}},
\bauthor{\bsnm{Bonanomi}, \binits{N.}},
\bauthor{\bsnm{Fable}, \binits{E.}},
\bauthor{\bsnm{Schneider}, \binits{P.A.}},
\bauthor{\bsnm{Tardini}, \binits{G.}},
\bauthor{\bsnm{Luda}, \binits{T.}},
\bauthor{\bsnm{Staebler}, \binits{G.M.}},
\bauthor{\bsnm{Team}, \binits{t.A.U.}}:
\batitle{The dependence of tokamak {L}-mode confinement on magnetic field and
  plasma size, from a magnetic field scan experiment at {ASDEX} {Upgrade} to
  full-radius integrated modelling and fusion reactor predictions}.
\bjtitle{Nuclear Fusion}
\bvolume{63}(\bissue{5}),
\bfpage{056005}
(\byear{2023}).
\doiurl{10.1088/1741-4326/acc193}.
\bcomment{Publisher: IOP Publishing}.
2023-03-31
\end{barticle}
\endbibitem

\bibitem{Angioni2022}
\begin{barticle}
\bauthor{\bsnm{Angioni}, \binits{C.}},
\bauthor{\bsnm{Gamot}, \binits{T.}},
\bauthor{\bsnm{Tardini}, \binits{G.}},
\bauthor{\bsnm{Fable}, \binits{E.}},
\bauthor{\bsnm{Luda}, \binits{T.}},
\bauthor{\bsnm{Bonanomi}, \binits{N.}},
\bauthor{\bsnm{Kiefer}, \binits{C.}},
\bauthor{\bsnm{Staebler}, \binits{G.}},
\bauthor{\bsnm{ASDEX Upgrade~Team}, \binits{T.}},
\bauthor{\bsnm{Mst}, \binits{E.}}:
\batitle{Confinement properties of {L}-mode plasmas in {ASDEX} {Upgrade} and
  full-radius predictions of the {TGLF} transport model}.
\bjtitle{Nuclear Fusion}
\bvolume{62},
\bfpage{066015}
(\byear{2022}).
\doiurl{10.1088/1741-4326/AC592B}.
\bcomment{Publisher: IOP Publishing}.
2022-03-11
\end{barticle}
\endbibitem

\bibitem{wilson_steppathway_2020}
\begin{barticle}
\bauthor{\bsnm{Wilson}, \binits{H.}},
\bauthor{\bsnm{Chapman}, \binits{I.}},
\bauthor{\bsnm{Denton}, \binits{T.}},
\bauthor{\bsnm{Morris}, \binits{W.}},
\bauthor{\bsnm{Patel}, \binits{B.}},
\bauthor{\bsnm{Voss}, \binits{G.}},
\bauthor{\bsnm{Waldon}, \binits{C.}},
\bauthor{\bsnm{Team}, \binits{t.S.}}:
\batitle{{STEP}on the pathway to fusion commercialization}.
\bjtitle{Commercialising Fusion Energy: How small businesses are transforming
  big science}
(\byear{2020}).
\doiurl{10.1088/978-0-7503-2719-0ch8}.
2023-10-18
\end{barticle}
\endbibitem

\bibitem{Verdoolaege2021}
\begin{barticle}
\bauthor{\bsnm{Verdoolaege}, \binits{G.}},
\bauthor{\bsnm{Kaye}, \binits{S.M.}},
\bauthor{\bsnm{Angioni}, \binits{C.}},
\bauthor{\bsnm{Kardaun}, \binits{O.J.W.F.}},
\bauthor{\bsnm{Maslov}, \binits{M.}},
\bauthor{\bsnm{Romanelli}, \binits{M.}},
\bauthor{\bsnm{Ryter}, \binits{F.}},
\bauthor{\bsnm{Thomsen}, \binits{K.}}:
\batitle{The updated {ITPA} global {H}-mode confinement database: description
  and analysis}.
\bjtitle{Nuclear Fusion}
\bvolume{61},
\bfpage{076006}
(\byear{2021}).
\doiurl{10.1088/1741-4326/abdb91}.
2021-07-27
\end{barticle}
\endbibitem

\bibitem{battaglia_cfs-energycfspopcon_2023}
\begin{botherref}
\oauthor{\bsnm{Battaglia}, \binits{D.}},
\oauthor{\bsnm{Body}, \binits{T.}},
\oauthor{\bsnm{Creely}, \binits{A.J.}},
\oauthor{\bsnm{Hasse}, \binits{C.}},
\oauthor{\bsnm{Logan}, \binits{J.}},
\oauthor{\bsnm{Mumgaard}, \binits{R.T.}},
\oauthor{\bsnm{Rodriguez-Fernandez}, \binits{P.}},
\oauthor{\bsnm{Schneider}, \binits{N.}},
\oauthor{\bsnm{Sorbom}, \binits{B.N.}},
\oauthor{\bsnm{Tse}, \binits{C.}},
\oauthor{\bsnm{Vernacchia}, \binits{M.}},
\oauthor{\bsnm{Nelson}, \binits{A.O.}}:
cfs-energy/cfspopcon: v4.0.0
(2023).
\url{https://zenodo.org/records/10054880}
2023-11-22
\end{botherref}
\endbibitem

\bibitem{Angioni2007}
\begin{barticle}
\bauthor{\bsnm{Angioni}, \binits{C.}},
\bauthor{\bsnm{Weisen}, \binits{H.}},
\bauthor{\bsnm{Kardaun}, \binits{O.J.W.F.}},
\bauthor{\bsnm{Maslov}, \binits{M.}},
\bauthor{\bsnm{Zabolotsky}, \binits{A.}},
\bauthor{\bsnm{Fuchs}, \binits{C.}},
\bauthor{\bsnm{Garzotti}, \binits{L.}},
\bauthor{\bsnm{Giroud}, \binits{C.}},
\bauthor{\bsnm{Kurzan}, \binits{B.}},
\bauthor{\bsnm{Mantica}, \binits{P.}},
\bauthor{\bsnm{Peeters}, \binits{A.G.}},
\bauthor{\bsnm{Stober}, \binits{J.}}:
\batitle{Scaling of density peaking in {H}-mode plasmas based on a combined
  database of {AUG} and {JET} observations}.
\bjtitle{Nuclear Fusion}
\bvolume{47}(\bissue{9}),
\bfpage{1326}--\blpage{1335}
(\byear{2007}).
\doiurl{10.1088/0029-5515/47/9/033}
\end{barticle}
\endbibitem

\bibitem{Howard2021}
\begin{barticle}
\bauthor{\bsnm{Howard}, \binits{N.T.}},
\bauthor{\bsnm{Rodriguez-Fernandez}, \binits{P.} \bsuffix{P.}},
\bauthor{\bsnm{Holland}, \binits{C.}},
\bauthor{\bsnm{Rice}, \binits{J.E.}},
\bauthor{\bsnm{Greenwald}, \binits{M.}},
\bauthor{\bsnm{Candy}, \binits{J.}},
\bauthor{\bsnm{Sciortino}, \binits{F.}}:
\batitle{Gyrokinetic simulation of turbulence and transport in the {SPARC}
  tokamak}.
\bjtitle{Physics of Plasmas}
\bvolume{28},
\bfpage{072502}
(\byear{2021}).
\doiurl{10.1063/5.0047789}.
\bcomment{Publisher: AIP Publishing LLC}
\end{barticle}
\endbibitem

\bibitem{White2019}
\begin{barticle}
\bauthor{\bsnm{White}, \binits{A.E.}}:
\batitle{Validation of nonlinear gyrokinetic transport models using turbulence
  measurements}.
\bjtitle{Journal of Plasma Physics}
\bvolume{85}(\bissue{1}),
\bfpage{925850102}
(\byear{2019}).
\doiurl{10.1017/S0022377818001253}
\end{barticle}
\endbibitem

\bibitem{Rodriguez-Fernandez2018a}
\begin{barticle}
\bauthor{\bsnm{Rodriguez-Fernandez}, \binits{P.}},
\bauthor{\bsnm{White}, \binits{A.E.}},
\bauthor{\bsnm{Creely}, \binits{A.J.}},
\bauthor{\bsnm{Greenwald}, \binits{M.J.}},
\bauthor{\bsnm{Howard}, \binits{N.T.}},
\bauthor{\bsnm{Sciortino}, \binits{F.}},
\bauthor{\bsnm{Wright}, \binits{J.C.}}:
\batitle{{VITALS}: {A} surrogate-based optimization framework for the
  accelerated validation of plasma transport codes}.
\bjtitle{Fusion Science and Technology}
\bvolume{74}(\bissue{1-2}),
\bfpage{65}--\blpage{76}
(\byear{2018}).
\doiurl{10.1080/15361055.2017.1396166}.
\bcomment{Publisher: Taylor \& Francis}
\end{barticle}
\endbibitem

\bibitem{grierson_main-ion_2017}
\begin{barticle}
\bauthor{\bsnm{Grierson}, \binits{B.A.}},
\bauthor{\bsnm{Wang}, \binits{W.X.}},
\bauthor{\bsnm{Ethier}, \binits{S.}},
\bauthor{\bsnm{Staebler}, \binits{G.M.}},
\bauthor{\bsnm{Battaglia}, \binits{D.J.}},
\bauthor{\bsnm{Boedo}, \binits{J.A.}},
\bauthor{\bsnm{deGrassie}, \binits{J.S.}},
\bauthor{\bsnm{Solomon}, \binits{W.M.}}:
\batitle{Main-{Ion} {Intrinsic} {Toroidal} {Rotation} {Profile} {Driven} by
  {Residual} {Stress} {Torque} from {Ion} {Temperature} {Gradient} {Turbulence}
  in the {DIII}-{D} {Tokamak}}.
\bjtitle{Physical Review Letters}
\bvolume{118}(\bissue{1}),
\bfpage{015002}
(\byear{2017}).
\doiurl{10.1103/PhysRevLett.118.015002}.
\bcomment{Publisher: American Physical Society}.
2024-03-21
\end{barticle}
\endbibitem

\bibitem{hornsby_global_2018}
\begin{barticle}
\bauthor{\bsnm{Hornsby}, \binits{W.A.}},
\bauthor{\bsnm{Angioni}, \binits{C.}},
\bauthor{\bsnm{Lu}, \binits{Z.X.}},
\bauthor{\bsnm{Fable}, \binits{E.}},
\bauthor{\bsnm{Erofeev}, \binits{I.}},
\bauthor{\bsnm{McDermott}, \binits{R.}},
\bauthor{\bsnm{Medvedeva}, \binits{A.}},
\bauthor{\bsnm{Lebschy}, \binits{A.}},
\bauthor{\bsnm{Peeters}, \binits{A.G.}},
\bauthor{\bsnm{Team}, \binits{T.A.U.}}:
\batitle{Global gyrokinetic simulations of intrinsic rotation in {ASDEX}
  {Upgrade} {Ohmic} {L}-mode plasmas}.
\bjtitle{Nuclear Fusion}
\bvolume{58}(\bissue{5}),
\bfpage{056008}
(\byear{2018}).
\doiurl{10.1088/1741-4326/aab22f}.
\bcomment{Publisher: IOP Publishing}.
2024-03-21
\end{barticle}
\endbibitem

\bibitem{hornsby_effect_2017}
\begin{barticle}
\bauthor{\bsnm{Hornsby}, \binits{W.A.}},
\bauthor{\bsnm{Angioni}, \binits{C.}},
\bauthor{\bsnm{Fable}, \binits{E.}},
\bauthor{\bsnm{Manas}, \binits{P.}},
\bauthor{\bsnm{McDermott}, \binits{R.}},
\bauthor{\bsnm{Peeters}, \binits{A.G.}},
\bauthor{\bsnm{Barnes}, \binits{M.}},
\bauthor{\bsnm{Parra}, \binits{F.}},
\bauthor{\bsnm{Team}, \binits{T.A.U.}}:
\batitle{On the effect of neoclassical flows on intrinsic momentum in {ASDEX}
  {Upgrade} {Ohmic} {L}-mode plasmas}.
\bjtitle{Nuclear Fusion}
\bvolume{57}(\bissue{4}),
\bfpage{046008}
(\byear{2017}).
\doiurl{10.1088/1741-4326/aa5aa1}.
\bcomment{Publisher: IOP Publishing}.
2024-03-21
\end{barticle}
\endbibitem

\bibitem{Honda2018}
\begin{barticle}
\bauthor{\bsnm{Honda}, \binits{M.}}:
\batitle{Application of genetic algorithms to modelings of fusion plasma
  physics}.
\bjtitle{Computer Physics Communications}
\bvolume{231},
\bfpage{94}--\blpage{106}
(\byear{2018}).
\doiurl{10.1016/j.cpc.2018.04.025}.
\bcomment{Publisher: Elsevier B.V.}
\end{barticle}
\endbibitem

\bibitem{siena_predictions_2023}
\begin{barticle}
\bauthor{\bsnm{Siena}, \binits{A.D.}},
\bauthor{\bsnm{Rodriguez-Fernandez}, \binits{P.}},
\bauthor{\bsnm{Howard}, \binits{N.T.}},
\bauthor{\bsnm{Navarro}, \binits{A.B.}},
\bauthor{\bsnm{Bilato}, \binits{R.}},
\bauthor{\bsnm{Grler}, \binits{T.}},
\bauthor{\bsnm{Poli}, \binits{E.}},
\bauthor{\bsnm{Merlo}, \binits{G.}},
\bauthor{\bsnm{Wright}, \binits{J.}},
\bauthor{\bsnm{Greenwald}, \binits{M.}},
\bauthor{\bsnm{Jenko}, \binits{F.}}:
\batitle{Predictions of improved confinement in {SPARC} via energetic particle
  turbulence stabilization}.
\bjtitle{Nuclear Fusion}
\bvolume{63}(\bissue{3}),
\bfpage{036003}
(\byear{2023}).
\doiurl{10.1088/1741-4326/acb1c7}.
\bcomment{Publisher: IOP Publishing}.
2023-02-01
\end{barticle}
\endbibitem

\bibitem{mitim}
\begin{botherref}
\oauthor{\bsnm{Rodriguez-Fernandez}, \binits{P.}}, et al.:
MITIM: a toolbox for modeling tasks in plasma physics and fusion energy.
\url{https://github.com/pabloprf/MITIM-fusion}.
Version 1.1
(2024).
\url{https://mitim-fusion.readthedocs.io/en/latest/}
\end{botherref}
\endbibitem

\end{thebibliography}

\end{document}